\documentclass[12pt]{article}

\usepackage[english]{babel}
\usepackage[usenames]{color}
\usepackage[cp1250]{inputenc}
\usepackage{amsfonts}
\usepackage{amsthm}
\usepackage[bitstream-charter]{mathdesign}
\usepackage[T1]{fontenc}
\usepackage{graphicx}
\usepackage{epsfig}

\theoremstyle{plain}
\newtheorem{df}{Definition}

\newtheorem{tw}[df]{Theorem}

\newtheorem{sps}[df]{Observation}

\newtheorem{hyp}{Hypothesis}

\begin{document}
\newcommand{\bea}{\begin{eqnarray}}
\newcommand{\eea}{\end{eqnarray}}
\newcommand{\be}{\begin{equation}}
\newcommand{\ee}{\end{equation}}
\newcommand{\beas}{\begin{eqnarray*}}
\newcommand{\eeas}{\end{eqnarray*}}
\newcommand{\bs}{\backslash}
\newcommand{\bc}{\begin{center}}
\newcommand{\ec}{\end{center}}

\title{From Maximal Entropy Random Walk \\ to quantum thermodynamics}
\author{Jarek Duda}

\date{\it \footnotesize Jagiellonian University, Cracow, Poland, \\
\textit{email:} dudaj@interia.pl}
\maketitle

\begin{abstract}
There are mainly used two basic approaches for probabilistic modeling of motion: stochastic in which the object literally makes succeeding random decisions using arbitrarily chosen by us probabilities or ergodic in which we usually assume some chaotic classical evolution and probabilities appear while averaging over infinite trajectories. Both approaches assume we know the exact way the system evolves.

In contrast, in this paper we will focus on thermodynamical motion models: assuming maximal uncertainty. Specifically, in the space of possible choices of transition probabilities, we take the optimizing entropy or free energy one. Equivalent condition appears to be calculating transition probabilities as proportions between single steps in canonical ensemble of trajectories going through a given point. It makes that these probabilities depend on the whole space - the walker cannot directly use them. This model is thermodynamical: only we use it to predict the most probable behavior.
Standard diffusion models like Brownian motion can be seen as obtained by locally maximizing uncertainty. For regular space it agrees with fully maximizing entropy choice of transition probabilities, but generally while local approximation leads to nearly uniform stationary probability, presented approach has strong localization property. Specifically, its stationary probability density is the square of coordinates of the minimal energy eigenvector/eigenfunction of Hamiltonian for given situation, like Bose-Hubbard or Schrödinger - finally getting agreement with thermodynamical predictions of quantum mechanics. It also provides natural intuition about the squares relating amplitudes and probabilities.

We will mainly focus on deep understanding of the discrete case, which is mathematically simpler: the space is a graph and the question is how to assign probabilities to its edges. The basic Maximal Entropy Random Walk choice will be derived and discussed in general form - including asymmetric graphs, multi-edge graphs, periodic graphs and various transition times.

Later it will be first expanded to emphasize some paths by using potentials and then after making infinitesimal limit we will get the Schrödinger's case. Considering time dependent potential will lead to similar as in quantum mechanics probability current, or thermodynamical analogues of Ehrenfest equation, momentum operator and Heisenberg principle. Then we will naturally generalize to multiple particle case by considering ensembles of histories of configurations instead of trajectories. We will first focus on fixed number of particles and then by introducing creation/annihilation operators we will get to the Bose-Hubbard Hamiltonian for various numbers of particles.
\end{abstract}
\section{Introduction}
There are mostly used two probabilistic approaches to modeling the motion. From one side there are diffusion/stochastic approaches in which we assume that the object literally makes succeeding random decisions, accordingly to local transition probabilities we arbitrarily choose. From the other side there are classical chaos models, in which we usually assume some deterministic evolution and probability density appears on ergodic level: while averaging position over infinite trajectory. These models assume that we know and control the exact way the system evolves, while in real physics there is usually additional large number of degrees of freedom, hidden for us, which in practice can be considered only as thermal fluctuations.

Above approaches use strong assumption that we know the exact evolution model. In contrast, in thermodynamics we assume maximal uncertainty - for example if there is no base to emphasize some scenarios, we should assume uniform probability distribution among possibilities. So thermodynamics is not able to predict the exact situation, but only the most probable set of probabilistic parameters like density function. Standard application of this philosophy is the static picture - canonical ensemble of possible configurations in a single moment.

\begin{figure}[b]
    \centering
        \includegraphics{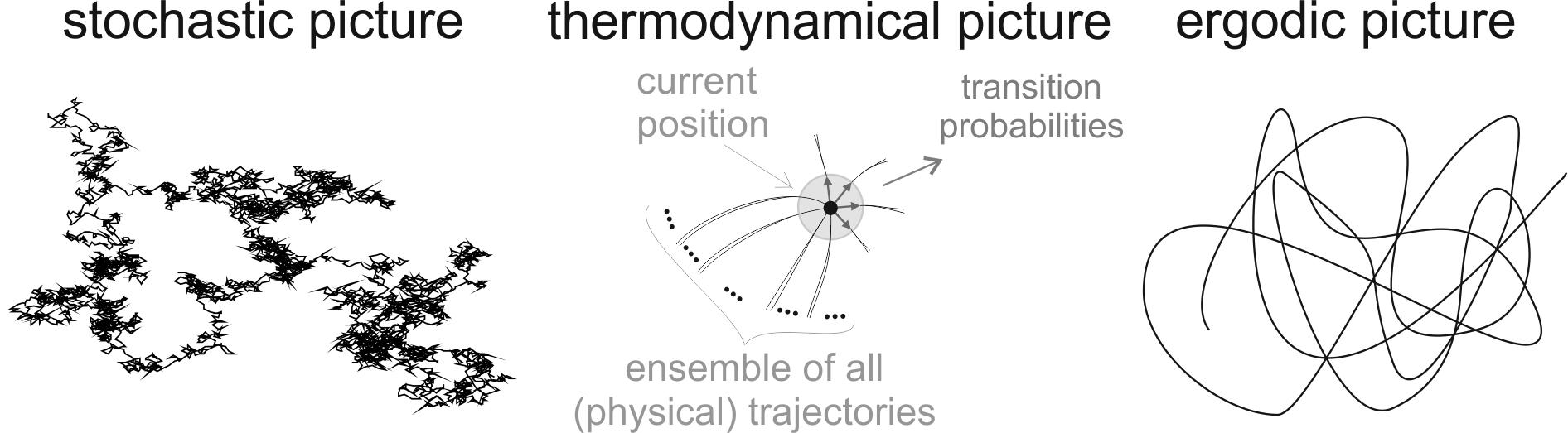}
        \caption{Different philosophies of probabilistic approaches to mmotion.}
        \label{pictures}
\end{figure}

In this paper thermodynamical approach is applied to model motion - to find the most probable probabilistic description of dynamics in situations when there is no base for strong assumptions, like for models which use diffusion or chaos approaches. Our considerations will be based on thermodynamical principles like maximizing entropy production or generally minimizing free energy. This condition appears to be equivalent to assuming canonical ensemble of possible scenarios, which this time are not static, but dynamical instead - we will assume Boltzmann distribution among dynamical scenarios, like trajectories or histories of configuration.

We base our considerations on local transition probabilities like it is in diffusion models. However, there are essential differences between values and interpretations of both approaches. This time the local probabilistic rules are not arbitrarily chosen as usually, but they are found accordingly to thermodynamical principles - as a proportion between infinitesimal steps in canonical ensemble of possible paths going through a given point, like in Fig. \ref{pictures}. Considering ensemble of whole paths requires to know the whole space - in opposite to diffusion approach, this time the object cannot have this nonlocal knowledge. Generally direct use by the object of calculated probabilities is not the essence of thermodynamical models - the latter assume that the object just chooses a trajectory in too complex or uncontrollable way, so we should assume uniform or Boltzmann distribution among possible trajectories which the object could choose. The obtained probabilities are only to be used by us to find the most probable behavior.

\begin{figure}[t]
    \centering
        \includegraphics{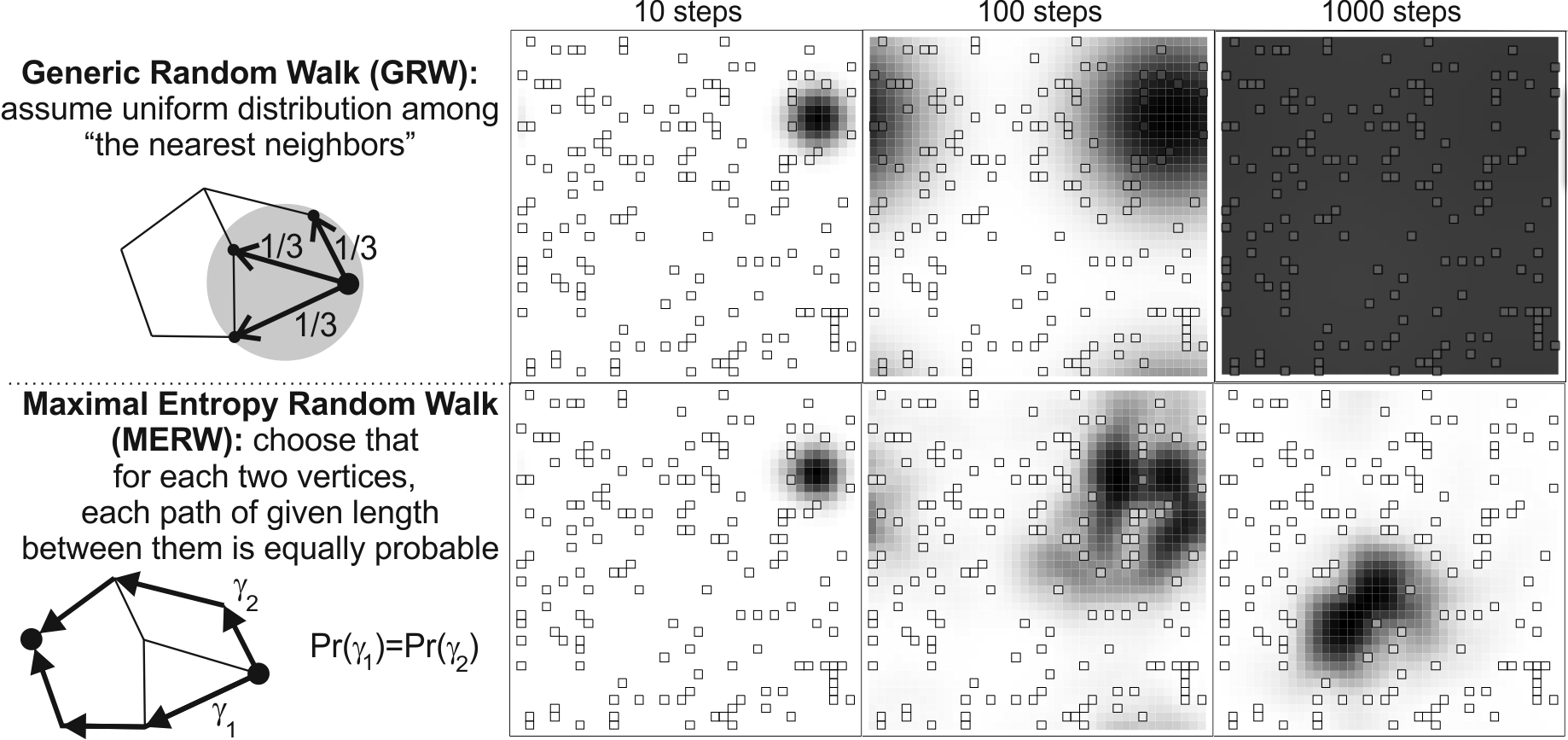}
        \caption{Two of possible ways for choosing transition probabilities on given graph and example of probability density evolution they produce for 2D lattice with cyclic boundary condition, in which all vertices but marked defects have additional self-loops (edge to itself).}
        \label{intr}
\end{figure}

We will see that the standard "static" statistical physics picture and diffusion models can be seen as local approximation of maximal uncertainty principle. In many situations, like regular space or lattice, both approaches lead to the same predictions, but irregularities make that while locally they might look similar, they usually have drastically different global behavior - for example, while diffusion leads to nearly uniform stationary density, densities in fully maximizing entropy models usually strongly localize in the largest defect-free region. Figure \ref{intr} shows example of such surprising difference for two basic models we will consider - Generic Random Walk(GRW) as a representant of standard approach locally maximizing uncertainty (leading to Brownian motion in infinitesimal limit) and Maximal Entropy Random Walk(MERW) as the basis of all thermodynamical motion models we will consider.

The natural question is: which approach better corresponds to the reality? If theoretical reasoning is not convincing enough, let us compare this huge difference in predicted thermal equilibrium with expectations of another basic tool used to model reality, namely the quantum mechanics. It predicts that a system in rest releases abundant energy and finally deexcitates to the ground state thermal equilibrium. We will see that the stationary probability densities predicted by the MERW-based models are squares of coordinates of the lowest energy eigenvector/eigenfunction of the Hamiltonian for given situation. For example, in opposite to standard approach, stationary probability density agrees with thermodynamical predictions of quantum mechanics for the Bose-Hubbard or Schr\"{o}dinger cases. In analogous experimental situation, strong localization property can be seen for example in recent STM measurements of electron density in semiconductor defected lattice \cite{frac}. The general conclusion is that if we want to get agreement between statistical physics and thermodynamical predictions of quantum mechanics, we should not use ensemble of static scenarios, but dynamical ones.\\

\begin{figure}[t]
    \centering
        \includegraphics{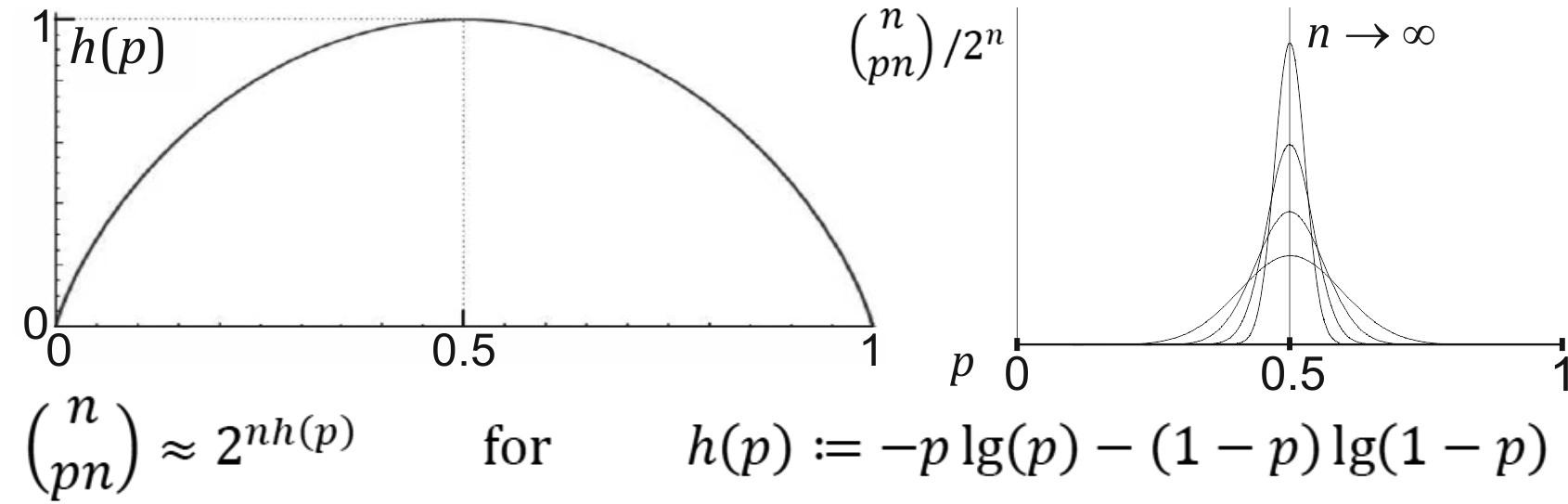}
        \caption{Left: Shannon entropy for $(p,\ 1-p)$ probability distribution ($\lg(x):=\log_2(x)$). Right: schematic distribution of subset size while restricting set of length $n$ sequences of 0/1 to having $p$ of "0" - Gaussian distribution degenerates to Dirac delta in $n\to \infty$ limit.}
        \label{entint}
\end{figure}

The base of such approaches is the maximum uncertainty principle - that when we have no additional information, we should assume uniform probability distribution among possible scenarios. If we would like to model our system using some parameterized family of statistical models, this principle translates to that we should use the maximizing entropy set of parameters. For example if there is some completely unknown length $n$ sequence of 0/1 symbols, the number of possibilities is $2^n$. Restricting to sequences such that $p\in[0,1]$ of symbols are "0", asymptotic behavior of their number is:
$${n\choose pn}\approx 2^{nh(p)} \qquad\qquad \mathrm{where}\qquad\qquad h(p):=-p\lg(p)-(1-p)\lg(1-p)$$
is Shannon's average entropy production and has single maximum: 1 (bit of information per symbol) for $p=1/2$ and we will use $\lg(x):=\log_2(x)$ notation. So if among all possible 0/1 sequences, we restrict to only those having $p$ very near 1/2, this looking generic subset in fact asymptotically contains practically all sequences. Assuming a different probability or some unjustified correlations would reduce the average entropy production, which is parameter in the exponent above - statistical model which maximize entropy asymptotically completely dominates all the others. Such universal purely combinatorial domination is much stronger than only representing our knowledge - if there are no physical reasons to emphasize some patterns, complex uncontrolled evolution should with the same probability lead to any of possible sequences. For example while counting patterns in some created by nature sequence of noninteracting objects, average number of patterns should asymptotically lead to conclusion that the sequence is uncorrelated (so called asymptotic equipartition principle). The situation becomes more complicated if there is dynamics involved - we will see that what standard approach to stochastic modeling unknowingly do, is analogue to assuming here not $p=1/2$ but an approximate value.

We will start our considerations with discrete situation, obtained for example by discretization of a continuous system, like assigning vertices to subsets of possibilities and choosing adjacency matrix describing possible transitions ($M_{ij}\in\{0,1\}$). For this graph, we would like to choose transition probabilities - for each allowed transition $(i,j)$: $M_{ij}=1$, choose a probability $S_{ij}$, normalized for each vertex ($\sum_j S_{ij}=1$). Obviously there is large freedom in choice of this matrix $S$. Standard approach maximizes uncertainty locally by assuming that for each vertex, each outgoing edge is equally probable - this choice is sometimes called "a drunken sailor", here we will call it Generic Random Walk (GRW). In infinitesimal limit it leads to the Brownian motion. It can be seen that for each vertex, we maximize entropy production for the next choice. However, it appears that this local approximation does not maximize average entropy production $H(S):=-\sum_i \pi_i \sum_j S_{ij}\lg(S_{ij})$, where $\sum_i \pi_i S_{ij}=\pi_j$ is the stationary probability distribution which this stochastic process leads to. $H(S)$ can be seen as average entropy per step in ensemble of paths produced by this choice of transition probabilities. So maximizing $H(S)$ in the space of all possible $S$ for a given graph denotes choosing probabilities such that all possible paths on this graph become equally probable. We will see that, like in Fig. \ref{pictures}, we can find this $S$ also by direct calculation of proportions of single steps inside uniform ensemble of full paths - infinite in both directions. Such choice of $S$ will be called Maximal Entropy Random Walk (MERW) and it can be determined for example by condition that for each two points, each path of given length between them is equally probable.

So while we should use GRW only if the walker indeed uses exactly given transition probabilities, MERW should be used (by us only) if there is no base to assume any local probabilistic rules. There are obvious cases that it is not always true, like if the walker indeed throw a dice in each intersection in order to use GRW directly. Generally this "no contraindications" condition is extremely subtle and there are rather no simple rules to answer if there are no hidden local probabilistic rules involved. One suggestion when to use maximal uncertainty is to compare its results with predictions of other theories, like the mentioned agreement with thermodynamical equilibrium of quantum mechanics suggests to use it for quantum scale objects. Another criterion can be using that while GRW emphasizes a concrete discrete distance to the neighboring vertices, we will see that MERW can be derived as its scale-invariant limit in which this characteristic length goes to infinity. So if the walker is a person, he among other thinks in category of single discrete choices, suggesting to shift toward GRW-like local models. From the other side, an example is provided by an electron in a crystal lattice - it behaves mainly accordingly to electromagnetic field generated by all atoms, so even if there is a discrete lattice there, the system remains deeply continuous, suggesting to use the MERW-like approach. Of course there remains a large spectrum of possibilities between these extremal choices, for example we could maximize entropy under some local probabilistic constrains to model some concrete situation.

We need to have in mind that assuming such transition probabilities does not mean that the walker directly uses them - it could even choose the path in some deterministic way. This model is thermodynamical - only represents our knowledge to predict the most probable evolution accordingly to information we have.\\

Abstract ensembles of four-dimensional scenarios also bring natural intuition about Born rule: the squares relating amplitudes and probabilities while focusing on constant-time cut of such ensemble. In given moment, there meets past and future half-paths of abstract scenarios we consider - we will see that the lowest energy eigenvector of Hamiltonian (amplitude) is the probability density on the end of separate one of these past or future ensembles of half-paths. Now the probability of being in given point in that moment is probability of reaching it from the past ensemble, multiplied by the same value for the ensemble of future scenarios we consider - is the square of amplitude.

In physics, uniform distribution among scenarios is usually replaced by Boltzmann distribution - there will be introduced potential to the graph for this purpose. Thanks of it, while taking infinitesimal limit of graphs being regular lattices, the Hamiltonian becomes the standard from Schr\"{o}dinger's equation. So the model for example says that from purely thermodynamical point of view, while considering corpuscular electron in proton's potential, the best assumption is dynamical equilibrium state having probability density of the quantum ground state.

This consequence of assuming only canonical ensemble of possible trajectories rightly bring in mind Feynmann's euclidean path integrals (\cite{pathint}). While they are mathematically very similar, there are also differences. One of them is the philosophy behind - they are imagined  as obtained by assuming axioms of quantum mechanics and then making philosophically problematic Wick rotation of time into the imaginary axis. From the other side, the presented approach uses only mathematically universal principles of thermodynamics - does not assume axioms of quantum mechanics, but derive their thermodynamical consequences. Another difference from path integral approach is that these considerations start with continuous physics, while here we rather focus on the discrete case, what allows for additional intuitions and understanding of mathematical nuances. There is also essential mathematical difference between propagators of these approaches - the one from eucliedean path integral is not properly normalized to be stochastic propagator. In presented approach there appears required additional term ($\psi_0(y)/\psi_0(x)$) carrying nonloacality of this effective model: depending on the ground state eigenfunction, which depends on the information about the whole system. Besides nonlocality, there appears also other looking problematic effects from quantum mechanics, like retrocausality in recently confirmed (\cite{whe}) Wheeler's experiment. We need to remember that these models are effective - only represent our knowledge and so we cannot imply that such effects came directly from the underlying physics. Nonlocality/retrocausality of a model representing our knowledge denotes only that some near experience may bring information we were missing about some distant/past situation.

Different concept which might seem connected is Nelson's stochastic interpretation of quantum mechanics (\cite{nels}). I would like to distinct considered here models from such ambitious approaches to recreating the whole quantum mechanics. The goal of this paper is only to improve stochastic modeling by not arbitrarily choosing transition probabilities as usual, but finding them accordingly to thermodynamical principles instead. Resulted models are in agreement with predictions of quantum mechanics only when their areas of focus intersect (thermal equilibrium), but generally there are essential differences between them, for example deexcitation is continuous process here. Mathematically closer is so called "Euclidean quantum mechanics" of Zambrini (\cite{zamb}) - there can be found similar formulas as here for single particle in time independent continuous case. There are also essential differences, mainly similar to Nelson's interpretation, motivation is resemblance to quantum mechanics and that instead of standard evolution there is used so called Bernstein process: situation in both past and future (simultaneously) is used to find the current probability density.\\

The disagreement of standard stochastic models (approximating thermodynamical principles) is one of many reasons of reluctance for imagining electron as a particle - undividable charge carrier, of radius so small that it is practically unmeasurable in particle colliders. Orthodox view on quantum mechanics leads to that physicists often try to forget about this half of wave-particle duality. The need for seeing electron as only a wave does not longer apply to macroscopic physics - for example in defected lattice of semiconductor or optical lattice, there is some concrete spatial density of particles - we should be able to imagine electrons or atoms hopping between sites like in Bose-Hubbard model. And so there should be also some stochastic description of such hopping, finally naturally appearing using presented dynamic thermodynamical approach.

While there are constructed and conducted experiments requiring simultaneous presence of both natures (e.g. Afshar experiment), orthodox view on wave-particle duality is that the particle has just one of these natures in given moment. However, there are only vague conditions which one exactly, like that electron is a wave near nucleus or while traveling through an unknown path. It is corpuscle if we know something about this path to prevent interference. Even more difficult would be the question of mechanism of changing this nature in continuous physics. Much less problematic view was started by de Broglie (\cite{debr}) in his doctoral paper: that with particle's energy ($E=mc^2$), there should come some internal periodic process ($E=\hbar \omega$) and so periodically created waves around - adding wave nature to this particle, so that it has simultaneously both of them. Such internal clock is also expected by Dirac equation as Zitterbewegung (trembling motion). Recently it was observed by Gouanere (\cite{clock}) as increased absorbtion of 81MeV electrons, while this "clock" synchronizes with regular structure of the barrier. Similar interpretation of wave-particle duality (using external clock instead), was recently used by group of Couder to simulate quantum phenomena with macroscopic classical objects: droplets on vibrating liquid surface. The fact that they are coupled with waves they create, allowed to observe interference(\cite{cou1}) in statistical pattern of double slit experiment, analogue of tunneling(\cite{cou2}): that behavior depends in complicated way on the history stored in the field and finally quantization of orbits (\cite{cou3}) - that to find a resonance with the field, while making an orbit, the clock needs to make an integer number of periods.

Like for tunneling in Couder's paper, such waves works also as practically unpredictable for us fundamental noise - thermodynamical models are used to handle such situations. The proper constant to get Schr\"{o}dinger equation from MERW was obtained by the choice of proportion between time and space lattice steps while the infinitesimal limit - not requiring some specific thermodynamical beta. It could be misleading, but similarity to quantum formalism for time dependent considerations, suggests to choose $\beta=1/\hbar$. In thermodynamics $\beta$ is related to temperature ($T$): $\beta=\frac{1}{k_B T}$. In standard view this temperature describes average energy of microscopic degrees of freedom as $\frac{1}{2}k_BT$. In our case, it is not standard energy, but energy of path (action): multiplied by time. For example if we choose this time as period ($1/\nu$) of some periodic process like internal clock, we get average energy as $\frac{1}{2}\hbar\nu$ - the level of uncertainty provided by the wave nature of particles. Surprising observation is that while these thermodynamical models completely ignore the wave nature which seems to be required for orbit quantization condition, they already "see" the structure of eigenstates.\\

The basic MERW formulas were known at least since 1984 to generate uniform path distribution required in Monte Carlo simulations (\cite{mocar}). However, using them for just stochastic modeling seems to appear in recent years (\cite{me}, \cite{loc}, \cite{varfac}). A simplified derivation for basic expansions: adding potential and making infinitesimal limit to get Schr\"{o}dinger equation, can be found in \cite{me}. Some discussion about its connection with quantum mechanics can be found in \cite{me1}. In present paper the considerations are made in more formal way and there are discussed some generalizations - for multi-edge graphs, directed graphs, periodic graphs, various transition times, time dependent case and multiple particles.

The second section contains preliminary definitions for graphs, stochastic models on them and the Frobenius-Perron theorem with discussion of periodic graphs. It also introduces to convenient interpretation of multi-edge and weighted graphs, in which the number of paths can be defined in two ways, called paths or pathways for distinction.

Section 3 concentrates on the basic MERW and its comparison with GRW. It contains two different derivations of MERW - as scale-invariant limit of GRW and by assuming uniform probability distribution among possible paths. There will be discussed combinatorial entropy, especially from the point of view of random walks. A convenient way to see the essential difference of behavior of these two approaches to random walk, is through their localization properties - there are presented and discussed numerical simulations for defected lattices. These examples also introduces potential in combinatorial way, mainly to prepare for more physical way in succeeding sections. To make this section purely combinatorial, it is the only one which uses multi-edge interpretation of weighted-graphs.

In comparison, section 4 introduces more physical interpretation of weighted graph, which will be also used in later sections: as assuming Boltzmann distribution among possible paths. It considers lattice graphs with physical potential to make infinitesimal limit, deriving deexcitation to the ground state probability density of the Schr\"{o}dinger equation.

Section 5 generalize these considerations to time-dependent case. It starts with discrete ones: using time-dependent eigenvector analogues and then there is discussed infinitesimal limit. While rapid potential changes, there appears difference between past and future amplitudes. Like in stationary case, these amplitudes should be nearly equal while relatively slow evolution, maintaining thermal equilibrium - we will call such assumption as adiabatic approximation. Time evolution allows to define thermodynamical analogue of the momentum operator ($\hbar\nabla$), which is not self-adjoined this time. While considering Ehrenfest equations, there appeared very surprising result - that we get second Newton's law, but with opposite acceleration. Fortunately, it appears to be natural in thermodynamical case: if probability density needs to get to a different potential minimum, it first has to accelerate uphill the potential, than decelerate downhill to finally stop in this new global minimum equilibrium state. In adiabatic approximation we can also introduce analogue of Heisenberg uncertainty principle.

While previously there were considered single particle in the space, in section 6 there are discussed generalizations to multiple particles. Assuming approximation that these particles does not interact with each other, obtained probability density is also expected actual density of such large number of particles. Interaction appears in analogue way as in quantum mechanics. The fact that amplitudes are real and positive now, make that we cannot make antisymmetrization to directly include Pauli exclusion principle. However, Coulomb repelling itself is enough to forbid particles to choose the same state of dynamical equilibrium. There is also presented combinatorial view on annihilation/creation operators to finally recreate the Bose-Hubbard model. Taking infinitesimal limit should lead to quantum field theory analogues as further perspective.

The last section briefly concludes the results and suggests ways for further development. While quantum mechanics focuses on wave nature of particles practically ignoring corpuscular one, presented approach do exactly oppositely - there will be also briefly discussed a way to combine both pictures using soliton particle models with topological charges as quantum numbers.

\section{Preliminaries}
\subsection{Basic definitions and properties of graphs}
We will start our considerations with the general discrete case: the walker makes succeeding transitions on some discrete set of locations. Generally this set could be infinite like for a lattice, but for simplicity let us assume that it is finite, like a part of lattice with cyclic boundary conditions. Time required for different transitions generally could be various, but for simplicity let us assume for this moment that it is constant, so we can describe time as the set of integer numbers ($t\in \mathbb{Z}$).\\

Let us assume that we have a graph $(\mathcal{V}, \mathcal{E})$ with some finite number of vertices $\mathcal{V}: \#\mathcal{V}=N\in\mathbb{N}$ identified by their number and some set of edges $\mathcal{E}\in\{1,2,..,N\}^2$. Generally we will allow to put real positive weights on these edges - natural numbers can represent multiple edges between given vertices. Later there will be introduced potential of vertices by using edge weights like $e^{-V_i}$.

In any case, we will identify the graph with real positive $N\times N$ matrix $M$. \emph{Adjacency matrix} of graph $M$ is defined as:
\be A_{ij}:=\left\{ \begin{array}{l}
                 0  \qquad\textrm{if }M_{ij}=0\qquad ((i,j)\notin \mathcal{E};\ \textrm{there is no edge from }i \textrm{ to }j)\\
                 1  \qquad\textrm{if }M_{ij}>0\qquad ((i,j)\in \mathcal{E};\ \textrm{there is an edge from }i \textrm{ to }j).
               \end{array}
               \right.
\ee
We will generally distinguish three types of graphs:
\begin{itemize}
  \item \emph{simple graphs} for which there can be only single edge between vertices:\\ $A_{ij}=M_{ij}\in \{0,1\}$,
  \item \emph{multi-edge graphs} for which there are also allowed multiple edges between two vertices: $M_{ij}\in\mathbb{N}$,
  \item \emph{weighted graphs} for which $0\leq M_{ij}\in\mathbb{R}$.
\end{itemize}
Mathematical formalism will be general, so this distinction has practically only interpretational meaning. Weights being natural numbers can be seen as the number of edges, but we will see that general weights can also be imagined this way.

Transition from $i$ to $j$ vertex in multi-edge graphs can be made through one of $M_{ij}$ edges - edge $(i,j)$ corresponds to $M_{ij}$ ways of transiting through it. To handle with such situations, we will distinguish \emph{paths} made on adjacency matrix from \emph{pathways} corresponding to the number given path can be realized:
\begin{df}$\ $\\
 $(\gamma_i)_{i=0}^l$ \emph{is length }$l$ path \emph{or} pathway \emph{on graph} $M$, \emph{if} $\forall_i M_{\gamma_i\gamma_{i+1}}>0$,\\
 $(\gamma_i)_{i=0}^l$ \emph{path} contains $M_{\gamma_0\gamma_1}M_{\gamma_1\gamma_2}..M_{\gamma_{l-1}\gamma_l}$ \emph{pathways}.\\
\end{df}
\textbf{Notation:} The index range in obvious cases will be omitted.
\begin{sps} $\ $\\
$(A^l)_{ij}$ \emph{ is the number of length} $l$ \emph{paths from} $i$ \emph{to} $j$,\\
$(M^l)_{ij}$ \emph{ is the number of length} $l$ \emph{pathways from} $i$ \emph{to} $j$.
\end{sps}
For example $(M^l)_{ij}=\sum_{\gamma_2,\gamma_3,..,\gamma_l} M_{i \gamma_2}...M_{\gamma_l j}$.

For simple graphs there is no difference between path and pathway. In opposite to multi-edge graphs, for weighted graph above interpretation seems strained, but still it will lead to self-consistent mathematics.

Above definitions for length $l$ path ($(\gamma_i)_{i=0}^l$) for time from 0 to $l$ can be naturally extended to different time segments, like $[t,t+l-1]$ and also to infinite paths: \emph{one-sided} infinite to the past ($[-\infty,t]$) or to the future ($[t,\infty]$) and finally \emph{full paths} ($[-\infty,\infty]$).\\

Let us define the basic concepts for graphs:
\begin{df} $\ $\\
  \emph{Graph is called} indirected, \emph{if} $\forall_{ij}\ M_{ij}=M_{ji}$,\\
  Neighbors \emph{of vertex} $i$ \emph{are} $N(i):=\{j:M_{ij}>0\}$,\\
  Degree \emph{of vertex }$i$ is $d_i :=\sum_j M_{ij}$,\\
  $j$ is accessible from $i$, \emph{if} $\exists_l (M^l)_{ij}>0$,\\
  Distance \emph{from} $i$ \emph{to accessible} $j$ \emph{is the minimal} $l\in\mathbb{N}\ :\ (M^l)_{ij}>0$,\\
  $(\gamma_i)_{i=0}^l$ \emph{path is length} $l$ loop, \emph{if} $\gamma_0=\gamma_l$,\\
  Self-loop \emph{is length 1 loop},\\
  \emph{Graph is called} strongly connected, \emph{if for all} $i,j$, vertex $i$ \emph{is accessible from} $j$,\\
  Period \emph{of strongly connected graph is the greatest common divisor of} $\{l:(M^l)_{ii}>0\}$,\\
  $i$ \emph{and} $j$ \emph{are in the same} periodic component, \emph{if their distance is divided by period} $p$,\\
  \emph{Vector} $v$ \emph{is called} nonegative ($v\geq 0$), \emph{if} $\forall_{i} 0\leq v_i\in \mathbb{R}$,\\
  \emph{Vector} $v$ \emph{is called} positive ($v>0$), \emph{if} $\forall_{i} 0<v_i\in \mathbb{R}$,\\
  \emph{Matrix} $M$ \emph{is called} nonegative ($M\geq 0$), \emph{if} $\forall_{ij}\ 0\leq M_{ij}\in \mathbb{R}$,\\
  \emph{Matrix} $M$ \emph{is called} irreducible, \emph{if} $\exists_n \forall_{ij}\ (M^n)_{ij}>0$,\\
  \emph{Graph is called} irreducible, \emph{if is strongly connected and has period 1 or equivalently if its adjacency matrix is irreducible}.
\end{df}

Restrictions for \emph{self-loops} are not required - there can be allowed transitions from vertex directly to itself, adjacency matrix may have nonzero values on the diagonal.

We will consider general \emph{directed} graphs: in which edge can work in both or single direction, but it is worth to distinguish \emph{indirected graphs}, for which if there is transition from $i$ to $j$, there is also transition from $j$ to $i$ - the adjacency matrix is symmetric. For simplicity we will use stronger condition: that $M$ is symmetric. This symmetry simplifies the situation: among others it means that the space of paths is time symmetric, $M$ matrix is diagonalizable, Markov process will fulfill detailed balance condition.

Another important graph property is \emph{connectiveness} - that for each two vertices, there exists a path between them. Situation is simple for undirected graph - path from $i$ to $j$ means that backward path is from $j$ to $i$. If such graph is not connected, random walk would remain in maximal connected subset (connected component) - we could divide the graph into such independent connected components and consider them separately.

Situation is more complex for directed graph - path from $i$ to $j$ does not imply existence of path from $j$ to $i$. In this case there can be vertices from which the walker should finally get to a subset, from which he cannot return to the initial state. We will be mainly interested in probabilistic equilibriums, so we can forget about these transient vertices he cannot return to - their probability will quickly drop to zero. So without loss of generality, we can focus on strongly connected graphs, for example chosen as maximal strongly connected subgraph of the original graph, which is called its strongly connected component.

More complex property which also can be removed without the loss of generality is graph \emph{periodicity}: the greatest common divisor of $\{n:(M^n)_{ii}>0\}$ is called the period of vertex $i$. In strongly connected graph all vertices have the same period $p\in \mathbb{N}$, so we just talk about the period of graph - the length of each loop in this graph is a natural multiplicity of $p$. Standard example is bipartite graph - the set of vertices can be divided into two disjoined subsets, such that edges are only between these subsets (no internal edges), so all its loops have even length ($p=2$). For indirected graphs each edge can be seen as length 2 loop, so they cannot have larger period than 2, in which case it is bipartite graph.

\begin{figure}[t]
    \centering
        \includegraphics{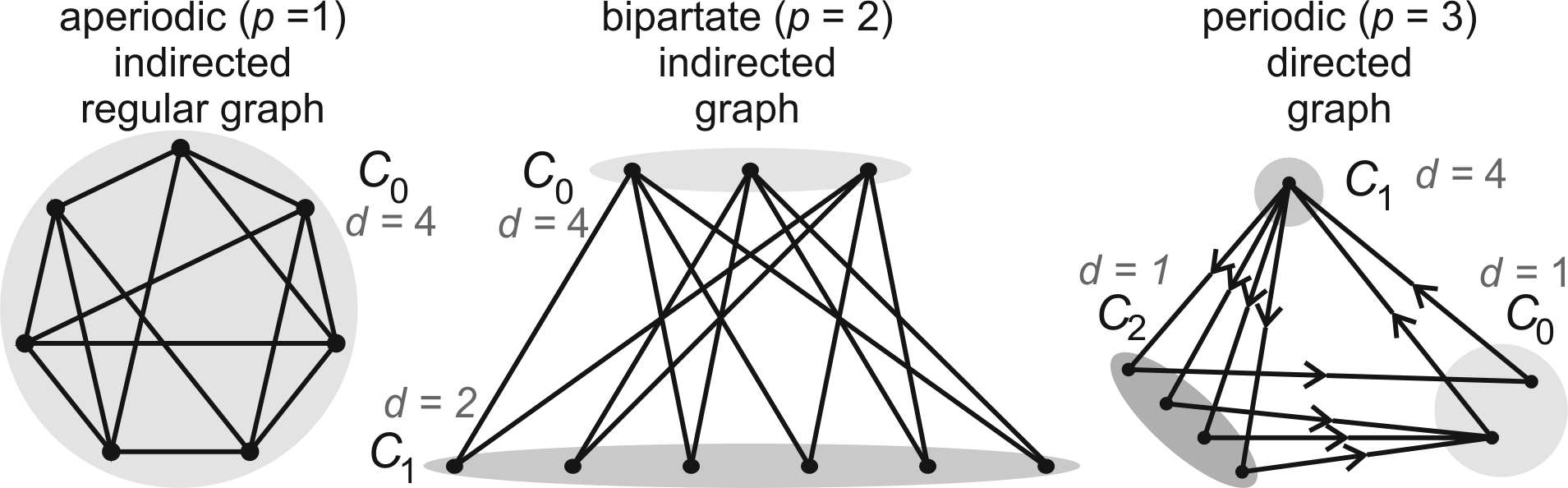}
        \caption{Some examples of graphs divided into periodic components. We will later see that constant vertex degrees ($d$) inside components will make that GRW and MERW is the same on these graphs (generally not true).}
        \label{exreg}
\end{figure}

Generally, if $p>1$ we can divide the graph into disjoined subsets (\emph{periodic components}) of the same distance modulo $p$ from any fixed vertex $v$:
\be C_i:=\{u: \exists_n (M^n)_{vu}>0 \ \wedge\ n\equiv i \textrm{ mod }p\} \label{cycpart}\ee
So while making single step from $C_i$, the walker gets to $C_{i+1 (\mathrm{mod}\ p)}$. By focusing on a single periodic component and using $M^p$ matrix instead, these components can be treated independently: we get $p$ separate multi-edge/weighted aperiodic graphs. We will use this reduction to be able to focus only on irreducible graphs. Using the original $M$ matrix later, we can connect back the behavior of these components.\\

We are now ready to remind the basic theorem for our considerations - about the dominant eigenvector of $M$. It was first proven by Perron (\cite{per}) for positive matrices and later generalized by Frobenius (\cite{frob}) for nonnegative ones. In this case, uniqueness requires that graph is strongly connected and aperiodic - fulfilling these both conditions is called irreducibility or primitiveness in literature. We will use the first name here:
\begin{tw}
  Perron-Frobenius theorem (PF): for nonnegative irreducible square matrix $M$, the dominant eigenvalue (having largest absolute value) is nondegenerated and the corresponding eigenvector can be chosen as positive.
\end{tw}
If a matrix fulfills these conditions, they are fulfilled also by its transposition, which has the same set of eigenvalues. Finally for the largest $\lambda>0$, there exist exactly one positive normalized right and left eigenvectors:
\be  M\psi = \lambda \psi \qquad\qquad\qquad \varphi^T M=\lambda \varphi^T \ee
If the matrix is symmetric, $\psi=\varphi$. For asymmetric matrices it is more convenient to use $\varphi^T\psi=1$ normalization.

The fact that the other eigenvalues have smaller absolute value, allows to use approximation:
\be M^l\approx \lambda^l \psi\varphi^T  \quad \textrm{for }l\to\infty \qquad\qquad
(\lambda^l \psi\varphi^T\cdot\psi=\lambda^l\psi\ ,\ \varphi^T\cdot\lambda^l \psi\varphi^T=\lambda^l\varphi^T)\label{eigapp}\ee

Situation for periodic graphs is more complicated. Like previously, instead of the original matrix, let us use $M^p$ first. The graph becomes aperiodic, but looses connectivity. So we can use PF theorem for its single connected components, getting unique eigenvalue $\lambda^p$ for some $\lambda>0$ and corresponding eigenvectors ($\psi^j$) on each of these subsets:
$$\forall_j\
M^p\psi^j = \lambda^p \psi^j \qquad (\varphi^j)^T M=\lambda^p (\varphi^j)^T\qquad\qquad (i\notin C_j \Rightarrow \psi^j_i=\varphi^j_i=0)$$
Any linear combination of these right/left eigenvectors would be corresponding eigenvector of $M^p$. Returning to the original $M$ determines the connection between these components: $\psi^j=\frac{M^j}{\lambda^j} \psi^0,\ (\varphi^j)^T=(\varphi^0)^T \cdot\frac{M^j}{\lambda^j}$. Now
\be\psi:=\sum_{j=0}^{p-1} \frac{M^j}{\lambda^j} \psi^0\qquad\qquad\varphi:= \sum_{j=0}^{p-1} (\varphi^0)^T \cdot\frac{M^j}{\lambda^j} \label{eigcon}\ee
are corresponding eigenvectors of $M$, for example
$$\frac{M}{\lambda}\sum_{j=0}^{p-1} \frac{M^j}{\lambda^j}\psi^0=\sum_{j=1}^{p} \frac{M^j}{\lambda^j}\psi^0=\sum_{j=0}^{p-1} \frac{M^j}{\lambda^j}\psi^0$$
Combinations (\ref{eigcon}) for $\lambda$ being different complex $p$-th root of $\lambda^p$ would be also eigenvector of $M$ to this eigenvalue - in periodic case there are $p$ dominant eigenvalues (with the same absolute value), but there is only one real positive. By writing dominant positive eigenpair, we will refer to this one.

\subsection{Markov process on a graph}
Let say we would like to model some system using a graph - divide the space of possibilities into disjoined subsets, assign a vertex to each of them and choose edges accordingly to possible transitions. For example we have a semiconductor lattice of atoms and we would like to imagine electrons jumping between such potential wells - one way to represent it as a graph could be assigning a vertex to each atom and connect it with its neighbors. We could also choose different discrimination, like into larger regions the electron could be in given moment.

We rather cannot precisely say to which region given electron will jump now - the complexity makes that the only reasonable approach seems to be some stochastic. The question is how to choose probabilities of transitions between vertices of such discretised system. Direct measurement of these probabilities is usually difficult, so let us assume that our knowledge is only the precise structure of such graph - we would like to find the most appropriate stochastic process for it.

In such situations of limited knowledge, there is used maximum uncertainty principle - among all probability distributions we could assume, the most appropriate is the one maximizing entropy. In simple words: which assume as little as possible. If we know only the graph, we rather do not have a base to assume some dependence of the history - entropy is maximized for Markov presses: in which probability of transition depends only on the vertex/state the walker is currently in. We also usually have also no base to assume that such probabilities vary with time, so we should focus on time homogeneous processes: these probabilities are chosen as time independent.

In this paper we will mainly focus on time homogeneous Markov processes. Analysis of entropy of more complicated stochastic processes on graphs can be found for example in \cite{szp}.

\begin{df} $\ $\\
$S$ \emph{matrix is called} stochastic on graph $M$, \emph{if}\ $\forall_{ij}\ 0\leq S_{ij}\leq 1$\emph{ and } $\forall_i \sum_j S_{ij}=1$ \emph{ and } $M_{ij}=0 \Rightarrow S_{ij}=0$,\\
\emph{Nonnegative vector} $p=(p_i)_{i=1}^n$ \emph{is} probability density \emph{on this graph, if} $\sum_i p_i =1$,\\
\emph{Probability density} $\pi$ \emph{is} stationary \emph{for stochastic matrix} $M$\emph{, if} $\forall_j \sum_i \pi_i S_{ij}=\pi_j$.
\end{df}

$S_{ij}$ is the probability that while being in vertex $i$, the walker will choose to jump to vertex $j$. The second condition above normalizes the probabilities and the third one restricts transitions to edges of the graph. The knowledge of the walker's position is usually incomplete, so we need to work on probability density representing our knowledge. It usually reduces while time passes and it should approach some limit - stationary probability density in given connected component, which is eigenvector of $S$ to eigenvalue 1. We would like to use PF theorem to get this uniqueness. For this purpose we will require that vertex accessibility of the stochastic matrix is the same as for the original one ($(M^k)_{ij}>0\Rightarrow(S^k)_{ij}>0$):
\begin{df}
  \emph{Stochastic matrix} $S$ \emph{on} $M$ \emph{is} nondegenerated, \emph{if} $\ \forall_{ij}\ M_{ij}>0 \Rightarrow S_{ij}>0$.
\end{df}
To handle with situation that the graph is periodic, like previously let us consider its partition into periodic components - disjoined subsets of fixed distance modulo period ($p$) from some chosen vertex (\ref{cycpart}) - probability density visits these subsets cyclically. As previously, using stochastic matrix $S^p$ instead, the walker remains in a single component - stochastic matrix restricted to such subset is aperiodic, so we can use PF theorem to get some unique stationary probability density. Let $\pi^0$ be stationary probability density on the first component ($\pi^0S^p=\pi^0)$. Now $\pi^0 S^{p+1}=\pi^0 S$ is the unique stationary probability density on the second subset and so on ($\lambda=1$). Finally $$\pi=\frac{1}{p}\sum_{i=0}^{p-1} \pi^0S^i$$ is the unique stationary probability density on the whole graph:
\begin{sps}
  Nondegenerated stochastic matrix on strongly connected graph has unique stationary probability density.
\end{sps}

\section{Derivations and properties of MERW}
Let us assume that there is a strongly connected graph and without any additional knowledge, we would like to choose a stochastic matrix on it. The standard approach is that the walker chooses where to jump with uniform probability distribution among possible single transitions. We will call this choice Generic Random Walk:
\begin{df} Generic Random Walk (GRW) \emph{on graph} $M$ \emph{is called stochastic process given by }
\be \left(S_{ij}^{\mathrm{GRW}(M)}\equiv\right)\qquad S^{\mathrm{G}}_{ij}:=\frac{M_{ij}}{d_i}\qquad\qquad\left(d_i =\sum_j M_{ij}\right) \ee
\end{df}
If the graph is default($M$), we will use an abbreviation $S^\mathrm{G}$ for GRW and $S^\mathrm{M}$ for MERW, in other case we will use full notation like above.
\begin{sps}
  For symmetric $M$ (indirected graph), stationary probability density of GRW is
  \be \left(\pi_i^{\mathrm{GRW}(M)}\equiv\right)\qquad\pi^\mathrm{G}_i=\frac{d_i}{\sum_j d_j} \ee
\end{sps}
Proof: $\sum_i d_i S_{ij}=\sum_i d_i\frac{M_{ij}}{d_i}=\sum_i M_{ij}=\sum_i M_{ji}=d_j,\ \sum_i\frac{d_i}{\sum_j d_j}=1$.

\subsection{MERW as scale invariant limit of GRW} \label{limder}
The walker in GRW makes random decisions accordingly to the knowledge about the nearest neighbors - GRW emphasizes distance corresponding to a single transition. The graph we are using could be created as discretization of a continuous system, which usually does not have such characteristic lengths - we would rather expect scale-invariant model. Here we will find such limit of GRW and call it MERW - later on we will see that it also maximizes entropy among all possible random walks on a given graph.

\begin{figure}[t]
    \centering
        \includegraphics{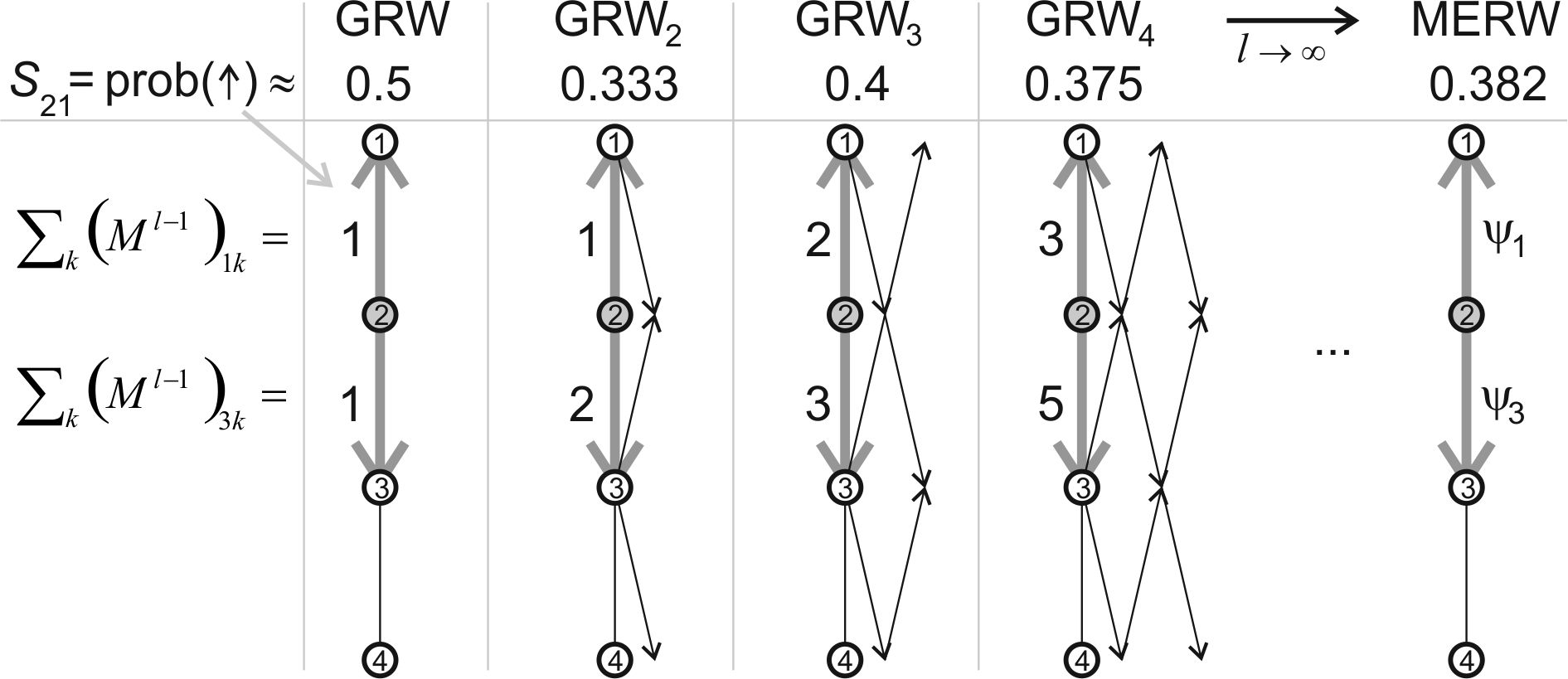}
        \caption{Example of generalizations of GRW - the number of length $l$ paths starting from given edge is written on its left. Above the graph there are written approximate probabilities of going up from vertex 2, obtained by normalization of the numbers of paths. The length $l$ paths from vertex 2 are symbolized on the right side of graphs.}
        \label{grwlim}
\end{figure}

We will start from a generalization of GRW in which instead of assuming uniform probability distribution among single edges (length 1 paths), we will choose uniform distribution among length $l$ paths and call it GRW$_l$, like in Fig. \ref{grwlim}:
\begin{df}
  $S^{\mathrm{GRW}_l(M)}_{ij}:=\frac{M_{ij}\sum_k (M^{l-1})_{jk}}{\sum_{j'}M_{ij'}\sum_k (M^{l-1})_{j'k}}\qquad$\emph{ for }$l\in \mathbb{N}^+$
\end{df}
We would like to calculate these probabilities for $l\to\infty$ limit. For this purpose we need asymptotic behavior of  $\sum_k (M^{l-1})_{jk} =(M^{l-1}\cdot (1,1,..,1)^T)_j$ for all vertices $j$. For irreducible matrix we can directly use (\ref{eigapp}):
\begin{sps} \label{onedir}
  For strongly connected aperiodic graph, the normalized number of one-sided infinite pathways from $j$ to the future (or past) is proportional to $\psi_j$($\varphi_j$):
  \be \lim_{l\to\infty} \frac{\sum_k (M^l)_{jk}}{\sum_k (M^l)_{j'k}}=\frac{\psi_j}{\psi_{j'}}\qquad\qquad\qquad
  \left(\lim_{l\to\infty} \frac{\sum_k (M^l)_{kj}}{\sum_k (M^l)_{kj'}}=\frac{\varphi_j}{\varphi_{j'}}\right) \label{owlim}\ee
  where $M\psi = \lambda \psi,\  \varphi^T M=\lambda \varphi^T$ is the dominant positive eigenpair.\\
 If the graph has period $p>1$, equation (\ref{owlim}) is fulfilled if $j$ and $j'$ are in the same periodic component ($p$ divides their distance).
\end{sps}
For periodic graph, as previously, we take $M^p$ adjacency matrix first. As long as $j$ and $j'$ are in the same periodic component, we can use equation (\ref{owlim}) for $M^p$ aperiodic matrix. This way we have shown the above limit (\ref{owlim}) for $l$ being natural multiplicities of $p$. For a general $l$, let us observe that we can write $\sum_k (M^{ap+b})_{jk}=(M^{ap}\cdot(M^b(1,1,..,1)^T))_j$, which leads to some dominant eigenvector of $M$. There are $p$ of them (formula (\ref{eigcon})), but division of their coordinates inside a single periodic component does not depend on this choice of the eigenvector.\\

Returning to the scale-free limit of GRW, all neighbors of given vertex are in the same periodic component, so we can use above observation: in the $l\to \infty$ limit, probability of jumping from vertex $i$ to vertex $j$ is proportional to $M_{ij}\psi_j$. The normalization is $\sum_j M_{ij}\psi_j=\lambda \psi_i$, so finally we obtain the stochastic matrix:
\begin{sps}
  For strongly connected graph, in the limit $l\to\infty$ of $GRW_l$ we get
  \be \left(S^{\mathrm{MERW}(M)}_{ij}\equiv\right)\qquad S^{\mathrm{M}}_{ij}=\frac{M_{ij}}{\lambda}\frac{\psi_j}{\psi_i} \label{merwform}\ee
  \be \left(\pi^{\mathrm{MERW}(M)}_i\equiv\right)\qquad\pi^{\mathrm{M}}_i=\varphi_i\psi_i\qquad\qquad(=\psi_i^2\ \ \mathrm{for\ symmetric\ }M)\ee
  where $M\psi=\lambda\psi,\ \varphi^T M=\lambda\varphi $ are the dominant positive eigenpairs, $\sum_i \varphi_i\psi_i=1$.
\end{sps}
Let us check that above $\pi^{\mathrm{M}}$ is the unique stationary probability distribution:
\be\sum_i \pi^{\mathrm{M}}_i S^{\mathrm{M}}_{ij}=\sum_i \varphi_i\psi_i \frac{M_{ij}}{\lambda}\frac{\psi_j}{\psi_i}=\frac{\psi_j}{\lambda}\sum_i \varphi_i M_{ij} = \varphi_j\psi_j=\pi^{\mathrm{M}}_j\label{check}\ee
This time we have guessed this density, but it will be derived while considering ensembles of full paths.

We can now calculate stochastic propagator: if the walker is in vertex $i$, probability that after $l$ steps it will be in vertex $j$ is
\be \left(\left(S^{\mathrm{M}}\right)^l\right)_{ij}=
\sum_{\gamma_1..\gamma_{l-1}}
\frac{M_{i\gamma_1}}{\lambda}\frac{\psi_{\gamma_1}}{\psi_{i}}\cdot\frac{M_{\gamma_1\gamma_2}}{\lambda}\frac{\psi_{\gamma_2}}{\psi_{\gamma_1}}
\cdot ... \cdot \frac{M_{\gamma_{l-1} j}}{\lambda}\frac{\psi_{j}}{\psi_{\gamma_{l-1}}}=\frac{(M^l)_{ij}}{\lambda^l}\frac{\psi_j}{\psi_i} \label{dprop}\ee
It can be imagined that there are $(M^l)_{ij}$ pathways and each of them has $\frac{1}{\lambda^l}\frac{\psi_j}{\psi_i}$ probability.
\begin{figure}[t]
    \centering
        \includegraphics{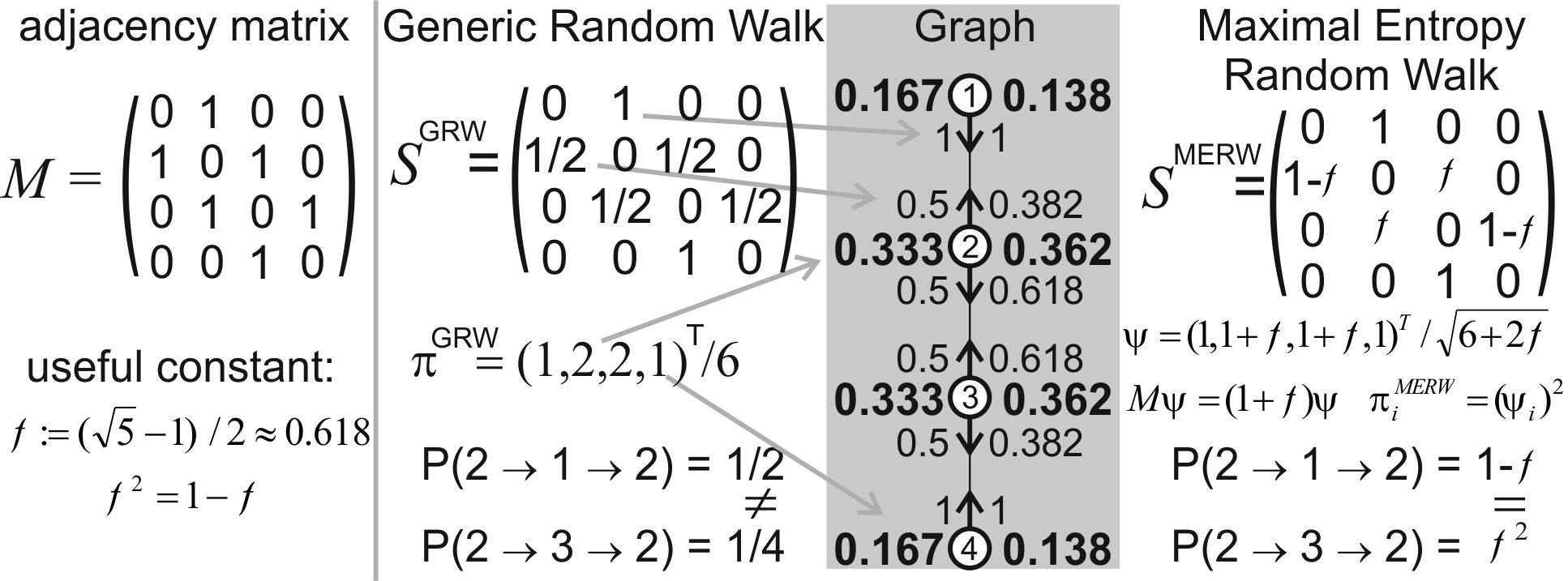}
        \caption{MERW and GRW for simple graph. Probabilities of paths $2\rightarrow 1\rightarrow 2$ and $2\rightarrow 3\rightarrow 2$ are equal in MERW and in GRW the first one is twice more probable.}
        \label{exgraph}
\end{figure}

While in GRW the walker can choose transition probabilities using only local knowledge, the $\frac{\psi_j}{\psi_i}$ term in MERW probability transition formula depends on the situation of the whole system - this effective model is nonlocal. It does not mean that the walker directly uses these nonlocal rules, but they are used only by us: to make the best predictions, we need to know the whole space of possibilities.
\subsubsection{Equally probable pathways}
Calculating MERW probability of $(\gamma_i)_{i=0}^l$ pathway, we get interesting observation that it does not depend on internal vertices:
\be S^{\mathrm{M}}_{\gamma_0\gamma_1}S^\mathrm{M}_{\gamma_1\gamma_2}..S^\mathrm{M}_{\gamma_{l-1}\gamma_l}=
\frac{M_{\gamma_0\gamma_1}}{\lambda}\frac{\psi_{\gamma_1}}{\psi_{\gamma_0}}\cdot\frac{M_{\gamma_1\gamma_2}}{\lambda}\frac{\psi_{\gamma_2}}{\psi_{\gamma_1}}
\cdot ... \cdot \frac{M_{\gamma_{l-1} \gamma_l}}{\lambda}\frac{\psi_{\gamma_l}}{\psi_{\gamma_{l-1}}}=
\frac{M_{\gamma_0\gamma_1}..M_{\gamma_{l-1}\gamma_l}}{\lambda^l}\frac{\psi_{\gamma_l}}{\psi_{\gamma_0}} \label{pathpr}\ee
For simple graph it means that for fixed length and ending points, all paths of this length between them are equally probable ($\frac{1}{\lambda^t}\frac{\psi_j}{\psi_i}$). For multi-edge (and weighted) graphs, we have to remember that they consist of many pathways and so probabilities of paths should be proportional to these numbers of pathways:
\begin{df}
Pathways $(\gamma_0,..,\gamma_l)$ and $(\gamma'_0,..,\gamma'_l)$ are equally probable if
\be \frac{S_{\gamma_0\gamma_1}S_{\gamma_1\gamma_2}..S_{\gamma_{l-1}\gamma_l}}{S_{\gamma'_0\gamma'_1}S_{\gamma'_1\gamma'_2}..S_{\gamma'_{l-1}\gamma'_l}}=
\frac{M_{\gamma_0\gamma_1}M_{\gamma_1\gamma_2}..M_{\gamma_{l-1}\gamma_l}}{M_{\gamma'_0\gamma'_1}M_{\gamma'_1\gamma'_2}..M_{\gamma'_{l-1}\gamma'_l}}
\qquad\qquad(=1\textrm{ \emph{for simple graph}})\label{eqprob}\ee
\end{df}
\begin{sps} \label{twover}
  Maximal Entropy Random Walk is the only random walk in which for any length and two vertices, each given length pathway between them are equally probable.
\end{sps}
We already know that MERW fulfills the above condition. To see that the condition (\ref{eqprob}) determines stochastic process in an unique way, for each vertex ($i$) and its two outgoing edges (to $j,j'$), we should find a vertex ($k$) and length ($l$), such that there exists two length $l$ paths between $i$ and $k$: starting with the first and with the second edge. In such case, counting corresponding pathways and using the condition (\ref{eqprob}), we get unique $S_{ij}/S_{ij'}$ proportion.

Let $p\geq 1$ be the period of $M$. Now $M^p$ is irreducible inside each periodic component, so some its power ($M^{np}$) is positive inside all components. Now because $j$ and $j'$ are in the same component, taking $k$ as any point in this component and $l=np+1$, we get the existence of required paths.\\

Generic Random Walk is usually different than MERW and so the condition (\ref{eqprob}) is no longer valid - GRW prefers paths through vertices of lower degrees, like in Fig. \ref{exgraph}:
\be S^{\mathrm{G}}_{\gamma_0\gamma_1}S^{\mathrm{G}}_{\gamma_1\gamma_2}..S^{\mathrm{G}}_{\gamma_{l-1}\gamma_l}=
\frac{M_{\gamma_0\gamma_1}..M_{\gamma_{l-1}\gamma_l}}{d_{\gamma_0}d_{\gamma_1}...d_{\gamma_{l-1}}}\ee

\subsubsection{Renormalization}
\begin{figure}[t]
    \centering
        \includegraphics{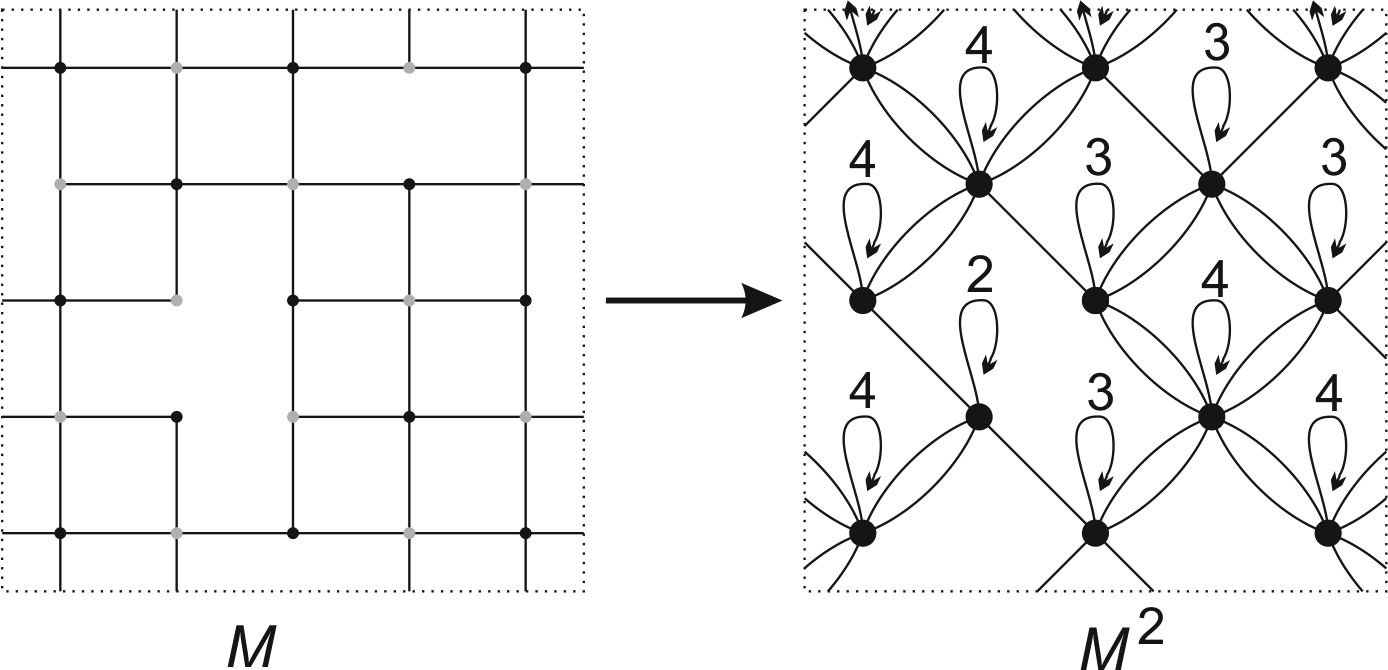}
        \caption{Renormalization of some defected lattice graph - the original simple graph is transformed into corresponding multi-edge
        graphs on sublattices of $\sqrt{2}$ times larger constant. Above self-loops there is written the number of them.
        In opposite to GRW, MERW is consistent with such change of discretization scale:
        $S^{\mathrm{MERW}(M^l)}=\left(S^{\mathrm{MERW}(M)}\right)^l$.}
        \label{renormal}
\end{figure}
Another view on scale invariance is some freedom in choosing spatial discretisation of continuous system, like in Fig. \ref{renormal}. Transforming the graph $M$ (for example representing single transitions) into multi-edge graph $M^l$ which edges
correspond to some fixed number of transitions, should not change the stochastic model:
\be\left(\left(S^{\mathrm{MERW}(M)}\right)^l\right)_{ij}=\sum_{\gamma_2,..,\gamma_l} \frac{M_{i\gamma_2}}{\lambda}\frac{\psi_{\gamma_2}}{\psi_i}\cdot
\frac{M_{\gamma_2\gamma_3}}{\lambda}\frac{\psi_{\gamma_3}}{\psi_{\gamma_2}}\cdot ... \cdot \frac{M_{\gamma_l j}}{\lambda}\frac{\psi_j}{\psi_{\gamma_l}}=
\frac{(M^l)_{ij}}{\lambda^l}\frac{\psi_j}{\psi_i}=\left(S^{\mathrm{MERW}(M^l)}\right)_{ij}\label{reneq}\ee

For GRW analogous relation usually is not satisfied: $\left(S^{\mathrm{GRW}(M)}\right)^l$ has stationary probability density $\pi^{\mathrm{GRW}}$, while
$S^{\mathrm{GRW}(M^l)}_{ij}=\frac{(M^l)_{ij}}{\sum_{j'} (M^l)_{ij'}}$ thanks to (\ref{eigapp}) for aperiodic strongly connected graph goes
to $\frac{\lambda^l \psi_i \varphi_j}{\sum_{j'} \lambda^l \psi_i \varphi_{j'}}\propto\varphi_j$ (it can be also seen from observation \ref{onedir}),
which is usually completely different stationary probability.

\subsubsection{When GRW=MERW?}
GRW and MERW are usually different, so let us now characterize cases they are the same:
\be \forall_{ij}\ \frac{M_{ij}}{d_i}=\frac{M_{ij}}{\lambda}\frac{\psi_j}{\psi_i}\qquad\Rightarrow\qquad \forall_{i,j:M_{ij}>0}\ \ \lambda\frac{\psi_i}{d_i}= \psi_j\label{condx}\ee
For vertex $i$, this condition has to be fulfilled for all its neighbors, so $\psi_j$ has to be constant inside neighborhood of any vertex. If neighborhoods of two vertices are not disjoined, $\psi$ has to be constant in their union and so on - we can expand this set with not disjoined neighborhoods of succeeding vertices. This way we get division of all vertices into disjoined components, such that the neighborhood of each vertex is a subset of one of them. Transitions from all vertices of such single component lead to the same different component, so above construction is exactly dividing the graph into periodic components (or we get single component for strongly connected aperiodic graph).

Knowing that $\psi$ has to be constant inside periodic components, (\ref{condx}) means that vertex degrees also have to be constant inside components.
Multiplying the eigenvector by $M^p$, the coordinates are multiplied by succeeding degrees, so the eigenvalue is
$$\lambda=\sqrt[p]{\prod_{i=1}^p d_{i\mathrm{-th\ component}}}\qquad\qquad(=d\ \ \mathrm{for\ regular\ graph})$$

For symmetric $M$, constant $\sum_j M_{ij}$ means that $\sum_j M_{ji}$ is also constant inside periodic components. For directed graphs the situation can be more irregular, like in Fig. \ref{exreg}. Finally
\begin{sps}
  GRW and MERW are the same for strongly connected graph, if this:\\
  - indirected graph is regular (has constant degrees) or bipartite with constant degrees inside both periodic components,\\
  - directed graph has constant $d_i=\sum_j M_{ij}$ inside each periodic component.
\end{sps}

\subsubsection{Detailed balance condition}
The probability that the walker uses $(ij)$ edge is the probability of being in $i$ vertex multiplied by probability of using $(ij)$ edge then: it is $\pi_i S_{ij}$ normalized to 1:
$$\sum_{ij}\pi_i S_{ij}=\sum_j \pi_j =1$$
We can now look at symmetry condition for stochastic matrix:
\begin{df} \emph{Stochastic matrix} $S$ \emph{with stationary probability density} $\pi$ \emph{fulfills} detailed balance \emph{condition iff}
$$\forall_{ij}\ \ \pi_i S_{ij} = \pi_j S_{ji} $$
\end{df}
It is natural for indirected graphs:
\begin{sps}   If $M$ is symmetric, $S^{\mathrm{GRW}(M)}$ and $S^{\mathrm{MERW}(M)}$ fulfills detailed balance condition.
\end{sps}
Proof: For symmetric $M$, $$\pi_i^\mathrm{G} S^\mathrm{G}_{ij}=\frac{d_i}{\sum_{j'} d_{j'}} \frac{M_{ij}}{d_i}=
\frac{M_{ij}}{\sum_{j'} d_{j'}}=\pi_j^\mathrm{G} S^\mathrm{G}_{ji},$$
$$\pi_i^\mathrm{M} S^\mathrm{M}_{ij}=\psi_i^2 \frac{M_{ij}}{\lambda}\frac{\psi_j}{\psi_i}=\frac{M_{ij}}{\lambda} \psi_i \psi_j=
\pi_j^\mathrm{M} S^\mathrm{M}_{ji}.$$
So if $M$ is symmetric, the walker uses edges equally frequent in both directions. It usually is not true for nonsymmetric $M$, for example the walker could prefer one circulation direction in ring-like graph.\\

For nonsymmetric $M$, there appears some imbalance of probability flow in stationary situation - in analogy to electric current, we can define antisymmetric \emph{probability current} describing resultant flow:
$$I_{ij}:=\pi_i S_{ij}-\pi_j S_{ji}=-I_{ji}$$
It vanishes for symmetric $M$ and generally fulfills analogue of the first Kirchoff law (continuity equation):
$$\sum_j I_{ij}=\sum_j\pi_i S_{ij}-\sum_j\pi_j S_{ji}=\pi_i-\pi_i=0$$

\subsection{Entropy of random walks} \label{entrrw}
Entropy can be seen as the amount of information required to describe given system. Quantitatively it can be represented in many units, usually multiplied by Boltzmann constant in physics. We will use it later, but for better intuition in this section we will count entropy in bits of information. The choice of one of $2^n$ elements generally requires $n$ bits of information, so in this section we use entropy as base 2 logarithm ($\lg\equiv \log_2$) of the number of possible choices (Boltzmann formula up to multiplicative constant).

Assume there is some long sequence of 2 symbols and we know the probability of the first one: $p\in [0,1],\ \tilde{p}:=1-p$. The number of such sequences behave asymptotically:
\begin{eqnarray*}
{n \choose pn}&=& \frac{n!}{(pn)!(\tilde{p}n)!}\approx
(2\pi)^{-1/2}\frac{n^{n+1/2}e^n}{(pn)^{pn+1/2}(\tilde{p}n)^{\tilde{p}n+1/2}e^n}=\\&=&
(2\pi np\tilde{p})^{-1/2}p^{-pn}\tilde{p}^{-\tilde{p}n}=(2\pi
np\tilde{p})^{-1/2}2^{-n(p\lg p+\tilde{p} \lg{\tilde{p}})}
\end{eqnarray*}
\be h(p):=-p\lg p-\tilde{p} \lg{\tilde{p}}=\lim_{n\to\infty} \frac{\lg{n \choose {pn}}}{n} \ee
where we've used the Stirling's formula: $\lim_{n\to\infty}\frac{n!}{\sqrt{2\pi n}\left(\frac{n}{e}\right)^n}=1$.

If we do not know anything about a length $n$ sequence of two symbols, the number of such sequences is $2^n$. We see that also while assuming $p=1/2$, we get the same asymptotic - these sequences completely dominate the space of all sequences like in Fig. \ref{entint}. It is an example of maximum uncertainty principle - that if we do not know anything about probability distribution among some events, the best is to assume uniform probability distribution. Generally average entropy is the coefficient in exponent, so again assuming probability distribution maximizing entropy (uncertainty), means focusing on sequences which asymptotically dominate the rest of them - almost all sequences fulfills maximizing entropy probability distribution. It is generally called Asymptotic Equipartition Property in information theory - for more information see e.g. \cite{szp}.

Analogously for more symbols/events with $(p_i)_i$ probability distribution, average entropy per symbol is:
\be h((p_i)_i)=-\sum_i p_i \lg (p_i) \ee
where we assume $0\lg(0)=0$.

Let us take it now to a stochastic process ($S$) on a simple graph ($M_{ij}\in\{0,1\})$): if the walker is in the vertex $i$, his next step will contain  $-\sum_j S_{ij} \lg (S_{ij})$ bits of information. The walker is in the vertex $i$ in asymptotically $\pi_i$ of cases, so finally average entropy production is $H(S)=-\sum_i \pi_i \sum_j S_{ij} \lg(S_{ij})$ per step. For multi-edge graph situation is a bit more complicated: now there are $M_{ij}\in \mathbb{N}$ identical edges from $i$ to $j$ of probability $S_{ij}/M_{ij}$. So $S_{ij}\lg(S_{ij})$ term in entropy formula changes into
$$\sum_{k=1}^{M_{ij}} \frac{S_{ij}}{M_{ij}}\lg\left(\frac{S_{ij}}{M_{ij}}\right)=S_{ij}\lg\left(\frac{S_{ij}}{M_{ij}}\right)$$

\begin{df}
  Average entropy production \emph{for stochastic process} $S$ \emph{with stationary probability} $\pi$ \emph{is}
  \be H(S)=-\sum_i \pi_i \sum_j S_{ij} \lg(S'_{ij})\ee
  \emph{where for simple graph }$S':=S$\emph{ and generally } $S'_{ij}:=\frac{S_{ij}}{M_{ij}}$\ \quad \emph{(=0 for} $M_{ij}=0):$
  \be H(S)=-\sum_i \pi_i \sum_j S_{ij} \lg\left(\frac{S_{ij}}{M_{ij}}\right)=-\sum_i \pi_i \sum_j S_{ij} \lg(S_{ij})+\sum_i \pi_i \sum_j S_{ij} \lg(M_{ij}).\label{entr}\ee
\end{df}
The last formula can be mathematically used also for weighted graph with $M$ having not natural values. In this case, we will see the additional term (with $\lg M_{ij}$) as the average energy and so the whole formula as minus average free energy per step.

To show that among all stochastic processes on given graph, the Maximal Entropy Random Walk is indeed the only one maximizing this formula, let us calculate entropy for probability distribution of length $l$ pathways expected in this stochastic process:
\begin{eqnarray*}
&-\sum_{(\gamma_i)_{i=0}^l}\pi_{\gamma_0} S_{\gamma_0\gamma_1}S_{\gamma_1\gamma_2}..S_{\gamma_{l-1}\gamma_l}
\left(\lg(S'_{\gamma_0\gamma_1})+\lg(S'_{\gamma_1\gamma_2}..S'_{\gamma_{l-1}\gamma_l})\right)=\\
&=-\sum_{\gamma_0\gamma_1}\pi_{\gamma_0}S_{\gamma_0\gamma_1}\lg(S'_{\gamma_0\gamma_1})\sum_{\gamma_2..\gamma_l}
S_{\gamma_1\gamma_2}..S_{\gamma_{l-1}\gamma_l}-\\
&-\left(\sum_{\gamma_0}\pi_{\gamma_0}S_{\gamma_0\gamma_1}\right)\sum_{\gamma_1..\gamma_l}S_{\gamma_1\gamma_2}..S_{\gamma_{l-1}\gamma_l}
\lg(S'_{\gamma_1\gamma_2}..S'_{\gamma_{l-1}\gamma_l})=\\
&=H(S)-\sum_{(\gamma_i)_{i=1}^l}\pi_{\gamma_1} S_{\gamma_1\gamma_2}..S_{\gamma_{l-1}\gamma_l}
\lg(S'_{\gamma_1\gamma_2}..S'_{\gamma_{l-1}\gamma_l}))=...=\\
&2H(S)-\sum_{(\gamma_i)_{i=2}^l}\pi_{\gamma_2} S_{\gamma_2\gamma_3}..S_{\gamma_{l-1}\gamma_l}
\lg(S'_{\gamma_2\gamma_3}..S'_{\gamma_{l-1}\gamma_l}))=...= l H(S)
\end{eqnarray*}
where $S'_{ij}:=S_{ij}/M_{ij}$ to include e.g. multi-edge graphs.

We see that the average entropy production of stochastic process is exactly the entropy growth per symbol of the probability distribution of pathways it generates. Without additional constrains, the only probability distribution maximizing entropy is the uniform distribution, so average entropy production is maximized only for stochastic process generating uniform probability distribution among pathways. For finite paths we already know from observation \ref{twover}, that MERW is the only random walk having uniform probability distribution among pathways of fixed length between fixed vertices. In the next section we will see that there is also analogous condition for infinite pathways.

Let us now find the maximal average entropy production available for a given graph and check that MERW really achieves it. Assume there is some set of pathways ending in given point, such that $v_i$ of them ends in vertex $i$. Expanding this ensemble a single step in all possible ways, we get $v^T M$ vector of number of pathways. So the maximal increase of the number of pathways per step is multiplying by the dominant eigenvalue($\lambda$) - their number asymptotically grows like $\lambda^l$. Uniform distribution among them maximizes the entropy, leading to upper boundary:
\begin{sps}
  For stochastic process $S$ on graph $M$,
  \be H(S)\leq \lg(\lambda) \label{theq}\ee
  where $\lambda$ is the positive dominant eigenvalue of $M$.
\end{sps}
Let us check that MERW indeed achieves this boundary:
\begin{eqnarray*}
H(S^\mathrm{M})&=&-\sum_i \pi^\mathrm{M}_i \sum_j S^\mathrm{M}_{ij}\lg\left(\frac{S^\mathrm{M}_{ij}}{M_{ij}}\right)=\\
&=&-\sum_i \varphi_i \psi_i \sum_j \frac{M_{ij}}{\lambda}\frac{\psi_j}{\psi_i}\lg\left({\frac{1}{\lambda}\frac{\psi_j}{\psi_i}}\right)=
\frac{-1}{\lambda}\sum_{ij}\varphi_i M_{ij} \psi_j\lg\left({\frac{1}{\lambda}\frac{\psi_j}{\psi_i}}\right)=\\
&=&\frac{\varphi^TM\psi}{\lambda}\lg{\lambda}+\frac{1}{\lambda}\sum_{ij}(\varphi_i
(\lg{\psi_i}) M_{ij}\psi_j-\varphi_i M_{ij}\psi_j(\lg{\psi_j}))=\\
&=&\lg{\lambda}+\frac{1}{\lambda}\sum_i(\varphi_i (\lg{\psi_i}) \lambda \psi_i-\varphi_i
\lambda \psi_i(\lg{\psi_i}))=\lg{\lambda}
\end{eqnarray*}

The fact that random walk cannot have larger entropy leads to interesting inequalities. For example for GRW while $M$ is symmetric:
$$H(S^\mathrm{G})=-\sum_i \frac{d_i}{\sum_{k} d_{k}} \sum_j \frac{M_{ij}}{d_i} \lg\left(\frac{1}{d_i}\right)=
\frac{\sum_i d_i\lg(d_i)}{\sum_{k} d_{k}}\leq \lg(\lambda)$$
for any nonnegative matrix $M$. Assuming uniform distribution among the nearest neighbors in GRW can be seen as local maximization of entropy - for each $i$ maximize $-\sum_j S_{ij} \lg(S_{ij})$, while in MERW we maximize the average entropy production.

In \cite{varfac} there are other useful inequalities between some effective degrees of graph:
\be \min_i d_i\leq \frac{\sum_i d_i}{N}\leq \exp\left(\frac{\sum_i d_i\ln(d_i)}{\sum_{k} d_{k}}\right)\leq \lambda\leq \max_i d_i \label{inequals}\ee
In (\ref{inequals}) the first and the fourth inequalities are trivial, the third is equation (\ref{theq}). The second inequality can be derived using convexity of $F(\beta):=\ln\left(\sum_i d_i^\beta\right)$
$$\frac{\sum_i d_i\ln(d_i)}{\sum_{k} d_{k}}=F'(1)\geq F(1)-F(0)=\ln\left(\frac{\sum_i d_i}{N}\right).$$
There was not required any additional assumptions for inequality (\ref{inequals}), so it is fulfilled not only for indirected simple graph like in \cite{varfac}, but also for general multiple-edge or weighted graphs.

\subsection{MERW from the point of view of full paths and Born rules } \label{ensder}
Up to now we were considering one-sided infinite paths, now we will look at ensembles of full paths - infinite in both directions: past and future. It leads to better understood derivation of MERW formulas. We will see why considering statistical ensemble of full scenarios leads to the Born rules for constant time cuts - that to translate amplitudes into probabilities we do need to "square" them. Intuitively, amplitudes correspond to probabilities on the end of past/future half-spaces and we need to multiply them to estimate probabilities of events in a given moment.

\begin{figure}[t]
    \centering
        \includegraphics{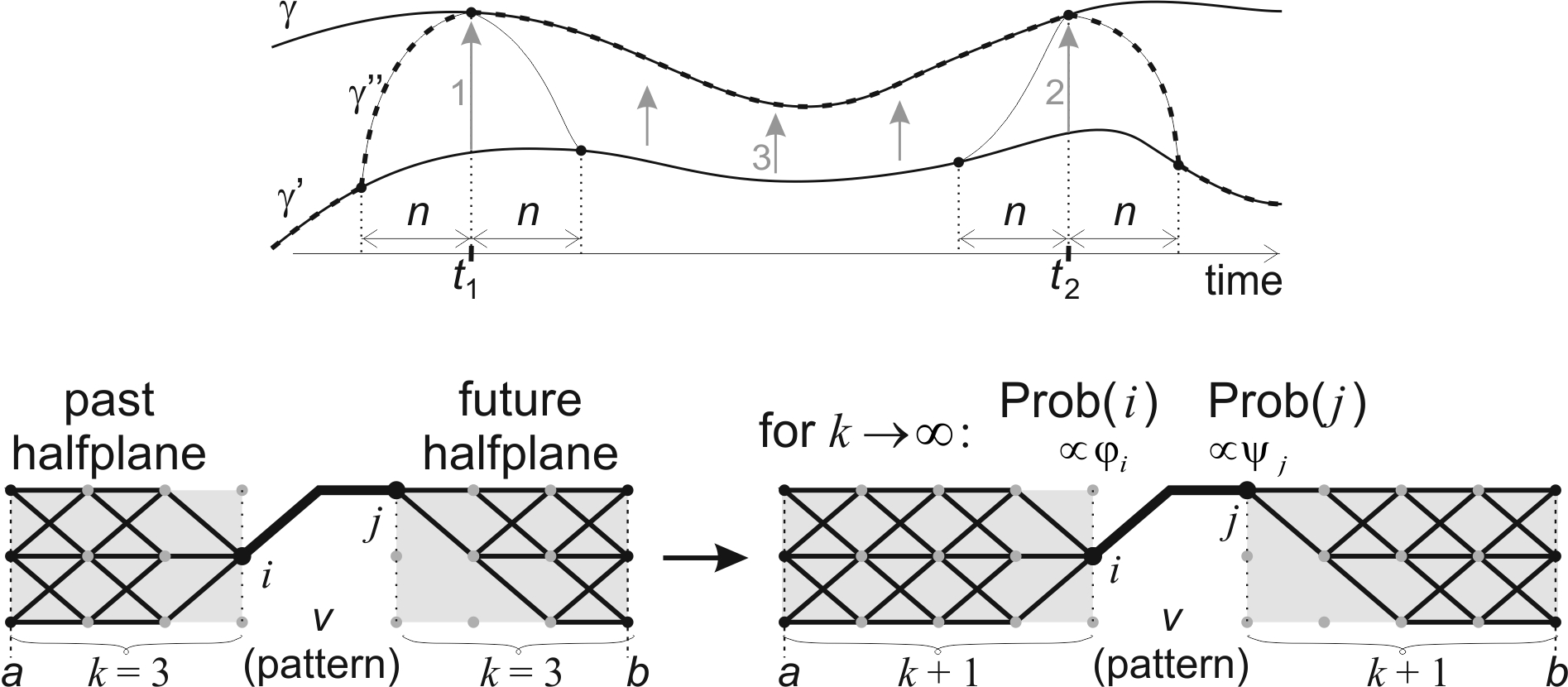}
        \caption{Top: schematic picture for Observation \ref{fullpath} of exchanging any large enough subpath using equivalence relation for finite intervals. Bottom: calculating probability of patterns from ensembles of all allowed paths on interval growing in both directions.}
        \label{bidir}
\end{figure}

Observation \ref{twover} says that MERW is the only random walk in which while fixing vertices on the ends of a time range, all pathways between them are equally probable. Let us extend it to full paths: $(\gamma_t)_{t=-\infty}^{\infty}$ and $(\gamma'_t)_{t=-\infty}^{\infty}$. If the graph is periodic, $\gamma_t$ and $\gamma'_t$ could belong to different periodic components for all $t$ - in such a case it is enough to shift indexes to synchronize them.
Let us make general observation implying that all full pathways on a strongly connected graph are intuitively equiprobable for MERW:
\begin{sps} \label{fullpath}
 Assume there is a strongly connected graph and an equivalence relation, in which subpaths are equivalent if they have the same length and ending points. For two full paths $(\gamma_t)_{t=-\infty}^{\infty}$ and $(\gamma'_t)_{t=-\infty}^{\infty}$, such that
 $\gamma_0$ and $\gamma'_0$ are in the same periodic component, for any large enough finite time interval $([t_1,t_2])$ we can find full path
 $((\gamma''_t)_{t=-\infty}^{\infty})$, which is equivalent to $(\gamma'_t)_{t=-\infty}^{\infty}$ and on the chosen time segment is equal to $(\gamma_t)_{t=-\infty}^{\infty}.$
\end{sps}
\textbf{Proof:} Let us focus first on aperiodic graph ($\gamma_0$ and $\gamma'_0$ are always in the same periodic component) - there is some $n\in \mathbb{N}$ such that $\forall_{ij} (M^n)_{ij}>0$. Let us choose any length $2n$ subpath of $(\gamma'_t)_{t=-\infty}^{\infty}$ - for its beginning point there exists a length $n$ path to any vertex and then there is always further path to the original ending point. So this subpath is equivalent to subpath having any vertex in the middle - let us choose it as the corresponding vertex of $(\gamma_t)_{t=-\infty}^{\infty}$ like in Fig. \ref{bidir}. Now doing it for two such subsets:
$[t_1-n,t_1+n]$, $[t_2-n,t_2+n]$  and using the equivalence relation third time (between their middle vertices), we get $(\gamma''_t)_{t=-\infty}^{\infty}$ as required (dashed line in Fig. \ref{bidir}).

For periodic graphs, thanks of that $\gamma_0$ and $\gamma'_0$ are in the same periodic component, we can make above construction for $M^p$.\\

We will now focus on the opposite route - assume that all full pathways are asymptotically equally probable to derive MERW formulas and get better intuition about them.
Let us imagine that everything is happening in discrete space-time: $\mathcal{V}\times \mathbb{Z}$ - the graph is our space and the time is the set of all integer numbers. We are interested in finding correlations - probabilistic dependence between situations in different times. Let us assume that there is a length $l\in\mathbb{N}$ segment of time (for example $[0,l]$) and we are interested in probability of situations on its endings. For $l=0$ it corresponds to probability distribution of events in single moment (measurement outcomes), for $l=1$ it corresponds to transition probabilities, which for Markovian process determine situation for larger $l$.

The situation looks like in Fig. \ref{bidir}: growing finite length ensembles of paths to estimate probability distribution of situations inside some  fixed length time segment. Let us choose some pattern (path $(v_t)_{t=0}^l$):
$$ (N_v^{kk'})_{ab}:=\sum_{(\gamma_t)_{t=-k}^{l+k'}\ :\ \gamma_0=v_0,\ \gamma_1=v_1,..,\gamma_l=v_l,\ \gamma_{-k}=a,\ \gamma_{l+k'}=b} M_{\gamma_{-k}\gamma_{-k+1}}M_{\gamma_{-k+1}\gamma_{-k+2}} ... M_{\gamma_{l+k'-1}\gamma_{l+k'}}$$
$$ N_v^{k+1,k'+1}=M\cdot N_v^{kk'}\cdot M \qquad \qquad\qquad\qquad  N_v^{kk'}=M^k\cdot N_v^{00} \cdot M^{k'} $$
Now for strongly connected graph, using Observation \ref{onedir} twice: to make $k\to\infty$ and $k'\to\infty$ limit for $v$ and some other pattern $(w_i)_{i=0}^l$, we get
$$\frac{\mathrm{Prob}(v)}{\mathrm{Prob}(w)}=\lim_{k,k'\to \infty} \frac{\sum_{ab}\left(N_v^{kk'}\right)_{ab}}{\sum_{ab}\left(N_w^{kk'}\right)_{ab}}=
\lim_{k,k'\to \infty} \frac{\sum_{ab}(M^k)_{av_0} M_{v_0 v_1} M_{v_1 v_2}..M_{v_{l-1} v_l} (M^{k'})_{v_l b}}
{\sum_{ab}(M^k)_{aw_0} M_{w_0 w_1} M_{w_1 w_2}..M_{w_{l-1} w_l} (M^{k'})_{w_l b}}=$$
$$=\lim_{k,k'\to \infty} \frac{\sum_a(M^k)_{av_0}}{\sum_a(M^k)_{aw_0}}\cdot
\frac{M_{v_0 v_1} M_{v_1 v_2}..M_{v_{l-1} v_l}}{M_{w_0 w_1} M_{w_1 w_2}..M_{w_{l-1} w_l}}\cdot
\frac{\sum_b (M^{k'})_{v_l b}}{\sum_b (M^{k'})_{w_l b}}=$$
\be = \frac{\varphi_{v_0}\cdot M_{v_0 v_1} M_{v_1 v_2}..M_{v_{l-1} v_l}\cdot \psi_{v_l}}
{\varphi_{w_0}\cdot M_{w_0 w_1} M_{w_1 w_2}..M_{w_{l-1} w_l}\cdot \psi_{w_l}}\label{pathprob}\ee
where $M\psi = \lambda \psi,\ \varphi^T M=\lambda \varphi^T$ are the dominant eigenvectors.\\
We see that probability of pattern $v$ is proportional to the number of pathways it contains $M_{v_0 v_1} M_{v_1 v_2}..M_{v_{l-1} v_l}$ (1 for simple graphs) and to probabilities of ending points of past/future halfplanes: $\varphi_i$ and $\psi_j$.

Previously stationary probability formula was guessed and checked (\ref{check}), now we can derive it using $l=0$ - past and future halfplanes "glues together":
\be \pi_i\propto \varphi_i\psi_i\qquad\qquad (\pi_i\propto \psi_i^2 \mathrm{\ for\ symmetric}\ M) \label{Born}\ee
For $l=1$ we get transition probabilities:
$$\mathrm{Prob}((ij))\propto \varphi_i M_{ij} \psi_j\qquad \Rightarrow \qquad S^M_{ij}\propto \frac{\mathrm{Prob}((ij))}{\pi_i}\propto M_{ij}\frac{\psi_j}{\psi_i}$$
from which using normalization condition ($\sum_j S_{ij}=1$) we get the missing $1/\lambda$ coefficient for MERW formulas - that assuming uniform probability distribution among full pathways, indeed unequally leads to MERW.

For $l>1$ we get probability of pathways as previously (\ref{pathpr}), so equiprobability assumption leads to Markovian process as expected.\\

The most interesting from this derivation seems to be clear understanding of Born formulas (\ref{Born}). In the next subsection we will see that $\psi$ corresponds to the ground state of discrete Schr\"{o}dinger's equation (or Bose-Hubbard Hamiltonian for single particle) and later of the original Schr\"{o}dinger equation after introducing potential and making infinitesimal limit.

The intuition about Born rules is that amplitudes describe probability distribution on the end of past/future halfplanes and if we want to translate them into probability distribution on constant time cuts, we need to multiply both amplitudes. Intuitively, to draw some event in given time, we have to draw it twice: from the past and from the future of abstract trajectories we consider in our ensemble. Time dependent case will bring more intuition.

\subsection{Examples and localization property}  \label{exsect}
For better intuition about MERW and its difference from GRW, we will now look at simple examples. For connection with physics, there will be used lattice-like graphs which can be e.g. imagined as crystal lattice or discretization of a continuous system. Standard lattice is regular graph, making that GRW and MERW are the same - to observe the difference we can remove the regularity by introducing some defects. We will see that in opposite to GRW, MERW has strong localization properties. Its stationary probability corresponds to quantum mechanical ground state probability distribution, for which Lifshitz argument \cite{lif} says that probability is localized in the largest defect-free spherical region. It was used to make some predictions of statistical behavior in \cite{loc} and \cite{varfac} and will be presented here briefly.

\subsubsection{One dimensional segment-like graph}
\begin{figure}[t]
    \centering
        \includegraphics{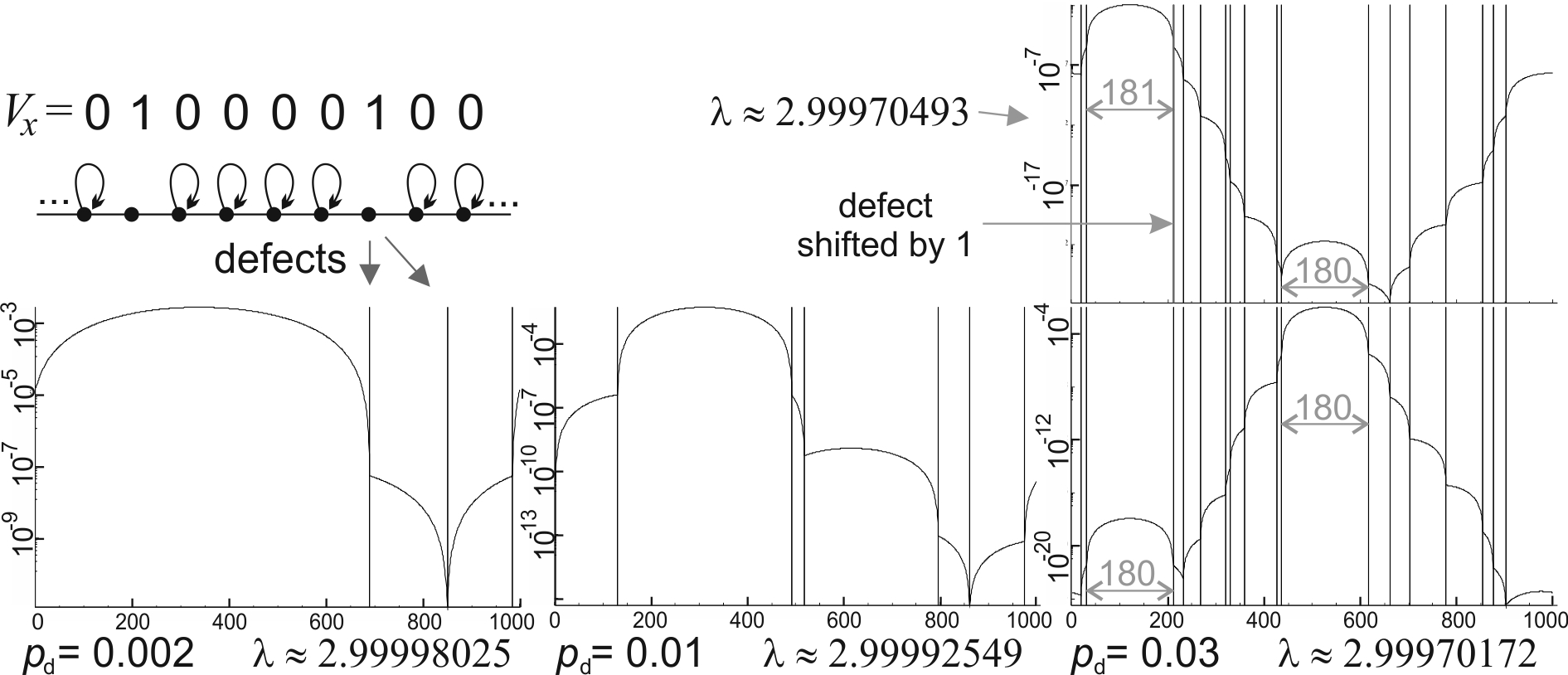}
        \caption{Stationary probability distribution in logarithmic scale for MERW on 1000 node segment-like graph with cyclic boundary conditions and self-loops from which some small portion was randomly removed: correspondingly 0.002, 0.01 and 0.03. Above the last one, there is presented completely different situation after shifting one defect a single position.}
        \label{1dex}
\end{figure}

Let us start with one-dimensional case: length $N$ segment-like graph for which we assume that in a single step the walker can jump to one of two neighboring nodes or stay in given position. The last possibility denotes that there are self-loops in vertices - we can introduce defects by removing some of them. In presented numerical simulations we will choose randomly the positions of these defects, like in Fig. \ref{1dex} - choose some probability $p_d$, which is independently used for each node as probability that it is defected. Physical intuition for such simplified model could be that there are randomly distributed two types of atoms in the lattice: most are potential well for electrons, while the defects are rather repelling. In the next section we will get freedom for choosing these potentials in more physical way.

Let us introduce the potential representing the positions of self-loops:
$$V_x=\left\{ \begin{array}{cc}
               0 \quad &\textrm{if self-loop at position }x\textrm{ is present,} \\
               1 \quad &\textrm{if self-loop at position }x\textrm{ is absent.}
             \end{array}\right.$$
The eigenequation becomes:
$$ (\lambda \psi)_x = (M \psi)_x = \psi_{x-1}+(1-V_x)\psi_x+\psi_{x+1} \qquad\qquad /-3\psi_x\qquad/\cdot -1$$
\be E\psi_x=-(\triangle\psi)_x+V_x\psi_x \label{dsch}\ee
where $(\triangle\psi)_x:=\psi_{x-1}-2\psi_x+\psi_{x+1}$ is standard discrete Laplacian and $E:=3-\lambda$ is analogue of quantum mechanical ground energy - maximizing $\lambda$ is equivalent to finding minimal eigenvalue of the found discrete analogue of Schr\"{o}dinger's operator ($-\triangle+V$). Like for the quantum ground state, stationary probability distribution is $\rho(x)=\psi^2(x)\ $ ($\psi\geq 0$).  Later we will use more physical potential and make infinitesimal limit, getting thermalization to the quantum ground state probability density of the standard continuous Schr\"{o}dinger's equation. The fact that obtained Hamiltonian is minus adjacency matrix up to linear transformation, allows to connect it also with Bose-Hubbard Hamiltonian for single particle without potential. We will look later at the general case, but for single particle the space of possibilities becomes the vertices of lattice/graph and so the Hamiltonian $-\sum_{(i,j)\in \mathcal{V}} \hat{a}^\dag_i\hat{a}_j+h.c.$ is equivalent to minus adjacency matrix.

The choice of value 3 in $E=3-\lambda$ formula was arbitrary - different choice would change the values of $V$, such that $E-V$ would remain the same. The reason for the used choice is to make that most of $V_i$ are zero and $E$ became a small positive number. For general lattice of dimension $D$ we will use for example $2D+1$ instead of 3.

The (\ref{inequals}) inequality allows to see $\lambda$ as some effective average of degrees of the graph. In our case:
$$3-p_d = \frac{\sum_i d_i}{N} \leq\lambda\leq \max_i d_i=3\qquad\qquad\qquad 0\leq E\leq p_d$$
$\lambda$ describes the optimal growth of the number of paths while elongation by a single step ($M\psi=\lambda\psi$). For paths ending in given vertex, this growth of their number is the degree of this vertex - vertices above this average ($d_i>\lambda$ or equivalently $V_i<E$) produce more paths than average. Intuitively it acts as there was attractive potential and repulsive for $V_i>E$ vertices.

For regions of constant potential larger than $E$, like in quantum mechanics the local solution of (\ref{dsch}) for such energy barrier has leading to tunneling-like exponential behavior:
$$\psi(x)=Ae^{Kx}+B^{-Kx}\quad\textrm{where}\quad K=\textrm{arccosh}\left(1+\frac{V-E}{2}\right)\approx \sqrt{V-E}+\frac{(V-E)^{3/2}}{24}$$
for some local parameters $A$ and $B$. Such situations would be natural for example in opposite model: in which most of vertices would not have self-loops.

In our case we are rather interested in the solution for regions of constant $V$ below $E$:
$$\psi(x)=A\cos(k(x-x_0))\quad\textrm{where}\quad k=\arccos\left(1-\frac{E-V}{2}\right)\approx \sqrt{E-V}-\frac{(E-V)^{3/2}}{24}$$
$A$ and $x_0$ are some local parameters, $x_0$ is the position of maximal value which is not necessarily in the region. The value of $\psi$ cannot drop below 0, so $x\in(x_0-\frac{\pi}{2k},x_0+\frac{\pi}{2k})$. For $E-V>2$, $k>\pi/2$ makes that the region of positive $\psi$ completely degenerates, so $E-V$ has to be smaller. For example in our case $E$ can be bounded from above asymptotically by $p_d$. The Lifshitz argument says that $\psi$ is approximately zero out of the largest defect-free region, like in Fig. \ref{1dex}. Let us denote the width of this Lifshitz region by $2R$. Probability of $2R$ succeeding nodes without defects behaves like $(1-p_d)^{2R}$. Such region could start in any node, so for the largest of them $N(1-p_d)^{2R}$ should be of order of unity, making $2R\approx \frac{\ln(N)}{|\ln(1-p_d)|}$. In this case $x_0$ is approximately the center of this region, $k R\approx \pi/2$, so
$$E\approx 2-2\cos\left(\frac{\pi}{2R}\right)\approx \frac{\pi^2}{4R^2}\approx \left(\frac{\pi|\ln(1-p_d)|}{\ln(N)}\right)^2$$

Stationary probability in GRW is proportional to the degree of given vertex, so here it would be just constant for most of vertices - it has practically no localization properties. We see that situation in MERW is completely different - each defect influence the whole system. The right hand graphs in Fig.  \ref{1dex} shows how strong this effect is by presenting surprising agreement with the Lifshitz argument while shifting one defect a single position. This rapid change seems nonintuitive, but it does not mean that the eigenvector changes so drastically, only that there was changed the order of eigenvalues of the first two eigenvectors.

\subsubsection{Two dimensional defected lattice}
\begin{figure}[t]
    \centering
        \includegraphics{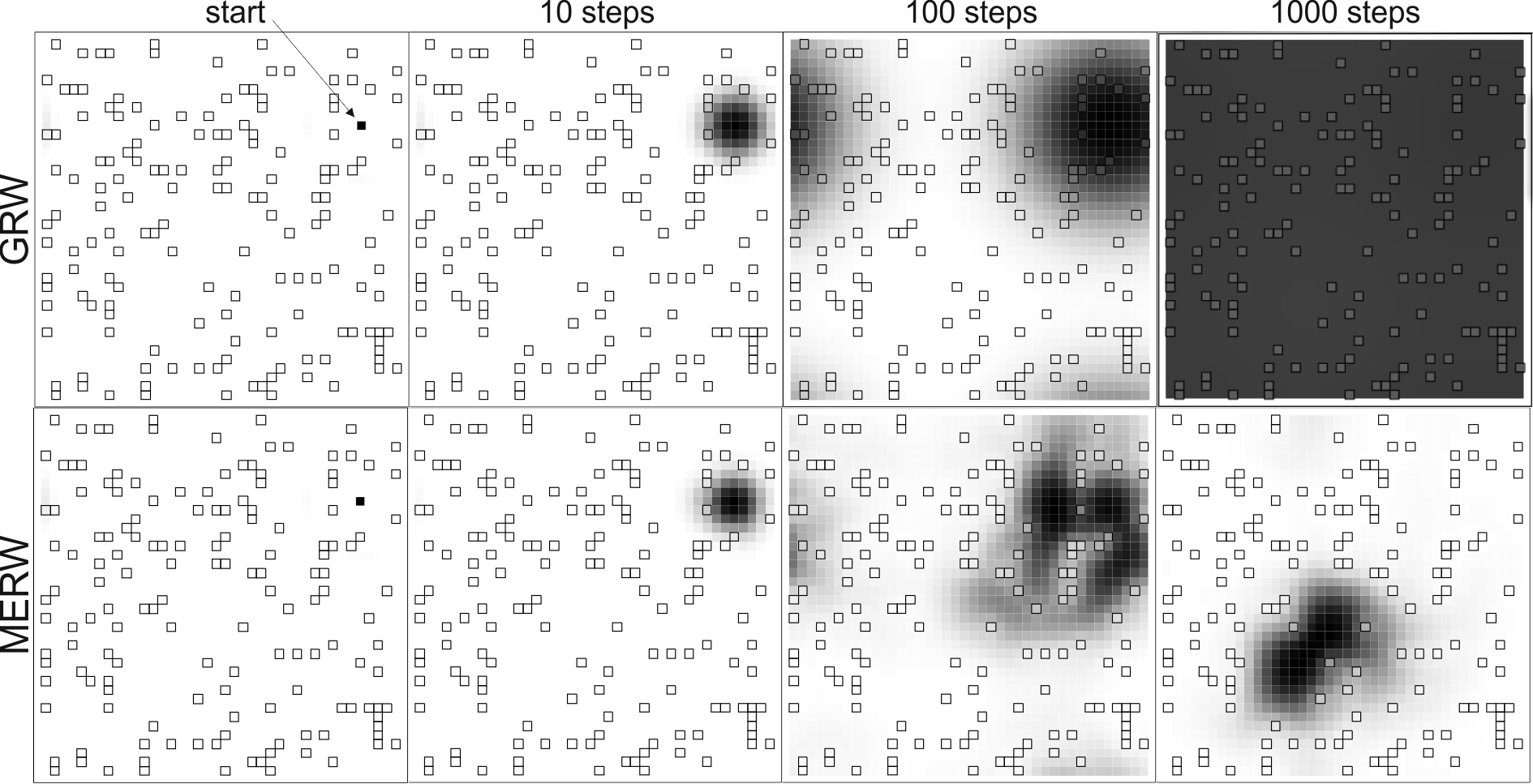}
        \caption{Comparison of evolution of probability density of GRW and MERW on $40\times 40$ lattice graph with cyclic boundary conditions and self-loops in all vertices but some randomly chosen 0.1 of them (represented by squares). The initial situation is probability localized in a single vertex (known position) and the graphs represent probability density after correspondingly 10, 100 and 1000 steps.}
        \label{2dex}
\end{figure}

Let us now look at constructed in analogous way two dimensional lattice with self-loops in all but some randomly chosen portion of vertices. The dominant eigenvector is again the ground state of the discrete Schr\"{o}dinger equation (\ref{dsch}), but using two-dimensional discrete Laplacian:
$$(\triangle\psi)_{x,y}:=\psi_{x-1,y}+\psi_{x,y-1}+\psi_{x+1,y}+\psi_{x,y+1}-4\psi_{x,y}$$
Lifshitz argument suggests that probability distribution should be localized in the largest defect-free sphere. From presented numerical results we see that situation is more complicated now, but intuitively it localizes in the largest nearly spherical defect-free region.

Two-dimensional example makes it more convenient to compare dynamics of GRW and MERW. In Fig. \ref{2dex} there is example of such comparison of evolution of probability density starting with known walker's position - probability density concentrated in a single point. We see that after ten steps for both models we can expect similar probability distributions - there is not large difference between their local behavior and transition probabilities (up to a few percent). However, while time passes the difference grows. GRW behaves like there was practically no defects and finally thermalize on nearly uniform distribution. Dynamics of MERW is much more complicated. The defects create some entropic landscape - the probability localizes first in some defect-free region of large local entropy production and soaks to finally get to the deepest entropic well. The discrete propagator (\ref{dprop}) can be written:
\be \left(S^{\mathrm{M}}\right)^t_{ij}=\frac{(M)^t_{ij}}{\lambda_0^t}\frac{\psi_{0,j}}{\psi_{0,i}} =\left(\sum_k\left(\frac{\lambda_k}{\lambda_0}\right)^t \varphi_{k,j} \psi_{k,i}\right)\frac{\psi_{0,j}}{\psi_{0,i}} \label{decom}\ee
where we have used eigenvalue decomposition of $M=\sum_k \lambda_k\cdot \varphi_k(\psi_k)^T$. Eigenvectors $\varphi_k, \psi_k$ are real and fulfill:
$$(\varphi_k)^T M=\lambda_k (\varphi_k)^T,\qquad M\psi_k=\lambda_k \psi_k,\qquad (\varphi_k)^T\cdot\psi_l=\delta_{kl},$$
$$\lambda=\lambda_0\geq \lambda_1 \geq ... \geq \lambda_{N-1},\qquad \psi=\psi_0,\qquad \varphi=\varphi_0$$
For symmetric $M$, $\varphi_k=\psi_k$. In quantum mechanics they would be stable eigenstates, while here higher states deexcite toward lower ones and finally thermalize in the ground state.

\begin{figure}[t]
    \centering
        \includegraphics{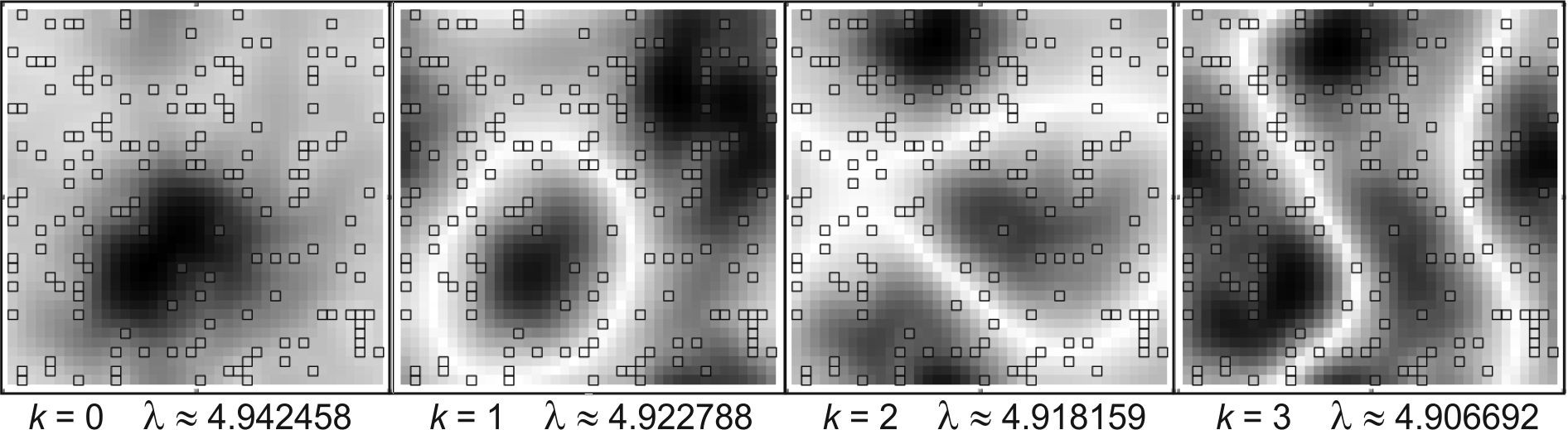}
        \caption{Plots of absolute values of the first four eigenvectors of graph from the previous figure. The first one is positive, while the rest of them have regions of constant sign, which are separated by boundaries of near zero values, represented by white color.}
        \label{exeig}
\end{figure}
For one dimensional defected lattice the second eigenvector was previously localized in the second largest defect-free region. Figure \ref{exeig} suggests that this intuition may continue to a few further eigenvectors - in this figure the first three eigenvectors visually correspond to three largest defect-free regions. However, the fourth one seems to disagree with this rule, so generally we should be careful about it. The intuition about MERW dynamics it provides is that temporary domination of given coordinate is responsible for localization in corresponding local entropic well - for example MERW density after 100 steps in Fig. \ref{2dex} is similar to $k=1$ eigenvector and finally later it deexcitate to the ground state. The initial coordinates in this eigenvalue decomposition depends on overlapping of given eigenvector with the initial probability distribution. While evolution, the speed of dominance of larger eigenvalue coordinates depends on proportion between eigenvalues. Finally, the intuition is that probability will first localize in the nearest defect-free region (local entropic well), then it will relaxate into succeeding larger Lifhsitz regions and finally thermalize in the ground state. If because of some additional constrains lower energy states are somehow restricted, presented picture suggests evolution should be "stochastically shifted" toward near (overlapping) eigenstate.

Assuming that the defect-free region is indeed approximately a sphere, like previously we would expect that $N(1-p_d)^{\pi R^2}$ is of order of unity:
$R\approx \sqrt{\frac{\ln(N)}{\pi|\ln(1-p_d)|}}$.\\
Eigenvector $\psi$ should have maximum near the center of this sphere and has approximately spherical symmetry - such local eigenfunction solution is approximately Bessel function $J_0$: $\psi(r)\approx J_0(jr/R)$, where $r$ is the distance from the center and $j\approx 2.404825$ is the first zero of $J_0$. Finally we obtain:
$$E\approx \left(\frac{j}{R}\right)^2\approx \frac{\pi j^2 |\ln(1-p_d)|}{\ln(N)}$$
For the general dimension of lattice $D$, skipping lattice-dependent multiplicative constants, the above estimates becomes:
$$R\sim \left(\frac{\ln(N)}{|\ln(1-p_d)|}\right)^{1/D}\qquad\qquad E\sim \left(\frac{|\ln(1-p_d)|}{\ln(N)}\right)^{2/D}$$

\begin{figure}[t]
    \centering
        \includegraphics{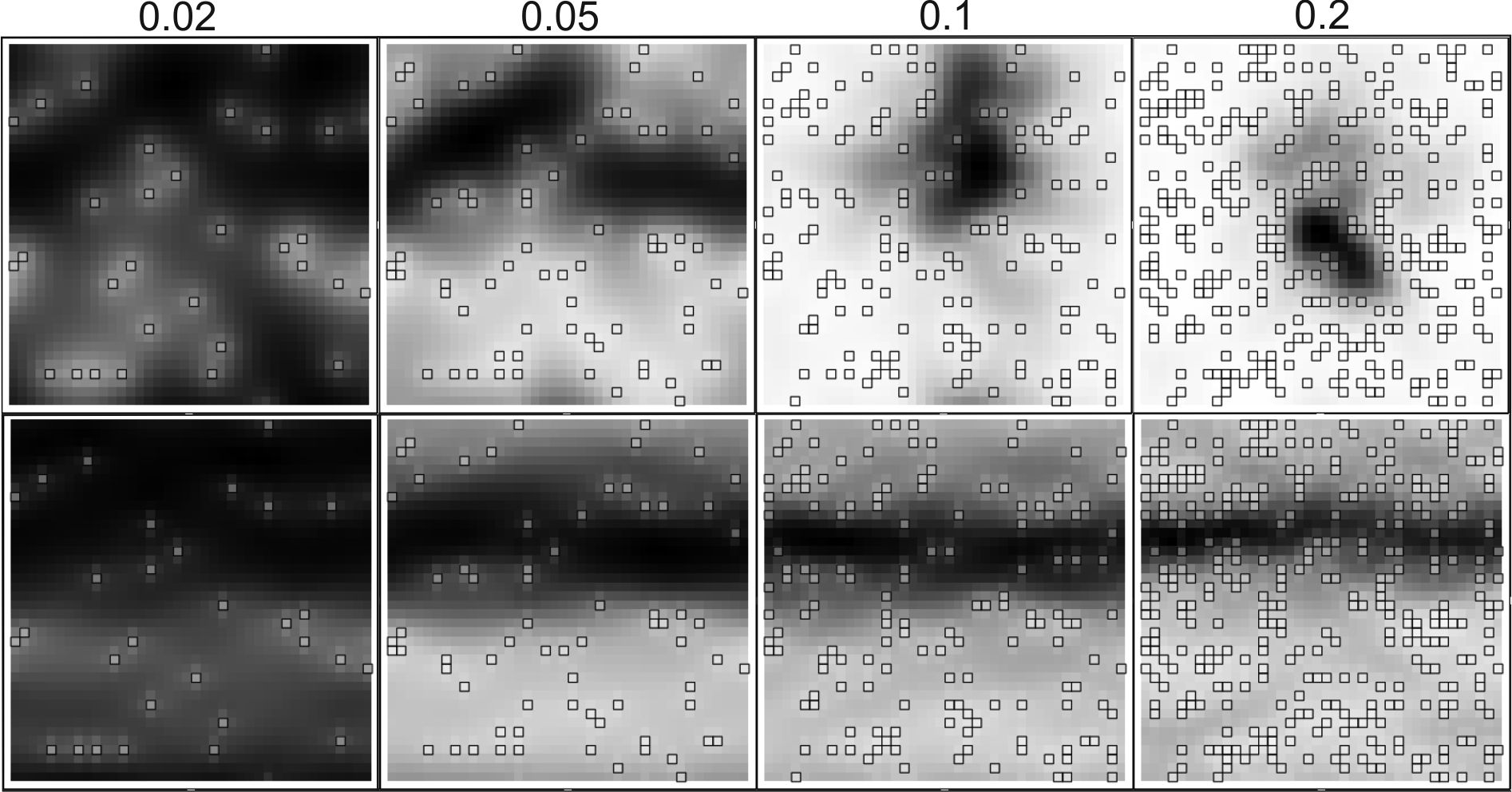}
        \caption{Stationary probability of MERW on $40\times 40$ lattice graph with cyclic boundary conditions and self-loops in all vertices but some randomly chosen of them: correspondingly 0.02, 0.05, 0.1, 0.2. In the upper row the graph is indirected, while in the lower row all horizontal indirected edges were replaced with directed edges toward right of the plot, to simulate conductance in this direction.}
        \label{2dexx}
\end{figure}

The previous examples were using indirected graph and so symmetric $M$. Let us now briefly look at MERW on modification of these graphs: each indirected horizontal edge is replaced by directed edge toward right hand side of the plot. Examples of numerical results are in the bottom row of Fig. \ref{2dexx}. The asymmetry makes that left and right eigenvectors are no longer the same (in practical cases their difference seems to be rather insignificant). Thanks of the previous symmetry, there was fulfilled detailed balance condition: for each edge, probability flow in both direction was equal. This time probability flows only in one horizontal direction - there is nearly uniform flow for low defect rate and it localizes near low defect paths for larger rates. The stationary flow allows to imagine this situation as simplified conductance model. While in GRW the flow would be nearly uniform, in MERW there appears some analogue of avalanche breakdown. For more realistic models, instead of forbidding some transitions, there should be introduced potential gradient. In the next section there will be introduced required methodology.

\subsection{Various transition times}
It was essential in the used formalism that all transitions last the same amount of time. We will not use it further, but for generality let us expand these considerations to situation in which edges could require different times of transition. There will be presented construction for transition times being natural multiplicities of some chosen unit time, so it can be also used for rational proportions by dividing the time unit by the lowest common denominator. Irrational proportions would rather require approximating by rational ones.

\begin{figure}[t]
    \centering
        \includegraphics{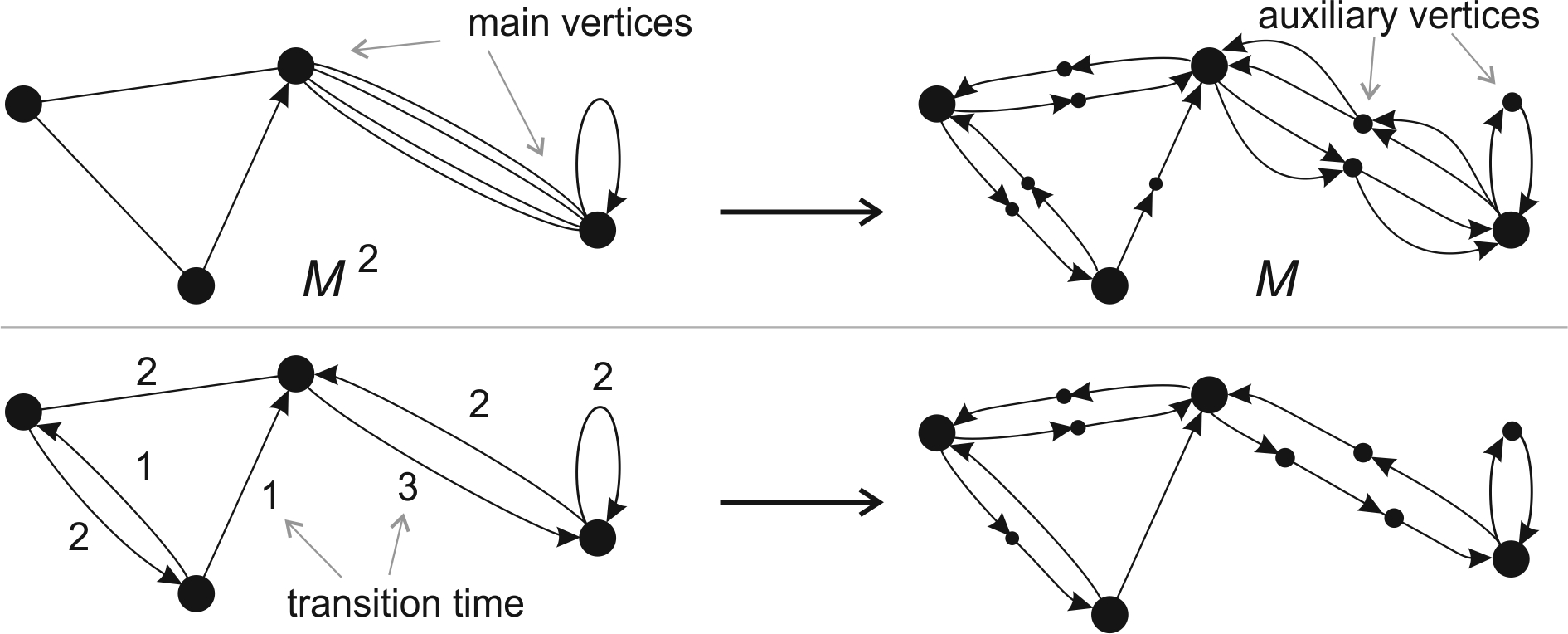}
        \caption{Top: example of construction of square root analogue of graph. Bottom: example of construction of graph to handle various transition times.}
        \label{trtime}
\end{figure}

Let us first look at the upper part of Fig. \ref{trtime} - for given graph we can construct period 2 (or $p$ generally) graph, which second ($p$-th) power is the original graph while restricting to the connected component of main vertices. For multi-edge or weighted graph, the weights on such path replacing an edge should be chosen to multiply to the original weight, for example as corresponding root of the weight of the original edge.

Like on the lower part of  Fig. \ref{trtime}, we can analogously replace edges with one-directional paths of chosen length and calculate MERW formulas for such extended graph. Now while fixing two vertices and some length, each pathway of this length on such extended graph is equally probable. Original edges correspond to subpaths on these paths, with lengths being their transition times - the pathway equiprobability condition become as we would expect. Obtained stationary probability distribution would have nonzero values in auxiliary vertices - it denotes probability of situations that specific transition was already chosen, but it was not yet finalized. If we are interested in probability distribution among the main vertices, we can for example interpret auxiliary vertices as preparation to transition and so move their stationary probabilities to the corresponding starting main vertex.

For multi-edge or weighed graph the weight of edges should be chosen to multiply to the original edge weight. There is some freedom of such choice, but these weights are not used independently - weights of full paths would not depend on this choice (with fixed product) and so derivation from \ref{ensder} would always lead to the same MERW formulas for all such weight distributions.

\subsection{Summary}
GRW is appropriate for a walker which indeed makes succeeding random decisions, using exactly uniform probability distribution among the nearest possibilities. MERW should be understood in completely different way: the walker does not have to make random decisions - the randomness represents only our lack of knowledge. The walker chooses a path in practically any allowed way (can be deterministic) and because we do not know which path he is choosing, we assume some natural thermodynamical ensemble of possible scenarios - paths.

Obtaining MERW transition probabilities requires knowing the eigenvector, which depends on the whole system - this effective model is nonlocal. It cannot be interpreted that there is required nonlocality in walker's behavior - the walker can choose the path in any way he want. Nonlocality is only a natural feature of models representing our knowledge - distant event may give us missing information, like thanks of angular momentum conservation, spin of one particle gives us information about the spin of a coupled one in EPR experiment. In MERW case, nonlocality means that to make the best predictions, we should know the whole space of possibilities. Later considering time dependent case we will see that we should also know future potential. Models representing our knowledge can have also retrocausality like in Wheeler's experiment - it only means that further event may give us missing information about past events.
\begin{center}
\begin{tabular}{c|c|c|}
    & GRW & MERW \\ \hline
  characteristic length& 1 & $\infty$ \\ \hline
  $S_{ij}=$ & $\frac{M_{ij}}{d_i}$ &  \raisebox{0cm}[5.5mm][3mm]{$\frac{M_{ij}}{\lambda}\frac{\psi_j}{\psi_{i}}$} \\ \hline
  $\pi_i=$ & complicated & $\varphi_i \psi_i$ \\
  for symmetric $M$: & $\frac{d_i}{\sum_j d_j}$ & $\psi_i^2$ \\ \hline
  $S_{\gamma_0\gamma_1}\cdot .. \cdot S_{\gamma_{l-1}\gamma_l}=$ &
  \raisebox{-2mm}{$\frac{M_{\gamma_0\gamma_1}\cdot..\cdot M_{\gamma_{l-1}\gamma_l}}{d_{\gamma_0}\cdot .. \cdot d_{\gamma_{l-1}}}$}&
  \raisebox{-2mm}{$\frac{M_{\gamma_0\gamma_1}\cdot..\cdot M_{\gamma_{l-1}\gamma_l}}{\lambda^l}\frac{\psi_{\gamma_l}}{\psi_{\gamma_0}}$} \\
  simple graph: & $\frac{1}{{d_{\gamma_0}\cdot .. \cdot d_{\gamma_{l-1}}}}$ & $\frac{1}{\lambda^l}\frac{\psi_{\gamma_l}}{\psi_{\gamma_0}}$  \\ \hline
  $\pi_{\gamma_0} S_{\gamma_0\gamma_1}\cdot .. \cdot S_{\gamma_{l-1}\gamma_l}=$ &
  \raisebox{0cm}[7.5mm][4mm]{$\frac{M_{\gamma_0\gamma_1}\cdot..\cdot M_{\gamma_{l-1}\gamma_l}}{d_{\gamma_1}\cdot .. \cdot d_{\gamma_{l-1}}\cdot\sum_j d_j}
  \quad$ (sym. $M$)} &
  $\varphi_{\gamma_0}\cdot\frac{M_{\gamma_0\gamma_1}\cdot..\cdot M_{\gamma_{l-1}\gamma_l}}{\lambda^l}\cdot\psi_{\gamma_l}$ \\ \hline
  $(S^l)_{ij}$= & complicated &  \raisebox{0cm}[5.5mm][3mm]{$\frac{(M^l)_{ij}}{\lambda^l}\frac{\psi_j}{\psi_i}$}  \\
  scaling, generally & $S^{\mathrm{GRW}(M^l)}\neq\left(S^{\mathrm{GRW}(M)}\right)^l$ & $S^{\mathrm{MERW}(M^l)}=\left(S^{\mathrm{MERW}(M)}\right)^l$ \\ \hline
  $H(S)=$ &  \raisebox{0cm}[6mm][3mm]{$\sum_i \frac{d_i\lg(d_i)}{\sum_{i'} d_{i'}}$} & $\lg(\lambda)$ \\ \hline
\end{tabular}\\
\end{center}
\section{Boltzmann paths and infinitesimal limit}
In this section we will make basic expansions of mathematical constructions from the previous section to make them more physical - add potential and then make infinitesimal limit.
\begin{figure}[t]
    \centering
        \includegraphics{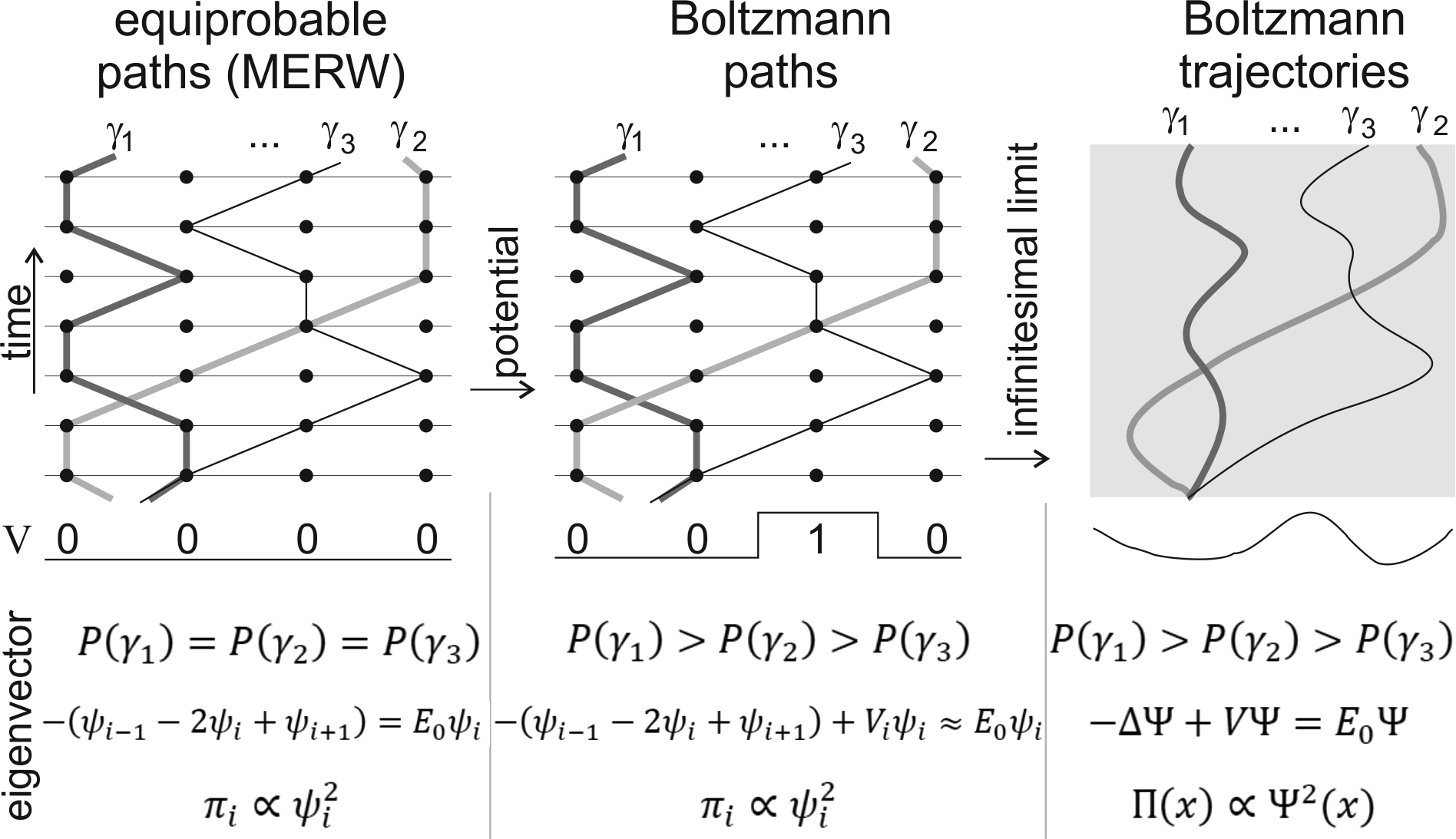}
        \caption{Expansion toward Schr\"{o}dinger equation - emphasizing paths using potential energy and making infinitesimal limit.}
        \label{fqm5}
\end{figure}

\subsection{Adding potential - Boltzmann paths}
If there is no reason to emphasize some of scenarios, the best assumption is to choose uniform probability distribution among them. Standard way of emphasizing some scenarios in physics, is by assigning them energy - for example trajectory remaining in potential well should be more probable than trajectory tunneling through a barrier, which should be still more probable than trajectory remaining on the top of this barrier, like in Fig. \ref{fqm5}. In such
situations we maximize entropy while fixing total energy, or equivalently: assume some compromise between maximizing entropy and minimizing average energy. It leads to Boltzmann distribution:
\be \max_{(p_i):\sum_i p_i=1}\left(-\sum_i p_i\ln(p_i)-\sum_i p_i \beta E_i \right)=\ln\left(\sum_i e^{-\beta E_i} \right)
\qquad\quad\mathrm{for}\ p_i\propto e^{-\beta E_i} \label{boltz}\ee
where $\beta=1/k_BT $, $k_B\approx 1.3807\cdot 10^{-23}J/K$ is Boltzmann constant and $T$ is temperature. $\beta$ controls this compromise:
the higher it is (the lower $T$), the more important choosing low energy is. In zero temperature there would be chosen only lowest energy states, while in infinite temperature energy differences would vanish. Minus maximized expression is free energy up to constant.

Let us define the weights using energy required for given transition:
\be M_{ij}=A_{ij}e^{-\beta V_{ij}} \label{trans}\ee
where $V$ is some potential - usually it will be scalar potential: depending only on the position, like $V_{ij}=(V_i+V_j)/2$,
but we can also use vector potential from electromagnetism, for which $V_{ij}$ could be essentially different from $V_{ji}$.
If we would be interested in random walk in phase space, this term could be used to additionally introduce kinetic energy to the considerations. Eventual lack of edge between some vertices can be seen as that there is infinite potential barrier.

Thanks of (\ref{trans}) convention, formula (\ref{pathpr}) for MERW probability of $(\gamma_i)_{i=0}^l$ path becomes:
\be S^M_{\gamma_0\gamma_1}S^M_{\gamma_1\gamma_2}..S^M_{\gamma_{l-1}\gamma_l}=
\frac{M_{\gamma_0\gamma_1}..M_{\gamma_{l-1}\gamma_l}}{\lambda^l}\frac{\psi_{\gamma_l}}{\psi_{\gamma_0}} =
\frac{e^{-\beta(V_{\gamma_0\gamma_1}+V_{\gamma_1\gamma_2}+..+V_{\gamma_{l-1}\gamma_l})}}{\lambda^l}\frac{\psi_{\gamma_l}}{\psi_{\gamma_0}} \ee
So in this interpretation, instead of calling it uniform probability among pathways, we have Boltzmann distribution among paths by using:
\begin{df} \label{enpath}
  Energy of path $(\gamma_i)_{i=0}^l$ \emph{is} $V_{\gamma_0\gamma_1}+V_{\gamma_1\gamma_2}+..+V_{\gamma_{l-1}\gamma_l}$.
\end{df}

Let us now look at the previous entropy formula (\ref{entr}):
$$H(S)=-\sum_{ij} \pi_i S_{ij} \lg(S_{ij})+\sum_{ij} \pi_i S_{ij}\lg(M_{ij})=-\sum_{ij} \pi_i \sum_j \lg(S_{ij})-\beta\lg(e)\sum_{ij}\pi_i S_{ij} V_{ij}$$
The left hand side sum is already entropy of given random walk. The right hand side sum was previously required to take into considerations that there can be multiple edges between given vertices: a choice of probability $S_{ij}$ was in fact $M_{ij}$ choices of $S_{ij}/M_{ij}$ probability.
For weighted graphs this interpretation is far-fetched, so we will further use more physical one (\ref{trans}) - that values of $M$ does not longer represent the number of edges, but correspond to the energy of given transitions.

In this interpretation transition probabilities correspond to single choices, so entropy production per step is just
$$\mathbf{S}:=-k_B \sum_i \pi_i \sum_j S_{ij} \ln(S_{ij})$$
where instead of previous base 2 logarithms, we have used more appropriate for physics Boltzmann's normalization.
$\pi_i S_{ij}$ is probability of $(ij)$ situation, so the second sum in $H(S)$ is average energy per step:
$$U:=\sum_{ij}\pi_i S_{ij} V_{ij}$$
Finally, we see that in this energy interpretation of weights (\ref{trans}), $H(S)$ is up to constant just minus average free energy per step:
\be F=U-T\mathbf{S}=-k_B T\ln(2) H(S) \ee
For the complete picture, let us look at the partition function
$$Z_l:=\sum_{(\gamma_i)_{i=0}^l} e^{-\beta(V_{\gamma_0\gamma_1}+V_{\gamma_1\gamma_2}+..+V_{\gamma_{l-1}\gamma_l})}=\sum_{\gamma_0, \gamma_l}
(M^l)_{\gamma_0 \gamma_l}$$
it is asymptotically proportional to $\lambda^l$, so we can check that the free energy per step is as expected in thermodynamics:
$$F=-\frac{1}{\beta} \lim_{l\rightarrow \infty} \frac{\ln(Z^l)}{l}=-k_B T \ln(\lambda)$$

For simplicity we will further call both approaches as MERW, but for better intuition here are gathered differences between these mathematically equivalent interpretations - for this and further sections we will use the second one:\\

\begin{tabular}{r|c|c|}

  &MERW& Boltzmann paths \\
  &maximizing entropy & minimizing free energy \\ \hline
  perfect for & multi-edge graphs & weighted graphs \\   \hline
  $M_{ij}$ & number of edges & energy of transition \\   \hline
  $i\rightarrow j$ transition & $M_{ij}$ choices of $S_{ij}/M_{ij}$ probability & choice of $S_{ij}$ probability \\ \hline
  $M_{\gamma_0\gamma_1}..M_{\gamma_{l-1}\gamma_l}$ & pathways in path & energy of path \\   \hline
  for full paths or& all pathways are  & assume Boltzmann  \\
  fixed endings and length & equally probable  & distribution among paths   \\      \hline
  $H(S)=\lg(\lambda)$ & entropy & minus free energy \\   \hline
  simple graph & single edges & potential is zero \\   \hline
  Sections & 3 & 4,5,6 \\   \hline
\end{tabular} \\

If instead we would construct GRW from $M$ $(S^G_{ij}=M_{ij}/d_i)$, the walker would also assume Boltzmann distribution - this time not among full paths(scenarios), but only among the nearest neighbors (single steps) - minimizing free energy locally, in a way depending on discretization. If $M$ is symmetric like $M_{ij}=e^{-\beta(V_i+V_j)/2}$, as previously there is simple formula for stationary probability distribution:
\be \pi^G_i\propto d_i=\sum_j e^{-\beta(V_i+V_j)/2}=e^{-\beta V_i/2}\sum_j e^{-\beta V_j/2} \label{pig}\ee

\subsection {Boltzmann paths on lattices} \label{bolpat}
In the previous section we have seen analogy between the dominant eigenvector and quantum ground state for lattice type graphs. Having energetic interpretation of weighted graphs, we can take it further. Such lattice can for example represent regular lattice of a crystal or defected lattice of a semiconductor. It can also represent discretization of a continuous system - later we will make infinitesimal limit to get to the continuous case. Lattices considered in practice are often finite - to approximate infinite lattice, there can be used finite one with cyclical boundary conditions. For simplicity let us assume we use a finite one, but the considerations can be also generalized to infinite graphs.

So let us assume that we want to model a part od $\mathbb{R}^D$, where usually dimension $D$ is 2 or 3. We cover its part by a lattice $\{0,..,m-1\}^D$ - it could overlap with a crystal lattice, or just represent discretiztion of a continuous problem. For simplicity let us assume that it is rectangular lattice with the same constants in all directions ($\delta>0$), so $(x_i)_{i=1}^D\in\mathbb{Z}^D$ represents for example $x:=(\delta x_i)_{i=1}^D\in \mathbb{R}^D$. Another simplifying assumption is that in a single step there is allowed transition to at most the nearest neighbors. Let as also assume cyclic boundary conditions - that $0$ and $m-1$ coordinates are adjacent, so "$+1$" and "$-1$" below
will be made modulo $m$. Finally all vertices have exactly $2D+1$ neighbors (including itself).

Now we have to choose the potential function - for this moment depending only on position. To model electron in crystal lattice, it may represent tendency to remain near given atom (like electronegativity). For discretization of a continuous system, it can be just the average potential in given cell (integral of potential divided by volume). Let us choose time discretization: single transition corresponds to $\epsilon>0$ time .
Finally allowing the walker also to remain in given vertex, for $D=1$ case weights can be for example chosen as:
\be (M_\epsilon)_{i,i+1}=(M_\epsilon)_{i+1,i}=e^{-\epsilon\beta \frac{V_i+V_{i+1}}{2}}\qquad\qquad\qquad (M_\epsilon)_{ii}=e^{-\epsilon\beta V_i} \ee
Where index arithmetics is modulo $m$. It was chosen to make $M$ symmetric - such that energy of path $\{\gamma_i\}_{i=0}^l$ is
$$\epsilon\left(\frac{V_{\gamma_0}}{2}+V_{\gamma_1}+V_{\gamma_2}+..+V_{\gamma_{l-1}}+\frac{V_{\gamma_l}}{2}\right)$$
For simplicity physical dimensions will be omitted in this paper, but physically $\epsilon$ is time, $V$ is energy, so Boltzmann distribution instead of energy uses energy multiplied by time (action) in this case. Analogously $\beta$ is not one over energy as usual, but one over action.

To find MERW in this case, let us look at eigenvector equations for this $M$:
\be \lambda_{\epsilon,\delta} \psi_i=(M_\epsilon \psi)_i=e^{-\beta\epsilon \frac{V_{i-1}+V_i}{2}}\psi_{i-1}+e^{-\beta\epsilon V_i}\psi_i+
e^{-\beta\epsilon \frac{V_i+V_{i+1}}{2}}\psi_{i+1} \label{lateig}\ee
We can make a few approximations like $e^{-\epsilon}\approx 1-\epsilon$ for small $\epsilon$:
$$ \lambda_\epsilon \psi_i \approx \psi_{i-1}+\psi_i+\psi_{i+1}-\epsilon\beta\left( \frac{V_{i-1}+V_i}{2}\psi_{i-1}+
 V_i\psi_i +\frac{V_i+V_{i+1}}{2}\psi_{i+1}\right)$$
Solving this kind of equations would be required for lattice of atoms in which the potential could vary from site to site. If the lattice was made to discretize a continuous system, for small lattice constant ($\delta$), in $\epsilon$ order we can assume that $V$ and $\psi$ are nearly constant:
\be \lambda_\epsilon \psi_i =(M\psi)_i\approx \psi_{i-1}+\psi_i+\psi_{i+1}-3\epsilon\beta V_i\psi_i\quad\qquad
/-3\psi_i\qquad/\cdot\frac{-1}{3\beta\epsilon}\label{appeig}\ee
$$\frac{3-\lambda_\epsilon}{3\beta\epsilon}\psi_i \approx -\frac{1}{3\beta}\frac{\psi_{i-1}-2\psi_i+\psi_{i+1}}{\epsilon} + V_i\psi_i$$
$\psi_{i-1}-2\psi_i+\psi_{i+1}$ is used as a discrete Laplacian. If we divide it by $\delta^2$, it becomes approximation of continuous Laplacian.
Average distance in diffusion grows with square root of passed time, so for infinitesimal limit there have to be assumed some relation between time and space step, like
\be\epsilon=\frac{\delta^2}{(2D+1)\alpha}\ee
where $\alpha>0$ is some parameter we can freely choose. Using this substitution and making above derivation for general dimension $D$, above "3" coefficient becomes $2D+1$ and finally we get:
\be
E_\epsilon \psi_x \approx
-\frac{\alpha}{\beta}\sum_{i=1}^D \frac{\psi_{(x_1,x_2,..,x_i-1,..,x_D)}-2\psi_x+\psi_{(x_1,x_2,..,x_i+1,..,x_D)}}{\delta^2}+
V_x\psi_x
\label{disc}\ee
where $x=(x_1,..,x_D)\in\mathbb{Z}^D$,
\be E_{\epsilon}:= \frac{2D+1-\lambda_{\epsilon}}{(2D+1)\beta\epsilon}\label{Edef}\ee

Because of multiplying by negative number, maximizing over $\lambda$ becomes minimizing over $E$ - for properly chosen $\alpha$ and $\beta$: such that $\frac{\alpha}{\beta}=\frac{\hbar^2}{2m}$, the eigenvector $\psi$ would be the ground state amplitude
of discretization of Schr\"{o}dinger's equation and the stationary probability density would be the same as for quantum mechanical ground state.

\subsection{Infinitesimal limit - Boltzmann trajectories} \label{inflim}
We would like now to make $\epsilon \rightarrow 0^+$ limit to get Boltzmann distribution among continuous trajectories:
$$P(\textrm{path}\ \gamma)\quad \textrm{is proportional to}\quad e^{-\beta\int V(\gamma(t))dt}$$
As it was mentioned, this time in Boltzmann distribution we use energy of path instead of energy - multiplied by time like for action. The choice of $\beta$ is arbitrary, but considering time dependent case, similarity to quantum formalism (\ref{aldef}) will suggest to use $\beta=1/\hbar$.

The eigenvector becomes a function such that
\be \Psi(X)\approx \psi_x\qquad \qquad \textrm{where }X:=(\delta x_i)_{i=1}^D\in \mathbb{R}^D\ee
The right hand side of (\ref{disc}) becomes
$$\hat{H}\Psi(X):=-\frac{\alpha}{\beta} \triangle \Psi(X) +V(X)\Psi(X)$$
where $\triangle=\sum_{i=1}^D \partial_{ii}$ is Laplacian. If we choose
\be \alpha=\frac{\hbar^2}{2m} {\beta} \label{beta}\ee
$\hat{H}$ becomes Hamiltonian of Schr\"{o}dinger equation ($-\frac{\hbar^2}{2m}\triangle+V$). \\
Finally eigenvector equation (\ref{disc}) becomes in the limit:
$$E\Psi=\hat{H}\Psi$$
where $E$ is the lowest possible eigenvalue (the ground state energy):
$$E=\lim_{\epsilon\rightarrow 0^+} E_{\epsilon} = \lim_{\epsilon\rightarrow 0^+} \frac{2D+1-\lambda_{\epsilon}}{(2D+1)\beta\epsilon}$$
The stationary probability density for such MERW limit is
\be \rho(X):= \Psi^2(X) \qquad\qquad \textrm{for normalized }\Psi:\ \int\Psi^2(X)=1\ee

Let us now find the continuous propagator. The $S_{ij}$ matrix looses its meaning in infinitesimal time step limit, but
we can use $(S^l)_{ij}=\left(\left(\frac{M}{\lambda}\right)^l\right)_{ij}\frac{\psi_j}{\psi_i}$ MERW propagator:
$$ (S_\epsilon^{1/\epsilon})_{ij}=\left(\left(\frac{M_\epsilon}{\lambda_{\epsilon}}\right)^{1/\epsilon}\right)_{ij}
\frac{\psi_j}{\psi_i}=
\left(\left(1+\epsilon\frac{M_\epsilon-\lambda_{\epsilon}}{\epsilon\lambda_{\epsilon}}\right)^{1/\epsilon}\right)_{ij}\frac{\psi_j}{\psi_i}\approx
\left(\exp\left(\frac{M_\epsilon-\lambda_{\epsilon}}{\epsilon\lambda_{\epsilon}}\right)\right)_{ij}\frac{\psi_j}{\psi_i}$$
In the $\epsilon\to 0^+$ limit, eigenvector $\psi$ becomes eigenfunction $\Psi$. Let us focus on $\frac{M_\epsilon-\lambda_{\epsilon}}{\epsilon\lambda_{\epsilon}}$ fraction using (\ref{appeig}) approximation for $D=1$:
$$\left(\frac{M_\epsilon-\lambda_{\epsilon}}{\epsilon\lambda_{\epsilon}}\psi\right)_i=
\frac{\psi_{i-1}+\psi_i+\psi_{i+1}-3\epsilon_{\epsilon}\beta V_i\psi_i-\lambda_{\epsilon}\psi_i}{\epsilon\lambda_{\epsilon}}=
\frac{\psi_{i-1}-2\psi_i+\psi_{i+1}}{\epsilon\lambda_{\epsilon}}+\frac{3-3\epsilon\beta V_i-\lambda_{\epsilon}}{\epsilon\lambda_{\epsilon}}\psi_i $$
using (\ref{Edef}) definition ($\lambda_{\epsilon}=3-3\beta\epsilon E_{\epsilon}$), the first fraction above leads to $\alpha\triangle$,
the second to
$$ \frac{3-3\epsilon\beta V_i-(3-3\beta\epsilon E_{\epsilon})}{\epsilon\lambda_\epsilon}=
\frac{3}{\lambda_{\epsilon}}\beta (E_{\epsilon}-V_i)$$
For general $D$, as previously above 3 changes into $2D+1$, so finally
$$ \frac{M_\epsilon-\lambda_{\epsilon}}{\epsilon\lambda_{\epsilon}}\psi\qquad \textrm{tends to}\qquad
\left(\alpha\triangle - \beta V +  \beta E\right)\Psi =\beta(E-\hat{H})\Psi $$
$$ \exp\left(\frac{M_\epsilon-\lambda_{\epsilon}}{\epsilon\lambda_{\epsilon}}\right) \qquad \textrm{tends to}\qquad e^{\beta(E-\hat{H})} $$
To obtain coordinates of a matrix, we can multiply it both sides by canonical vectors ($M_{ij}=e_i^T M e_j$ where $e_i=(\delta_{ik})_k$). In continuous limit we analogously multiply them by Dirac deltas - let us use notation from quantum mechanics to write the final propagator ($\langle\Psi|x\rangle=\Psi(x)$):
\be S^t(x,y):=\frac{\langle x|e^{-t\beta\hat{H}}|y\rangle}{e^{-t\beta E}}\frac{\Psi(y)}{\Psi(x)} \ee
It can be imagined that if we know that in given moment the walker has position $x$, probability density of that it will be in $y$ after time $t$ is $S^t(x,y)$. It seems that there is a problem with points where $\Psi$ vanishes, like inside an infinite energy barrier -
stationary probability density there is also zero and infinite propagator means that the walker would immediately escape from there.

The $\langle x|e^{-t\beta\hat{H}}|y\rangle$ term is the propagator from euclidean path integrals - it is called the kernel and it describes local evolution. The additional $\frac{\Psi(y)}{\Psi(x)}$ term is required to make the propagator stochastic and it depends on the whole system, making this model nonlocal - to make the best predictions, we should know the whole system. This stochastic model only represents our knowledge - the walker does not directly use it, there is no need for nonlocality governing its behavior.

The graph is regular, so for constant $V$ propagator becomes the same as for GRW - leading to the Brownian motion with $\alpha/\beta$ diffusion coefficient. Generally it has much stronger localization properties.

Let us check that this propagator is properly normalized, compose correctly and leads to the expected stationary probability density:
$$\int S^t(x,y)dy=\int\frac{\langle x|e^{-t\beta \hat{H}}|y\rangle}{e^{-t\beta E}}\frac{\langle y|\Psi\rangle}{\Psi(x)}dy=
\frac{\langle x|\Psi\rangle e^{-t\beta E}}{e^{-t\beta E}\Psi(x)}=1$$
$$\int S^t(x,y)S^s(y,z)dy=
\int \frac{\langle x|e^{-t\hat{H}}|y\rangle}{e^{-t\beta E}}\frac{\Psi(y)}{\Psi(x)}\frac{\langle y|e^{-t\beta \hat{H}}|z\rangle}
{e^{-s\beta  E}}\frac{\Psi(z)}{\Psi(y)}dy=S^{t+s}(x,z)$$
$$ \int \rho(x)S^t(x,y)dx =\int \Psi^2(x) \frac{\langle x|e^{-t\beta \hat{H}}|y\rangle}{e^{-t\beta E}}\frac{\Psi(y)}{\Psi(x)}dx=$$
$$=\int \frac{\langle\Psi|x\rangle\langle x|e^{-t\beta \hat{H}}|y\rangle\langle y|\Psi\rangle}{e^{-t\beta E}}dx=
\frac{e^{-t\beta E}\langle\Psi|y\rangle\langle y|\Psi\rangle}{e^{-t\beta E}}=\rho(y) $$
We have used that $\Psi$ is real function, what is also essential for $\rho (x)= \Psi^2(x)$ formula, which in opposite to quantum mechanics
does not require using absolute value. For this purpose there was previously used Frobenius-Perron theorem - let us take it to the continuous case.
This uniqueness and positiveness of the ground state eigenfunction can be found for example in Faris \cite{faris} - here is Theorem 10.3 from this book:
\begin{tw}
  Let $\mathcal{H}=L^2(M,\mu)$. Let $A:\mathcal{H}\rightarrow \mathcal{H}$ be a bounded self-adjoined operator. Assume that $A\leq a$ where $a$ is eigenvalue of $A$.
  Assume also that $A$ is positivity preserving. Then $A$ is indecomposable if and only if the eigenvalue $a$ has multiplicity one and the corresponding
  eigenspace is spanned by a function $u$ which is strictly positive almost everywhere.
\end{tw}
\emph{Positivity preserving} condition corresponds to nonnegativity of matrix: for all real $u\geq 0$, $Au$ is also real and $Au\geq 0$,\\
\emph{Indecomposability} of operator $A$ corresponds to connectiveness - there is no projection operator ($P^2=P$) onto a non-trivial closed subspace,
such that $AP=PAP$.\\
We are interested in dominant eigenvalue of the operator, so assumption that it is bounded is necessary.
This time there is no longer a problem with periodic graphs - discrete lattice can have period 2,
but intuitively it degenerates while taking infinitesimal limit.
In our case the operator is obtained as a limit of nonnegative matrices, so we automatically get the positivity preservation and so the main question
is about the indecomposibility condition. For example on page 72 of cited book, above theorem is used for the ground state of Schr\"{o}dinger equation as we require:
\begin{sps}
  Let $\mathcal{H}=L^2(\mathbb{R}^n,dx)$. Let $H_0=-\triangle$ and let $V\geq 0$ be a function on $\mathbb{R}^n$ which is locally integrable on the
  complement of a closed set $K$ of measure zero. Assume that the complement of $K$ is connected. Then the ground state of $H=H_0+V$ (if exists) is unique.
\end{sps}
So infinite energy barriers could generally divide the space into independent components, but inside them the ground state is unique and real, nonnegative.\\

On the end of this subsection, let us look at the propagator decomposed in the eigenbase of Hamiltonian (assuming discrete energy spectrum):
$$\hat{H}\Psi_i=E_i \Psi_i \qquad\qquad \mathrm{where} \qquad E=E_0\leq E_1\leq ..$$
$$\langle\Psi_i|\Psi_j\rangle=\delta_{ij} \qquad\qquad \sum_i |\Psi_i\rangle\langle\Psi_i|=\mathbf{1}$$
where all $\Psi$ can be chosen as real functions, but usually only $\Psi_0=\Psi$ is nonnegative. Now we can write
\be S^t(x,y)=\frac{\sum_i e^{-t\beta E_i} \langle x|\Psi_i\rangle\langle\Psi_i|y\rangle}{e^{-t\beta E_0}}\frac{\Psi_0(y)}{\Psi_0(x)}\ee
Let us imagine that there is some idealized model preferring some concrete solutions like orbits of classical mechanics. We would like to add to these simplified considerations small perturbations we cannot directly control, like caused by the wave nature of particles. Analogous natural thermodynamical approach would be instead of considering a single idealized solution, use canonical ensemble of perturbed ones. The presented simplified approach uses nowhere differentable diffusive trajectories, which are not very physical and in this moment do not allow for additional restrictions. There is required further work, but the general suggestion of above propagator is that if idealized solution is additionally somehow restricted, adding thermodynamical perturbations would stochastically shift it toward "near" low eigenstate: having relatively large projection on the probability density. This restriction could be also that lower eigenstates are already occupied by repelling particles, preventing from choosing these dynamical equilibriums by such thermodynamical analogue of Pauli exclusion principle.\\

Thought-provoking observation from above derivation of Schrödinger's Hamiltonian is that Laplacian term is not (like in its quantum mechanical interpretation) a result of kinetic energy, but of using lattice as discretization like in Bose-Hubbard model - corresponds only to freedom of moving in the space. We will later introduce momentum operator in analogy to quantum mechanics, but it describes only density flow, not the real momentum of the particle. To include kinetic energy, we would need to include velocity of particle first - consider random walk in phase space like in Langevin equation.
\subsection{Comparison of ensembles and interpretations}

\begin{figure}[t]
    \centering
        \includegraphics{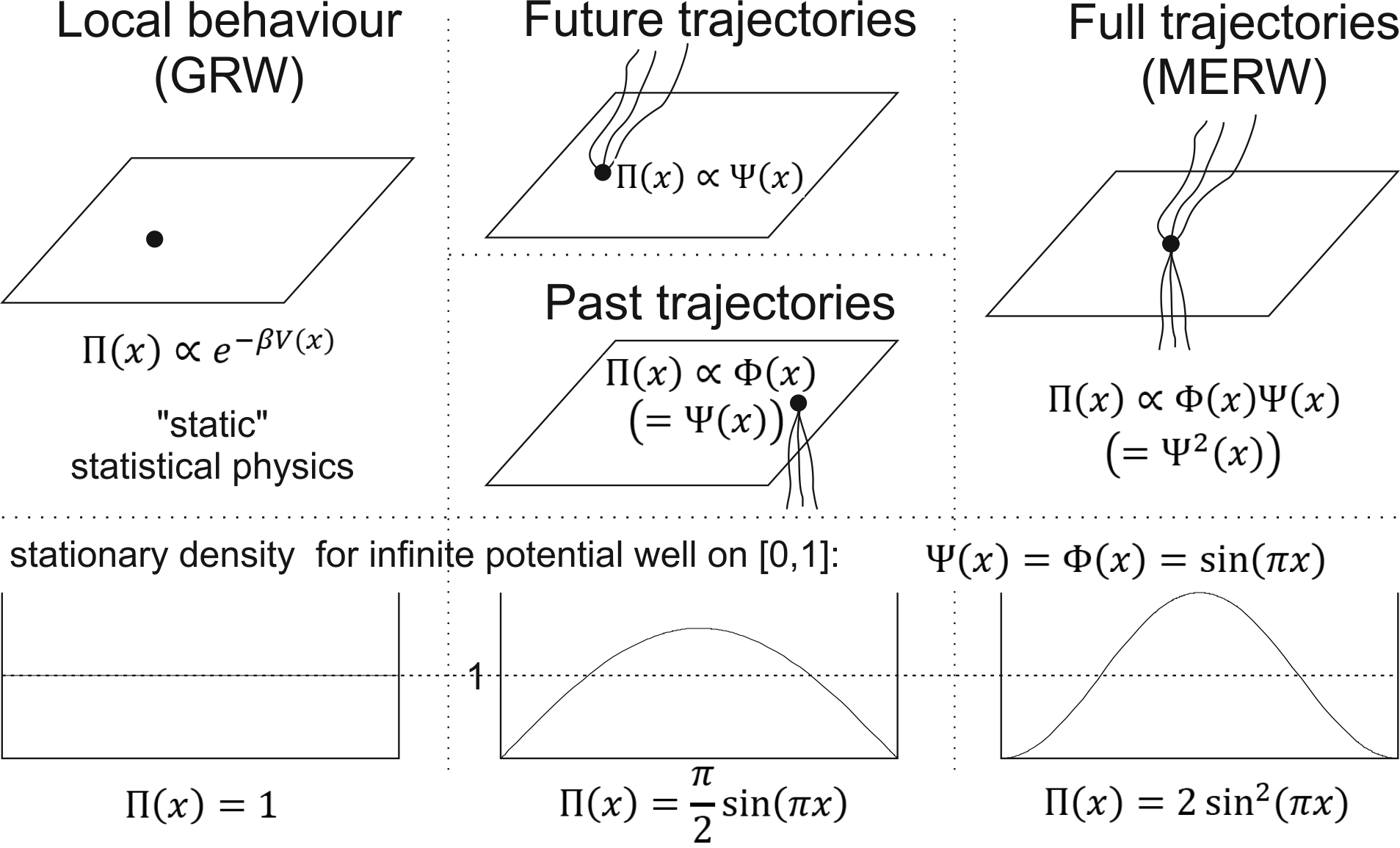}
        \caption{Comparison of different ensembles and their stationary probability densities for infinite potential well. Nonlocality required
        for such effective models does not imply nonlocality of the original model we are trying to predict. }
        \label{comp}
\end{figure}

If we make analogous infinitesimal limit of GRW instead, from (\ref{pig}) we get stationary probability density:
\be \rho^{GRW}(x) \propto e^{-\beta V(x)} \ee
While characteristic length in MERW remains infinite, in GRW case it is the distance corresponding to the nearest neighbor ($\delta$), so it drops to zero in infinitesimal limit - the walker makes succeeding random decisions accordingly only to situation in given point. In comparison, MERW randomness represents only our knowledge and the walker could even make decisions in an unknown complex deterministic way. There is also completely no need he knows this (nonlocal) model we use - it appeared by our assumption of thermodynamical ensemble of possible scenarios he could choose - asymptotically dominating all other assumptions we could make.

Figure \ref{comp} compares three basic approaches to statistical ensembles we could choose. The first one has practically no localization properties. The last one is MERW-based leading to probability density localized exactly as in quantum mechanics. The middle one has first power relating amplitudes and probabilities, so it would not violate Bell inequalities. However, such assumed ensemble changes while time passes, so there is no direct way to infer transition probabilities for it (connecting different ensembles). The MERW situation can be seen as two glued middle situations - leading to squares required to predict probability on constant time cut of such ensemble of abstract four-dimensional scenarios.\\

To conclude this section, let us think how to interpret obtained continuous probabilistic models. Standard view on Brownian motion is stochastic: that the object literally chooses succeeding steps of nowhere differentable trajectory using locally maximizing entropy transition probabilities, like a particle drifting in a fluid. On the other side there is ergodic picture of classical chaos - the object travels through some concrete trajectory, governed for example by classical mechanics and this trajectory effectively covers the whole space, allowing to introduce density function by averaging over infinite time.

Imagining a particle, it should travel through a differentiable and generally more determined than diffusive trajectory. Standard assumption of chaos theory that our model can fully describe the evolution is often also not appropriate, because there are usually plenty of hidden from us degrees of freedom there, which in practice can be included to considerations only as thermodynamical fluctuations. So what we would like is something intermediate - model physical trajectories and include thermodynamics.

The MERW-based approach is intended to be thermodynamical model - because we do not know what is exactly happening, we assume Boltzmann distribution among possibilities: trajectories. Like in stochastic picture, it is defined by local transition probabilities, but this time they are calculated not assumed - to fully optimize entropy/free energy. In subsection \ref{ensder} we have seen that MERW formulas can be obtained by calculating proportions between occurrences of patterns in ensemble of all full paths. These transition probabilities are no longer defined by only local conditions, but require the knowledge about the whole situation - they should not be interpreted that the walker uses them directly as in stochastic model, but only we use them to predict its probability density. Transition probabilities for single steps fully determined the process (Markov property), so it is enough to focus on finding probabilities for single steps - calculate transition probabilities as proportions between single steps in canonical ensemble of trajectories going through a given point like in Fig. \ref{pictures}.

The problem is that trajectories we use in the ensemble are diffusive, while we would rather expect restricting this ensemble to more physical ones - differentable, approximately preserving energy (up to thermodynamical fluctuations). Because in this moment we are interested only in probabilities of single steps, which become infinitesimally small in continuous limit, one could expect that such restriction to physical trajectories should practically not change local transition probabilities - let us call such assumption \emph{restricted ensemble hypothesis}:
\begin{hyp}
  Averaging over ensemble of all trajectories to obtain probability distribution of infinitesimal steps, practically does not change while restricting to physical trajectories.
\end{hyp}

There remain a question of defining what we mean by physical trajectories - we should not focus just on classical deterministic ones like in classical chaos, but remember that they are usually idealized: neglect many hidden degrees of freedom. However, including such thermodynamical perturbations should not affect proportions between infinitesimal steps while averaging over ensemble, leading again to above restricted ensemble hypothesis assumption.

The real difference between diffusive and physical trajectories appears on larger than infinitesimal time scale - physical trajectories in space do not fulfill Markov property, but their behavior depends on additional degrees of freedom like velocity. It could be improved using Markov process in phase space instead, getting more complicated optimizing entropy/free energy analogue of Langevin equation.

Observe that even if obtained model is not a Markov process, transition probabilities describe average behavior from a given point - while we should be careful about using Markov propagator, the stationary probability distribution is always the dominant eigenvector of stochastic matrix. The fact that the expected probability distribution agrees with thermodynamical equilibrium predicted by quantum mechanics, suggests that presented approach with above hypothesis is reasonable interpretation.

To summarize, using MERW-based models to generate stochastic trajectories is not the proper intuition. The transition probabilities should be rather seen as averaged local behavior over all possible scenarios. The stochastic propagator assumes Markov property, so generally we should be careful while interpreting it. The most essential conclusion from these models is that the equilibrium probability density is universal thermodynamical effect, especially because it is in agreement with thermodynamical equilibrium of quantum mechanics.
\section{Time dependence}
We will now focus on situation when the $M$ matrix can vary in time. There will be considered the general discrete case first. These considerations can be used if there are added or removed some graph edges while time passes. Later we will use them for lattice graphs and vary only weights of edges, representing evolution of potential. Finally we will look at continuous limit situation and find analogues of probability current, momentum operator, Ehrenfest equation and Heisenberg uncertainty principle. These considerations improve intuition about time symmetry of this thermodynamical model.

For simplicity in this section we assume that graph is aperiodic, but this restriction can be easily removed.
\subsection{General discrete case}
Let us generalize previous results to $M$ varying with time, what can represnt the change of potential like:
$$V_{ij}^t\equiv V_{ij}(t)\qquad\qquad M^t_{ij}=A_{ij} e^{-\beta V^t_{ij}}\qquad\qquad \left(\textrm{or symmetric } M^t_{ij}=e^{-\beta\frac{V^t_i+V^{t+1}_j}{2}}\right)$$
Previously powers of matrix represented behavior on time segments, now there are required time dependent analogues. In this section we will use upper index as time instead of power. For Boltzmann distribution among paths, by extending definition \ref{enpath} we would like that:
\be \textrm{energy of path }\ (\gamma_t,\gamma_{t+1},...,\gamma_{s})\ \textrm{ is }\ V^t_{\gamma_t \gamma{t+1}}+...+V^{s-1}_{\gamma_{s-1} \gamma{s}}\ee
where for this section we will assume that $s> t$.\\
For this generalization we can retrace one of the original derivations:\\
- by expanding path ensemble in both time directions, automatically getting normalization like in \ref{ensder}, or\\
- by expanding in single direction looking at its first step and then make normalization like in \ref{limder}.
\subsubsection{Generalized dominant eigenvectors}
In any case, we need analogues of right and left dominant eigenvectors (real, nonnegative):
probability distributions for final situation of ensembles of infinite one-sided paths - into the future ($\psi$) and eventually into the past ($\varphi$):
\be \varphi^t_j:=\lim_{l\to\infty} \frac{\sum_i (M^{t-l} M^{t-l+1}..M^{t-1})_{ij}}{\tilde{\mathcal{N}}^t(l)}   \qquad\qquad
\psi^t_i:=\lim_{l\to\infty} \frac{\sum_j (M^t M^{t+1}..M^{t+l-1})_{ij}}{\mathcal{N}^t(l)}\qquad(\geq 0)\label{defeig}\ee
where $\mathcal{N},\ \tilde{\mathcal{N}}$ are some normalizing functions - as previously $\varphi^t_i\psi^t_i$ will correspond to probability of $i$ (in time $t$), so it would be useful to make that that $\forall_t\ \sum_i\varphi^t_i \psi^t_i=1$.
It could be achieved using $\mathcal{N}^t(l)=\tilde{\mathcal{N}}^t(l)=\sqrt{\sum_{ij} M^{t-l}..M^{t+l-1}}$,
but it could make one eigenvector vanishing while the second goes to infinity. Anyway, these formulas using propagator from plus or minus infinity are rather impractical - more useful are analogues of
eigenvector equations:
\be (M^t\psi^{t+1})_i=\lim_{l\to\infty} \frac{\sum_j (M^{t}M^{t+1} M^{t+2}..M^{t+l})_{ij}}{\mathcal{N}^{t+1}(l)}=
\lambda^t \psi^t_i \label{tieig}\ee
\be((\varphi^t)^T M^t)_j=\lim_{l\to\infty} \frac{\sum_i (M^{t-l} M^{t-l+1}..M^{t-1}M^t)_{ij}}{\tilde{\mathcal{N}}^t(l)}
=\tilde{\lambda}^{t} \varphi^{t+1}_j \label{tieig2}\ee
where $\lambda^t=\lim_{l\to\infty}\frac{\mathcal{N}^t(l+1)}{\mathcal{N}^{t+1}(l)},\
\tilde{\lambda}^t=\lim_{l\to\infty}\frac{\tilde{\mathcal{N}}^{t+1}(l+1)}{\tilde{\mathcal{N}}^t(l)}$.

Let us assume that $\sum_i\varphi^t_i\psi^t_i$ remains constant for normalization:
\be (\varphi^t)^T\psi^t=(\varphi^t)^T\frac{M^t\psi^{t+1}}{\lambda^t}=(\varphi^t)^T M^t\frac{\psi^{t+1}}{\lambda^t}=
\frac{\tilde{\lambda}^t}{\lambda^t}(\varphi^{t+1})^T\psi^{t+1} \ee
So assumption that $(\varphi^t)^T\psi^t=\mathrm{const}$  is equivalent to $\tilde{\lambda}=\lambda$ as expected (also for different eigenvectors). Other view on this condition is through making below multiplication in two ways:
\be (\varphi^t)^T M^t \psi^{t+1}=\lambda^t \qquad\qquad(\textrm{generally:}\quad(\varphi_k^t)^T M^t \psi_k^{t+1}=\lambda_k^t)\label{tdlam}\ee
\subsubsection{Generalized further eigenvectors}
Choosing $M$ as constant in time, we see that these generalized eigenvector equations should have $N$ orthogonal solutions (for this short part, lower index denotes the number of solution of eigenequation):
$$M^t\psi_k^{t+1}=\lambda_k^t \psi_k^t\qquad\qquad (\varphi_k^t)^T M^t=\lambda_k^t(\varphi_k^{t+1})^T\qquad\qquad \textrm{for}\ \lambda^t=\lambda^t_0\geq \lambda^t_1 \geq ... \geq \lambda^t_{N-1}$$
which is fulfilled by decomposition analogous to the stationary case:
$$M^t=\sum_k \lambda_k^t \psi_k^t (\varphi_k^{t+1})^T\qquad\qquad \textrm{for orthogonal } \varphi, \psi:\ \forall_t\ (\varphi_k^t)^T \psi_l^t=\delta_{k,l}$$
For locally stationary situation we can use standard diagonalization, then there can be used the eigenvector equations to obtain their evolution. In our case we are only interested in the dominant ones: $\psi^t=\psi_0^t,\ \varphi^t=\varphi_0^t$. They can be also directly calculated using (\ref{defeig}), which can be seen as generalization of the power method for finding the dominant eigenvector.
\subsubsection{Time dependent MERW}
With $\tilde{\lambda}=\lambda$ assumption we know that $(\varphi^t)^T\psi^t$ is constant in time, so the exact choice of $\lambda$ only determines balance between
$\varphi$ and $\psi$, such that their product remain constant. We are mainly interested in $\pi$ vector depending on their product
and $S$ matrix depending on division of $\psi$ coordinates, so this balance between $\varphi$ and $\psi$ is in fact irrelevant -
this means we could choose practically any $\lambda$.
The only problem of choosing it in not optimal way, is that $\psi$ could grow exponentially to infinity while $\varphi$ would drop to zero or oppositely, what could be inconvenient in practical calculations.

There are some situations allowing to "calibrate" $\lambda$, for example if potential is practically constant on some time segment, both $\varphi$ and $\psi$ should tend to the stationary situation in which they are left and right dominant eigenvectors of this locally constant $M$ - choosing $\lambda$ as corresponding eigenvalue allows to make these eigenvectors constant. If potential evolution is slow enough, we could assume that such equilibrium is constantly maintained - in such \emph{adiabatic approximation} probability density in given time should be nearly the same as if potential would remain constant in time.

Now considering ensembles of paths growing in both time directions, (\ref{pathprob}) becomes:
\be \frac{\mathrm{Prob}((v_i)_{i=t}^s)}{\mathrm{Prob}((w_i)_{i=t}^s)}=
\frac{\varphi^t_{v_t}\cdot M^t_{v_t v_{t+1}} M^{t+1}_{v_{t+1} v_{t+2}}..M^{s-1}_{v_{s-1} v_s}\cdot \psi^s_{v_s}}
{\varphi^t_{w_t}\cdot M^t_{w_t w_{t+1}} M^{t+1}_{w_{t+1} w_{t+2}}..M^{s-1}_{w_{s-1} w_s}\cdot \psi^s_{w_s}}\ee
As previously, for $s=t$ we get dynamical analogue of stationary probability density - without additional knowledge, the optimal assumption of probability density in given moment is:
\be\pi^t_i= \varphi^t_i \psi^t_i\qquad\qquad\qquad\qquad \textrm{for }\ \sum_{i}\varphi^t_{i} \psi^t_{i}=1,\ \ \lambda=\tilde{\lambda} \ee
For $s=t+1$ we get time dependent $S$ matrix:
\be S^t_{ij}=\frac{M^t_{ij}}{\lambda^t}\frac{\psi^{t+1}_j}{\psi^t_i}\ee
Required normalization constant ($1/\lambda^t$) is consequence of (\ref{tieig}):
$$\sum_i \pi_i^t S^t_{ij} = \sum_i \frac{\varphi_i^t M^t_{ij}\psi^{t+1}_j}{\lambda^t} = \pi_j^{t+1} $$
The analogue of power of $S$ matrix now depends on time segment this propagator corresponds to:
\be (S^{ts})_{ij}:=(S^t S^{t+1}..S^{s-1})_{ij}=\frac{(M^t M^{t+1}..M^{s-1})_{ij}}{{\lambda^t\lambda^{t+1}..\lambda^{s-1}}}\frac{\psi^s_j}{\psi^t_i} \ee

The Botzmann distribution among finite length paths became:
\be S^t_{\gamma_t\gamma_{t+1}}S^{t+1}_{\gamma_{t+1}\gamma_{t+2}}..S^{s-1}_{\gamma_{s-1}\gamma_{s}}=
\frac{M^t_{\gamma_t\gamma_{t+1}}M^{t+1}_{\gamma_{t+1}\gamma_{t+2}}..M^{s-1}_{\gamma_{s-1}\gamma_{s}}}{\lambda^t\lambda^{t+1}..\lambda^{s-1}}
\frac{\psi^s_{\gamma_s}}{\psi^t_{\gamma_t}}=
\frac{e^{-\beta\left(V^t_{\gamma_t \gamma{t+1}}+...+V^{s-1}_{\gamma_{s-1} \gamma{s}}\right)}}{\lambda^t\lambda^{t+1}..\lambda^{s-1}}
\frac{\psi^s_{\gamma_s}}{\psi^t_{\gamma_t}} \ee

\begin{figure}[t]
    \centering
        \includegraphics{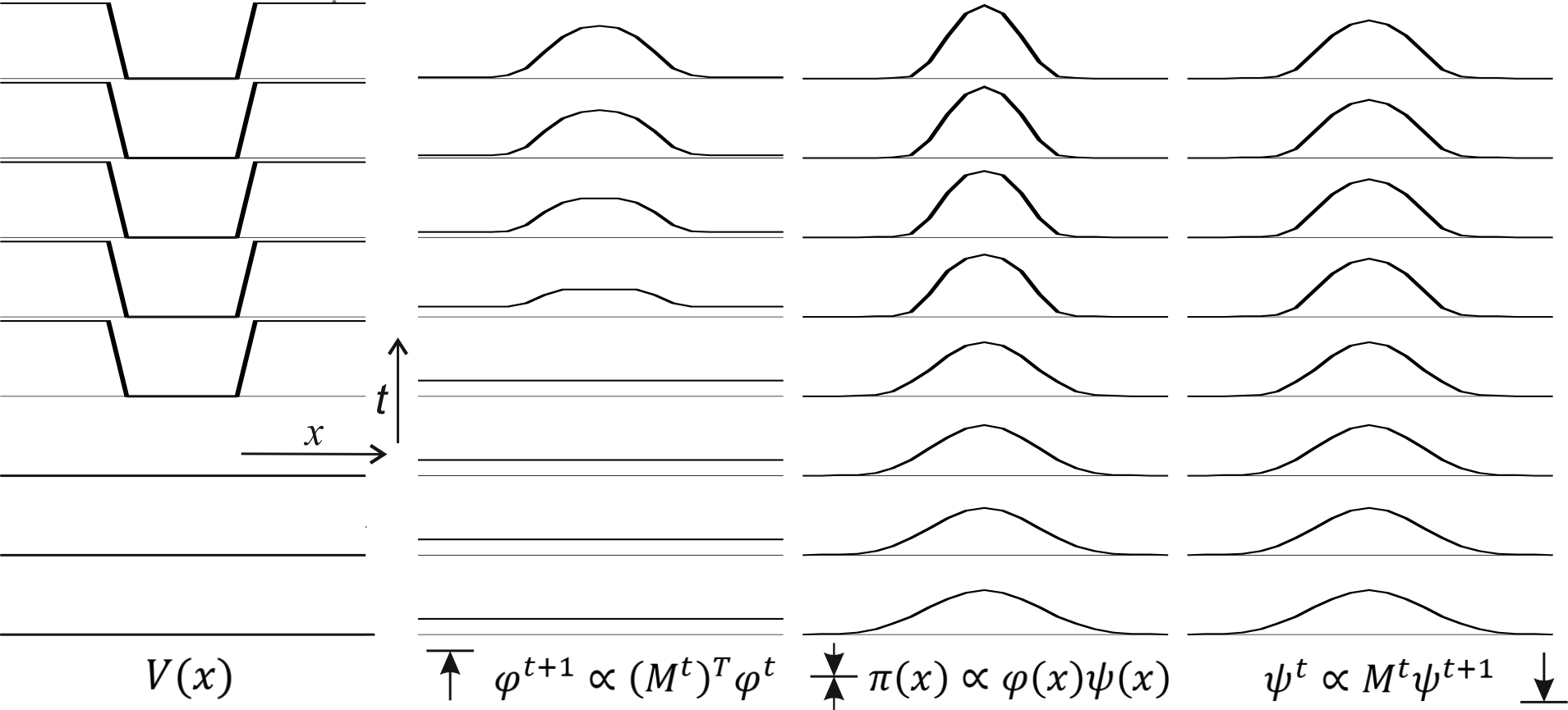}
        \caption{Example of segment graph with cyclic boundary condition on which there is switched potential well. Amplitudes are normalized to one, so they can be interpreted as densities on the end of past/future ensembles of one-sided paths like in Fig. \ref{comp}. Their non-adiabatic evolution start from stationary state in past/future and evolve toward future/past. }
        \label{tmdep}
\end{figure}

Fig. \ref{tmdep} brings some intuition about the situation. $\psi$ represents estimation of the future behavior and evolves backward in time and $\varphi$ represents knowledge about the past situation and evolves forward in time. This situation has time symmetry: transposing $M$ and negating time would switch $\psi$ and $\varphi$. The retrocausality of this effective model means only that to make the best estimation of the walker's position, we should know how the system will change in the future. It cannot be interpreted that the walker needs to know the future.

\subsection{Continuous limit}
As previously, we would like now to find the continuous limit for lattices. Knowing the dominant eigenvectors in some moment, we can use analogues of eigenvector equations to get their evolution:
\be M^t\psi^{t+1}=\lambda^t \psi^t,\qquad\qquad (M^t)^T \varphi^t  =\lambda^t \varphi^{t+1} \label{tddis}\ee
In addition to stationary situation, there appears time dependence of these generalized eigenvectors now - difference between succeeding ones will lead to time derivative in continuous limit.

The fact that $\psi$ represents ensemble of possible further evolutions allowed previously to understand the requirement of squares relating amplitudes and probabilities. In opposite to $\varphi$, natural direction of evolution of $\psi$ is into the past. We are usually interested in evolution into the future, what leads to some inconvenience - for this purpose we should iterate $M^{-1}$ matrix instead, but it diametrally changes the attractors ($\lambda \rightarrow 1/\lambda$). For example the attracting dominant eigenvector becomes the most repelling one - of the least absolute value. We will derive and use these unstable equations in theoretical considerations, but one has to be careful about using them especially in numerical calculations. For this purpose, it is better to start with some equilibrium in future and evolve $\psi$ backward in time like in Fig. \ref{tmdep}.

In opposite to thermodynamics, quantum mechanics has unitary evolution in which this inconvenience disappears. Instead of exponentially vanishing/exploding coordinates (eigenvalues on real axis), the absolute coordinates remain constant (eigenvalues on unit circle).

This time (\ref{appeig}) approximation for $D=1$ becomes (unstable):
$$ \lambda^t_{\epsilon} \psi^t_x =(M^t\psi^{t+1})_x\approx \psi^{t+1}_{x-1}+\psi^{t+1}_x+\psi^{t+1}_{x+1}-3\epsilon\beta V^t_x\psi^t_x\qquad\qquad
/-3\psi_x^{t+1}$$
The last $\psi$ should originally have $t+1$ upper index, but we are again interested in infinitesimal limit, in which terms of higher than epsilon order will vanish. As previously let us connect eigenvalue with energy: $\lambda^t_{\epsilon}=3-3\epsilon\beta E^t_{\epsilon}$, getting
$$3(\psi^t_x-\psi^{t+1}_x)-3\epsilon\beta E^t_{\epsilon}\psi^t_x \approx \psi^{t+1}_{x-1}-2\psi^{t+1}_x+\psi^{t+1}_{x+1}-3\epsilon\beta V^t_x\psi^t_x
\qquad\qquad /\cdot\frac{-1}{3\epsilon\beta}$$
$$\frac{1}{\beta}\frac{\psi^{t+1}_x-\psi^t_x}{\epsilon}+E^t_{\epsilon}\psi^t_x\approx
-\frac{1}{3\beta}\frac{\psi^{t+1}_{x-1}-2\psi^{t+1}_x+\psi^{t+1}_{x+1}}{\epsilon}+V^t_x\psi^t_x$$
Now as in \ref{inflim}, the infinitesimal limit for general $D$ becomes:
\be \frac{1}{\beta} \frac{d}{dT} \Psi(X,T)+E(T) \Psi(X,T)=\hat{H}(T)\Psi(X,T) \label{tdev}\ee
where $T=\epsilon t$, $X=\delta x$, $\epsilon=\frac{\delta^2}{(2D+1)\alpha}$, $E(T)=\lim_{\epsilon\to 0} E^t_{\epsilon}$ and as previously
$$ \hat{H}(T)\Psi(X,T) := -\frac{\alpha}{\beta}\triangle\Psi(X,T)+V(X,T)\Psi(X,T)$$
We can make the same route for $\varphi$ or just look at (\ref{tddis}) - now time derivative is with opposite sign and we should use transposed matrix instead. The matrix is real, so this transposition corresponds to conjugation of obtained Hamiltonian. It is usually self-adjoined ($\hat{H}^\dag=\hat{H}$), but let us look at the general situation:
$$ -\frac{1}{\beta} \frac{d}{dT} \Phi(X,T)+ E(t)\Phi(X,T)=\hat{H}^\dag(T)\Phi(X,T)$$
Finally we can write these equations for evolution probability densities for past and future half planes:
\be \frac{1}{\beta}\frac{d}{dt}\Phi=(E-\hat{H}^\dag)\Phi  \qquad\qquad\qquad
\frac{1}{\beta}\frac{d}{dt}\Psi=(\hat{H}-E)\Psi\ee

In quantum mechanics $\langle \psi|\psi \rangle=const$ for complex $\psi$ because "bra" rotates in one direction ($\langle\psi|\to e^{i t\hat{H}/\hbar} \langle\psi|$), while "ket" rotates in the other ($|\psi\rangle \to e^{-i t\hat{H}/\hbar} |\psi\rangle$). Here $\langle \Phi|\Psi\rangle=const$ for real positive $\Phi$, $\Psi$ because one drops exponentially while the other rises ($\langle\Phi|\to e^{-\beta t (\hat{H}-E)} \langle\Phi|,\ |\Psi\rangle\to e^{\beta t (\hat{H}-E)} |\Psi \rangle$).

As it was commented for the general time dependent case, the choice of $E$ does not affect probability density or propagator, so if we do not care that eigenvectors goes to zero and infinity, we could choose even $E=0$. For numerical simulations eigenvectors can be synchronized to $\Psi=\Phi$ for self-adjoined $H$ while locally stationary situations, by using $E$ as the ground state energy. As previously, we can also make adiabatic approximation - calculate $\Psi(t),\ \Phi(t)$ and $E(t)$ as the potential was not going to change:
$$E(t)=\langle \Phi(t)|\hat{H}\Psi(t)\rangle$$

Having $\Phi$ and $\Psi$ we can calculate the expected probability density:
\be \rho(x,t)=\Phi(x,t)\Psi(x,t)\qquad\qquad\textrm{if }\quad \int \Phi(x,t)\Psi(x,t)\ dx=1 \ee
Hamiltonians having different potentials generally do not commute, so formally like in quantum mechanics to calculate propagator there is required time-ordering operator ($\mathcal{T}$):
$$S^{T,S}(x,y)=\lim_{\delta\to 0} \int_{x_1 x_2 ..} S^{T,T+\delta}(x,x_1)S^{T+\delta,T+2\delta}(x_1,x_2)... dx_1 dx_2..=$$
$$=\frac{\lim_{\delta\to 0} \langle x|e^{-\beta\delta\hat{H}(T)}e^{-\beta\delta\hat{H}(T+\delta)}..e^{-\beta\delta\hat{H}(S-\delta)}|y\rangle}
{\mathcal{N}(T,S)}\frac{\Psi(y,S)}{\Psi(x,T)}
= \frac{\langle x|\mathcal{T}e^{-\beta\int_T^S \hat{H}(t)dt}|y\rangle}{\mathcal{N}(T,S)}\frac{\Psi(y,S)}{\Psi(x,T)}$$
where $\mathcal{N}$ can be obtained from $\int S^{T,S}(x,y) dy=1$ normalization condition.
\subsubsection{Probability current}
The probability current is much simpler to obtain:
$$\frac{d}{dt}\left(\Phi\Psi\right)=\beta\left(((E-\hat{H}^\dag)\Phi)\Psi+\Phi(\hat{H}-E)\Psi \right)=$$
$$=\alpha\left((\triangle \Phi)\Psi-\Phi(\triangle\Psi)\right)=
\alpha \nabla \cdot\left((\nabla  \Phi)\Psi-\Phi(\nabla \Psi)\right) $$
So probability current and continuity equation became:
\be J=\alpha \left(\Phi(\nabla \Psi)-(\nabla  \Phi)\Psi\right) \qquad\qquad\qquad   \frac{d}{dt}\rho=-\nabla \cdot J \label{flux}\ee

Let us find correspondence between this formula for real $\Phi$, $\Psi$ and its quantum mechanical analogue:
 $j=\frac{\hbar}{2mi}\left(\bar{\psi}\nabla \psi-\psi \nabla \bar{\psi}\right)$ for complex $\psi$. Substituting for example $\psi=\frac{e^{i\gamma}}{\sqrt{2}}(\Phi+i\Psi)$ for some phase $\gamma$, we get:
$$j=\frac{\hbar e^{i\gamma}e^{-i\gamma}}{4mi}\left((\Phi-i\Psi)\nabla (\Phi+i\Psi)-(\Phi+i\Psi)\nabla (\Phi-i\Psi)\right)=
\frac{\hbar}{2m}\left(\Phi\nabla \Psi-\Psi\nabla \Phi\right)$$
It is exactly the equation (\ref{flux}) obtained from MERW for $\alpha=\frac{\hbar}{2m}$. Generally above choice of $\psi$ changes probability density ($|\psi|^2\neq \Phi\Psi$), but we get equality if $\Psi=\Phi$, what can be obtained for self-adjoined $H$ when the system evolves so slowly, that we can assume constant equilibrium (adiabatic approximation).

\subsubsection{Ehrenfest equations}
Like in quantum mechanics, we can introduce operators acting on states in fixed time - we will look at their expected values here. This time they are not necessarily self-adjoined, so we need to clearly describe which side they apply to:
$$\langle \hat{O} \rangle :=\langle \Phi|\hat{O} \Psi \rangle = \int \Phi(x) (\hat{O} \Psi(x)) dx $$

Probability current allows to calculate time evolution of the expected position:
$$\frac{d}{dt}\int x\rho dx=\int x \frac{d\rho}{dt} dx =-\int x(\nabla \cdot J) dx=\int J dx=$$
$$=\alpha \int \Phi(\nabla \Psi)-(\nabla  \Phi)\Psi dx = 2\alpha \int \Phi(\nabla \Psi) dx = \frac{1}{m}\int \Phi (\hat{p} \Psi)=
\frac{\langle \hat{p}\rangle}{m}$$
where we have used partial integration twice, assuming vanishing at boundaries and
\be \hat{p} :=2m\alpha\nabla \qquad\qquad(=\hbar \nabla \quad \mathrm{for}\ \alpha=\frac{\hbar}{2m}) \ee
This time momentum operator is not self-adjoined, but antihermitean ($ \hat{p}^\dag =-\hat{p}$) - to apply it to $\Phi$ we would use $-\hbar \nabla$ instead. Intuitively $\Phi$ is the one evolving forward in time (stable) - the momentum operator can be imagined that the increase of density, brings thermodynamical flow in opposite direction to equilibrate densities.

For not self-adjoined $\hat{p}$, usually $\hat{p}^2$ also is not self-adjoined - we can repair it using $\hat{p}^\dag \hat{p}$ instead. Now the choice $\alpha=\frac{\hbar}{2m}$ as previously allows us to write our $H$ in more physical way:
\be \hat{H}=\frac{\hat{p}^\dag \hat{p}}{2m}+V \label{merwham}\ee
This choice of $\alpha$ means that
\be\beta=\frac{2m}{\hbar^2}\alpha=\frac{1}{\hbar}\qquad\qquad \textrm{for}\quad \alpha=\frac{\hbar}{2m} \label{aldef}\ee

Let us now find time evolution of the expected value of a general operator - analogue of Ehrenfest equation:
$$\frac{d}{dt} \langle \Phi|\hat{O}\Psi\rangle=\beta\langle \Phi|(E-\hat{H})\hat{O}\Psi\rangle+
\langle \Phi|\frac{\partial\hat{O}}{\partial t}\Psi\rangle+\beta\langle \Phi|\hat{O}(\hat{H}-E)\Psi\rangle$$
\be\frac{d}{dt} \langle\hat{O}\rangle =\left<\frac{\partial\hat{O}}{\partial t}\right>+\beta \left< [\hat{O},\hat{H}]\right> \ee
for example
\be [\hat{x},\hat{p}]=2m\alpha=\hbar \ee
$$[\hat{x},\hat{H}]=2\frac{\alpha}{\beta}\nabla\qquad\qquad \Rightarrow \qquad\qquad
\frac{d\langle\hat{x}\rangle}{dt}=\langle 2\alpha\nabla \rangle=\frac{\langle \hat{p} \rangle}{m}$$
$$[\hat{p},\hat{H}]=[\hbar \nabla,V]=\hbar \nabla V \qquad\qquad \Rightarrow$$
\be \frac{d}{dt} \langle\hat{p}\rangle = \beta \langle \hbar \nabla V \rangle = \langle  \nabla V \rangle = \int \rho(x) \nabla V(x) dx \ee
\begin{figure}[t]
    \centering
        \includegraphics{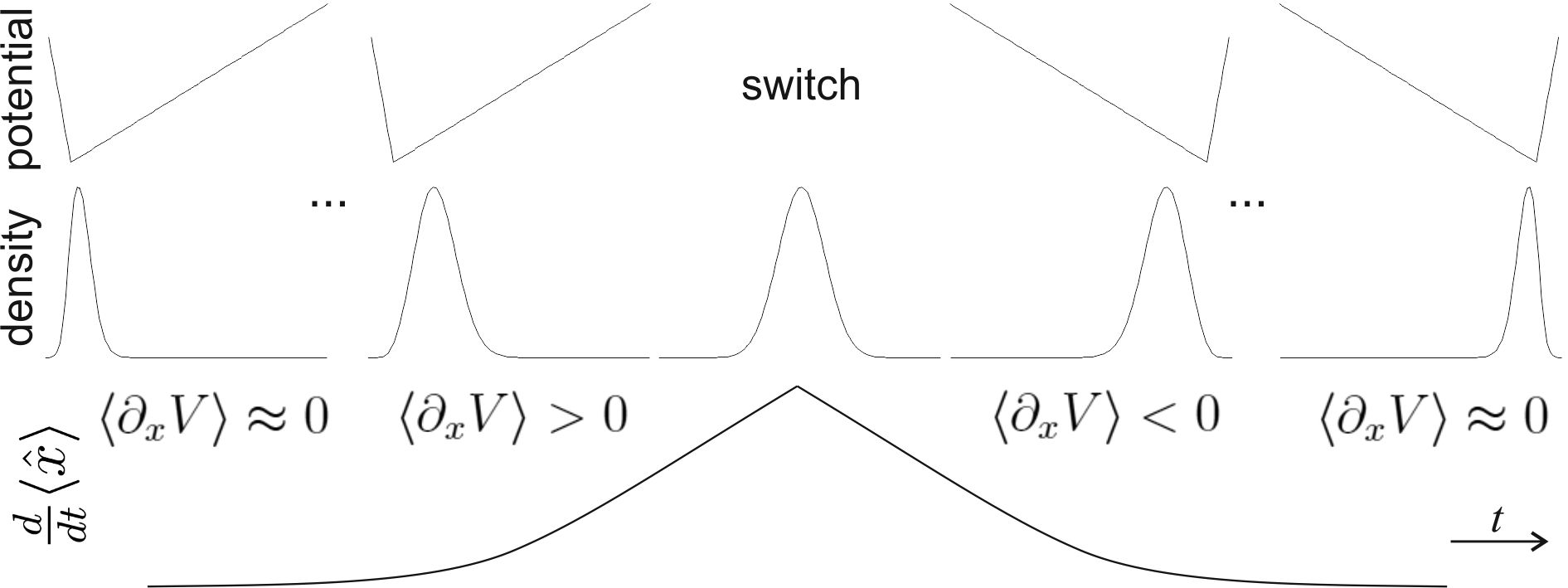}
        \caption{Simulation showing time-symmetry and that acceleration is opposite than in Newton's equations. While switching potential, probability density travels from thermodynamical equilibrium in the first well to the second one - first accelerating uphill the potential, then decelerating downhill. The acceleration grows with $\langle\nabla V\rangle$.}
        \label{ehr}
\end{figure}

Surprisingly, we get $$m\frac{d^2}{dt^2}\langle\hat{x}\rangle=\langle\nabla V \rangle$$ what is opposite acceleration than in Newton's law. Example in figure \ref{ehr} gives intuition that this looking contradictory result, is in fact expected. The system constantly searches for the thermodynamical equilibrium - while evolution is slow enough and the current entropy well remains optimal, probability density remains there. However, when a different local potential minimum starts dominating like in Fig. \ref{1dex}, probability density has to get out of the current potential well - accelerating uphill, then decelerating downhill to finally stop in the new optimal minimum.

This example also provides educative demonstration of time-symmetry of this thermodynamical model. Our best estimation of particle's probability density in plus or minus infinity is the ground state of corresponding potential well. The switch is symmetric, so our prediction of intermediate densities is also time-symmetric.
\subsubsection{Heisenberg uncertainty principle}
Having momentum operator, we can derive Heisenberg uncertainty principle analogue for adiabatic approximation ($\Phi=\Psi$). This time $\hat{p}$ is not self-adjoined, so $\langle \hat{p}^2 \rangle$ does not have to be nonnegative. Instead, like for Hamiltonian we need to use:
$$\langle \hat{p}^\dag\hat{p}\rangle=\langle \Psi |\hat{p}^\dag\hat{p}\Psi\rangle=\langle \hat{p}\Psi |\hat{p}\Psi\rangle$$
Now for any $\lambda$:
$$0\leq \langle (\hat{x}+\lambda\hat{p})\Psi|(\hat{x}+\lambda\hat{p})\Psi\rangle=
\langle\Psi|(\hat{x}-\lambda\hat{p})(\hat{x}+\lambda\hat{p})\Psi\rangle=
\langle \hat{x}^2\rangle+\lambda^2\langle \hat{p}^\dag\hat{p}\rangle-\lambda\hbar$$
This quadratic equation must have nonpositive discriminant, getting analogue of Heisenberg uncertainty principle:
\be \sqrt{\langle \hat{x}^2\rangle}\sqrt{\langle \hat{p}^\dag\hat{p}\rangle}\geq \frac{\hbar}{2} \ee

\section{Multiple particles}
In this section there will be discussed extensions of MERW methodology for multiple walkers on a single graph. It will be made in analogy to quantum mechanics, but these considerations are purely thermodynamical. The original graph of positions needs to be extended to graph of particle configurations on it - for distinction in this section we will call vertices of the original graph as \emph{nodes}, while of the extended graph as vertices.

\subsection{Noninteracting particles}
The stationary probability density from the previous sections was originally obtained as the best assumption we can make (maximizing uncertainty) for a single particle on the graph. It is also the dominant eigenvector of the stochastic matrix $S$, so in time-independent case it can be also seen as time average of particle's position over a long period.

Another general view is for multiple noninteracting particles - if they behave independently and each of them is expected to agree with this prediction, their actual density (current distribution among nodes) should asymptotically be near this expected probability density. More precisely, asymptotically this agreement improves exponentially with the number of particles and the coefficient is Kullback-Leibler distance between these distributions.

To see it, let us look at the simplest case of $n\in\mathbb{N}$ particles on 2 node graph: there is a random variable with binomial probability distribution $(p,1-p)$ and we are interested in asymptotic probability that if we use it $n$ times, the first possibility will happen $qn$ times ($p,q\in[0,1]$):
$${n \choose qn}p^{qn}(1-p)^{(1-q)n}\approx 2^{n(-q\lg(q)-(1-q)\lg(1-q)+q\lg(p)+(1-q)\lg(1-p))}=2^{-n\left(q\lg\left(\frac{q}{p}\right)+(1-q)\lg\left(\frac{1-q}{1-p}\right)\right)}$$
where we have used asymptotic approximation of binomial coefficients from subsection \ref{entrrw}. Straightforward generalization is: while using $n$ times probability distribution $(p_i)_i$, asymptotic probability of getting $(nq_i)_i$ distribution is:
\be \Pr(n,(q_i)||(p_i))=2^{-nD_{KL}((q_i)||(p_i))}\qquad\textrm{where}\qquad D_{KL}((q_i)||(p_i)):=\sum_i q_i \lg\left(\frac{q_i}{p_i}\right) \ee
is called Kullback-Leibler distance of these probability densities, but usually it is not symmetric. It is nonnegative and is zero only when $(p_i)$ and $(q_i)$ are equal - as expected, in $n\to\infty$ limit $(q_i)=(p_i)$ case completely dominates all the others.

Approximation that particles does not interact should be appropriate when their density is relatively low. In other case, there could be used some mean field approximation. For example: find the ground state probability density for single particle, assume such density of e.g. electrons to correspondingly modify the potential, then find the ground state for this modified potential and so on until stabilizing such iterative procedure.

\subsection{Fixed number of interacting particles}
Let us now extend the MERW methodology to directly work with fixed number of interacting walkers/particles.
\subsubsection{Distinguishable particles}
We will start these consideration with two distinguishable particles on the same graph, but it can be simply generalized for larger fixed amount. Position of such couple can be denoted as $(x_1,x_2)\in \mathcal{V}\times \mathcal{V}$. One way of choosing the adjacency matrix is that there is allowed transition from $(x_1,x_2)$ to $(y_1,y_2)$ vertices if and only if there is allowed transition from $x_1$ to $y_1$ and from $x_2$ to $y_2$ nodes. A different choice is used in Bose-Hubbard model: such transition is allowed only if a single particle jumps to its neighbor, while all the others remain in their positions.

\begin{figure}[t]
    \centering
        \includegraphics{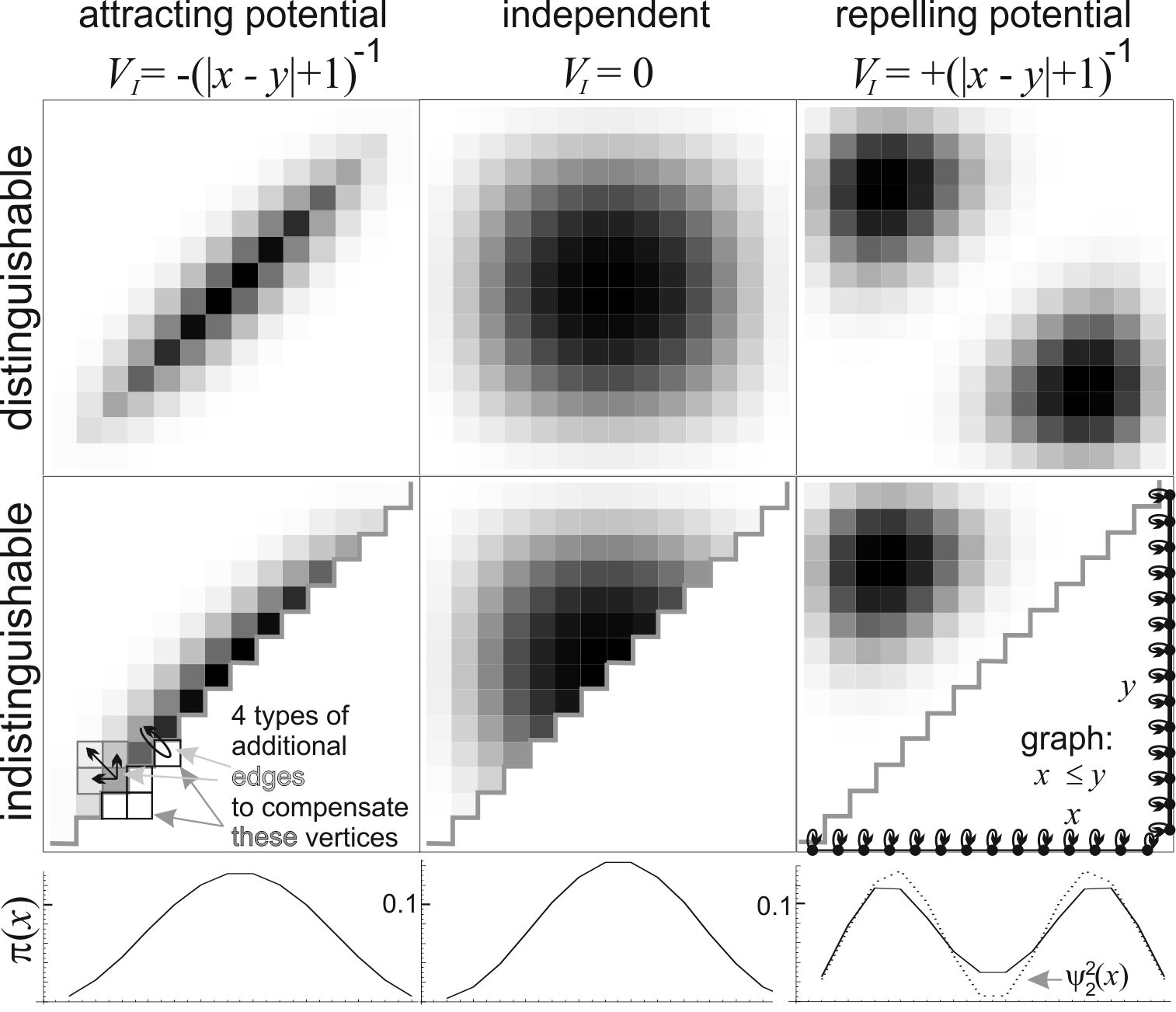}
        \caption{Example of stationary probability distributions for two particles on segment graph with self-loops ($\beta=1$). The bottom row shows $\pi(x)=\sum_y \pi(x,y)$ and its last graph has additional probability density for the second eigenvector to compare with Pauli exclusion principle.}
        \label{twopar}
\end{figure}

The next step is to assign energy to these transitions. If we would choose it as just the sum of energies of the two original edges, we would make these particles completely independent (assuming that simultaneously only one of them can make transition, would only change time scale). As in physics, to introduce interaction between them, we can use potential energy depending on both positions like Coulomb attraction/repulsion.
So finally the $\mathcal{M}$ matrix for this extended graph is not just the tensor product of two copies of the original matrix ($M\otimes M$), but it
additionally have term describing their interaction, like
\be \mathcal{M}_{(x_1,x_2),(y_1,y_2)}=e^{-\beta \left(\frac{V(x_1)+V(y_1)+V(x_2)+V(y_2)}{2}+V_I(x_1,x_2,y_1,y_2)\right)}\ee
where for example $V_I(x_1,x_2,y_1,y_2)\equiv\frac{V^0_I(x_1,y_1)+V^0_I(x_2,y_2)}{2}$ for some symmetric potential $V^0_I$.\\
Now random walk in such extended space ($\mathcal{V}\times \mathcal{V}$) corresponds to thermodynamics of coupled evolution of these two particles - the dominant eigenvector of $\mathcal{M}$ determines stationary probability in this extended space like in Fig. \ref{twopar}. If we are only interested in e.g. probability density of one of these particles, we can sum the joint probability density over some coordinates.

In continuous limit we would analogously use eigenfunction equation for $\Psi_{12}(x,y)$:
$$\hat{H} |\Psi_{12}\rangle = E|\Psi_{12}\rangle  \qquad\mathrm{where}\qquad
\hat{H}=\frac{\hat{p}_1^\dag\hat{p}_1}{2m}+\frac{\hat{p}_2^\dag\hat{p}_2}{2m}+V(x)+V(y)+V_I(x,y)$$
and lower index in momentum operators denotes variable it applies to. Like in quantum mechanics, for noninteracting particles the eigenfunction can be seen as tensor product ($\Psi_{12}(x,y)=\Psi_1(x)\Psi_2(y)$).

\subsubsection{Indistinguishable particles}
For particles we cannot distinguish, $(x,y)$ vertex is for us equivalent to $(y,x)$. It can be used to reduce considered number of vertices nearly twice or generally factorial of the number of particles times. For this purpose we would use only vertices of coordinates sorted in some linear order (like $x\leq y$) and identify them with all possible permutations. This reduction of the number of states in this effective model is not compulsory. Symmetrization is only abstract way for representing our information and it does not imply that underlying physics cannot distinguish particles having different positions.

In discrete case the number of corresponding permutations decreases when some coordinates equalize (e.g. $x=y$) - we need to be careful there. These vertices have neighbors of not ordered coordinates - the weights of edges to these vertices cannot just vanish. Instead these weights should be added to weights of edges to corresponding vertices of ordered coordinates, like in Fig. \ref{twopar}. Additionally, one need to remember that obtained probabilities of nondiagonal vertices correspond to both permutations ($n!$ generally) - to make that density is just restricted density of distinguishable case in Fig. \ref{twopar}, these diagonal probabilities were divided by 2.\\

The diagonal degenerates in continuous limit, so in this case we can neglect above complication. If Hamiltonian is invariant with respect to particle exchange, the fact that $\Psi_{12}(x,y)$  is the dominant eigenfunction (minimizing energy), implies that $\Psi_{12}(y,x)$ also fulfills eigenfunction equation to the same eigenvalue. This eigenfunction is unique, so $\Psi_{12}$ is symmetric $(\Psi_{12}(x,y)=\Psi_{12}(y,x))$. This symmetry is characteristic for bosons in quantum mechanics. Positivity of $\Psi$ coordinates in MERW does not allow for direct antisymmetrization required for fermion formalism, but we could for example artificially remove vertices with multiple particles in the same position (the diagonal for two particles) and edges to them.

However, even without artificially including some Pauli exclusion principle, the repulsion makes that the dynamical equilibrium state has already anti-correlated positions of these two particles. It applies to both MERW approach and two particle quantum wavefunction ($\psi(x,y)$) - it is usually approximated by tensor product of single particle states, but in fact the repulsive potential makes that the complete two particle wavefunction avoids configurations with e.g. electrons being close to each other (barriers of six dimensional potential) - probability density should be much larger when they are on opposite sides of the nucleus. Placing three electrons this way seems to be much less stable, especially remembering about magnetic moments - suggesting that even without Pauli principle, one of them should search for a different orbital.
\subsection{Harmonic oscillator and creation/annihilation operators}
Let us briefly look at the standard harmonic oscillator example - one dimensional continuous model with $V(x)=\frac{1}{2}m\omega^2 x^2$ potential for some angular frequency $\omega$. Its simplicity, existence of analytic solutions and equidistance between energy levels made it standard way to introduce multiple particle formalism, usually called second quantization. For example for lattice we could model its nodes as approximated by such potential. Like in quantum mechanics, in MERW formalism the harmonic oscillator Hamiltonian can be decomposed into first order annihilation($\hat{a}_i$)/creation($\hat{a}^\dag_i$) operators:
$$\hat{H}=\frac{p^\dag p}{2m}+\frac{1}{2}m\omega^2\hat{x}^2=\frac{1}{2}\hbar\omega\left(\hat{a}^\dag \hat{a}+\hat{a}\hat{a}^\dag \right)=\hbar\omega\left(\hat{a}^\dag \hat{a} +\frac{1}{2}\right)$$
$$\textrm{for real}\qquad \hat{a}=\sqrt{\frac{m\omega}{2\hbar}}\left(\hat{x}+\frac{1}{m\omega}\hat{p}\right)\qquad\qquad
\hat{a}^\dag=\sqrt{\frac{m\omega}{2\hbar}}\left(\hat{x}-\frac{1}{m\omega}\hat{p}\right)$$
because in our case $\hat{p}=\hbar\nabla=-\hat{p}^\dag$. \\
Eigenfunctions of this Hamiltonian are known:
$$ \langle x|n\rangle=\frac{\pi^{-1/4}}{\sqrt{2^n n!}}e^{-x^2/2}\mathcal{H}_n(x) \qquad\qquad
\hat{a}|n \rangle=\sqrt{n}\ |n-1\rangle\qquad\qquad \hat{a}^\dag|n-1\rangle=\sqrt{n}\ |n\rangle $$
where $\mathcal{H}_n$ is $n$-th Hermite polynomial. This time for single particle only the lowest state is thermodynamically stable ($|0\rangle$) - excited states would deexcite to this ground state. However, when there are multiple repelling particles, they should choose distant thermodynamical equilibriums. For example figure \ref{twopar} suggests that for two repelling particles, probability density for each of them is similar to of the second lowest energetic eigenstate ($|1\rangle$). However, we should have in mind that quantum or MERW amplitudes for interacting particles are a bit different than for noninteracting ones.

Annihilation/creation operators and $|n\rangle$ states can be also used as purely abstract way to work on multiple particle states/nodes. So let us disregard underlying physics for now and find the universal combinatorial coefficients by considering a graph of single node which can contain various number of particles ($n$). For clearer picture, let us assume for a moment that they are distinguishable - transition to from $n$ to $n-1$ particle state can be made by removing one of $n$ particles - in $n$ ways. Transforming back one of these subsets into the original $n$ particle state can be made in only one way:
$$\hat{a}|\overline{n}\rangle=n|\overline{n-1}\rangle\qquad\qquad \hat{a}^\dag|\overline{n-1}\rangle=|\overline{n}\rangle$$
where $|\overline{n}\rangle$ is nonstandard normalization of $n$ particle vertex - this time it represents the sum of $n!$ permutated states. However, it already leads to the proper commutation relations:
\be \hat{n}|\overline{n}\rangle :=\hat{a}^\dag \hat{a} |\overline{n}\rangle\ = n |\overline{n}\rangle\qquad\qquad \hat{a}\hat{a}^\dag |\overline{n}\rangle\ = (n+1) |\overline{n}\rangle\qquad\qquad
[\hat{a},\hat{a}^\dag]=1 \ee
Standard normalization of these states is $\frac{1}{\sqrt{n!}}$ times the sum of $n!$ permutation states: $|\overline{n}\rangle=\sqrt{n!}|n\rangle$, leading to relations used in quantum mechanics:
\be \hat{a}|n \rangle=\sqrt{n}\ |n-1\rangle\qquad\qquad \hat{a}^\dag|n-1\rangle=\sqrt{n}\ |n\rangle \qquad\qquad (\hat{a}^\dag)^n|0\rangle=\frac{1}{\sqrt{n!}}|n\rangle\ee

\subsection{Various number of particles and Bose-Hubbard model}
To consider $n$ particles on graph $(\mathcal{V},\mathcal{E})$, there can be used $\mathcal{V}^n$ extended graph with correspondingly defined transitions. To include various number of particles, the final graph becomes union of such graphs, between which we need to define allowed transitions (e.g. using $\hat{a}^\dag/\hat{a}$ formalism). Then the potential allows to choose new $M$ matrix for such extended graph. Additionally, there could be required chemical potential (or rest mass in QFT) to include the change of energy of system while the number of particles increases. Finally, for example the dominant eigenvector of $M$ determines stationary probability distribution among all vertices of this union of $\mathcal{V}^n$ .
\begin{figure}[t]
    \centering
        \includegraphics{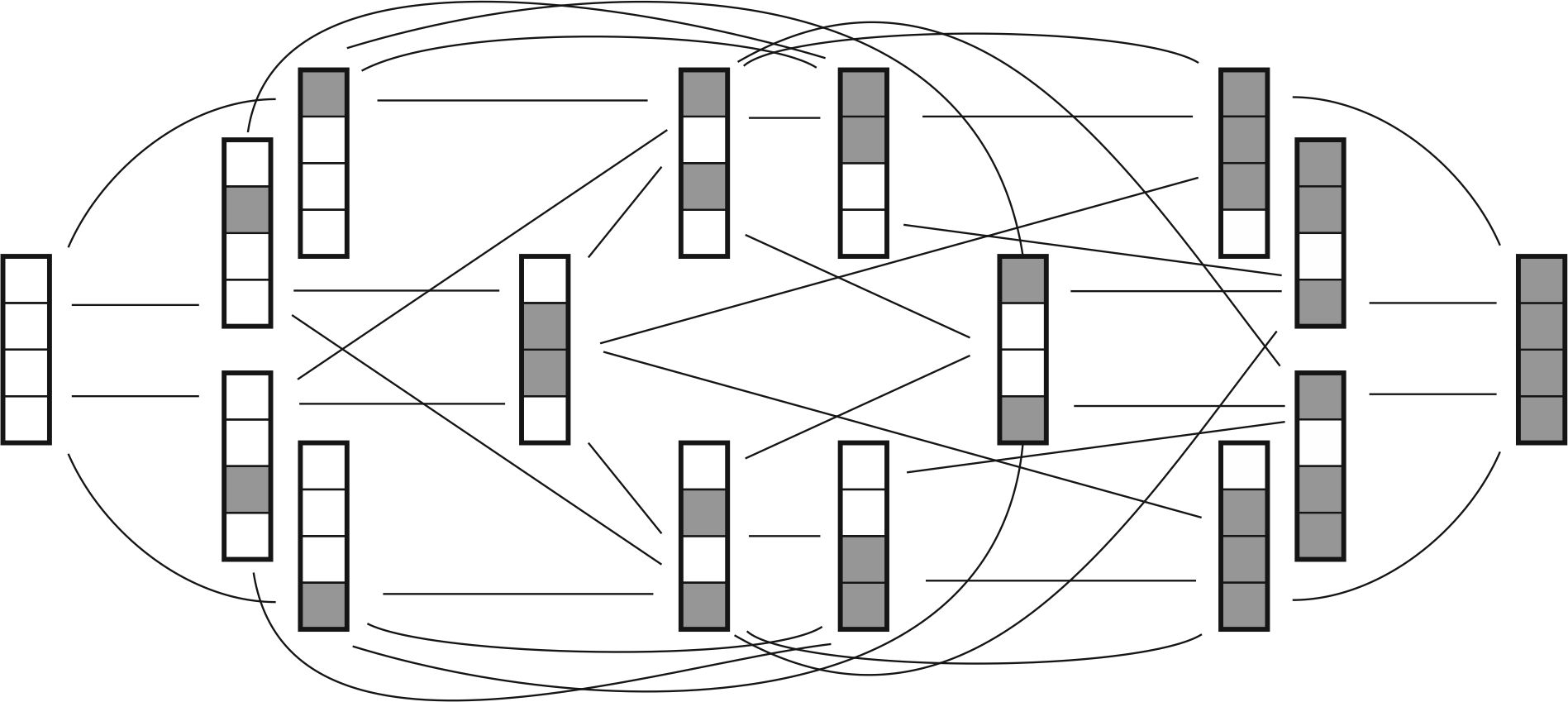}
        \caption{Example of extended graph for fermions on 4 node segment-like graph, such that in one step single particle can move by one position or be annihilated/created. Each of $2^4=16$ vertices represent situation in some moment, stationary probability distribution is among these 16 vertices.}
        \label{vpar}
\end{figure}

For indistinguishable particles there appears more convenient way of representing multiple particle states: by storing only the number of particles in each vertex - vertices of such extended graphs are functions from $\mathcal{V}$ to the set of possible number of particles in a single vertex. For fermions there should be at most one ($\mathcal{V}\rightarrow \{0,1\}$), for bosons there can be any natural number of particles in a node ($\mathcal{V}\rightarrow \mathbb{N}$). The boson picture is also used with repulsive interaction to approximate fermion behavior.

We are now prepared to look at the Bose-Hubbard model \cite{hubb} proposed and used for example for solid state physics or optical lattices. The main interest is the only stable: ground state to which other would deexcite. The simplest used Hamiltonian for $(\mathcal{V},\mathcal{E})$ underlying graph (usually a regular lattice) is:
\be \hat{H}_{BH}:=-t\sum_{(i,j)\in \mathcal{E}}\hat{a}_i^\dag\hat{a}_j+
\frac{U}{2}\sum_{i\in \mathcal{V}} \hat{n}_i\left(\hat{n}_i-1\right) \ee
Model parameter $t$ chooses transition probability (the larger $t$, the less important are other terms), $U$ is additional energy for two particles being in a single node. There is usually also added Hermitian conjugate of the first term - it can be made by just using indirected graph ($(i,j)\in \mathcal{E} \Rightarrow (j,i)\in \mathcal{E}$), so presented form is more general. This first term defines adjacency matrix of the extended graph of possible situations - there are allowed transitions shifting single particle to a neighboring node.

The second term introduces local repulsion for this bosonic model (used as approximation for fermions) - that there is required additional energy to place more than one particle in a single node. There is often added chemical potential term to compensate for various number of particles, but the fact that the number of creation and annihilation operators in each term is equal, makes that the number of particles is constant in this case. This model can be extended for example by including local energy of nodes ($V(i)$) or interaction between particles ($V_I(i,j)$):
$$\sum_{i\in \mathcal{V}}V(i) \hat{n}_i + \sum_{i,j\in \mathcal{V}}V_I (i,j) \hat{n}_i \hat{n}_j$$

Let us now look at the situation from MERW perspective. If there is only a single particle, the graph of possibilities is exactly the original graph $(\mathcal{V},\mathcal{E})$. In such case interactions vanish and the original Bose-Hubbard Hamiltonian becomes equivalent to $-tM$, where $M$ is standard adjacency matrix used in the basic MERW and this purely thermodynamical model also predicts the statistical equilibrium to be as for the quantum ground state of the Hamiltonian. The other terms do not change the vertex - their direct interpretation is self-loop like in subsection \ref{exsect}: that using only transition terms costs zero energy, but each time the particle would like to stay in a vertex, the system would have to pay in energy.

MERW approach suggests to realize these terms in a bit different way. Specifically, Boltzmann distribution among paths requires multiplying Hamiltonian transition terms by $e^{-\epsilon\beta V}$, where $\epsilon$ is time step and $V$ generally may depend on the whole configuration of particles before and after the transition, for example expressed using $\hat{n}$ operators like interactions in the Bose-Hubbard model. In analogy to making continuous limit of lattice in subsection \ref{bolpat}, for small lattice constant we can approximate it in $\epsilon$ order to Bose-Hubbard Hamiltonian:
 $$ \hat{H}_{MERW}\propto - \sum_{(i,j)\in \mathcal{E}}\hat{a}_i^\dag\hat{a}_j e^{-\epsilon\beta V(\textrm{configuration before and after transition})}\approx$$
 $$\approx- \sum_{(i,j)\in \mathcal{E}}\hat{a}_i^\dag\hat{a}_j + \epsilon\beta d\sum_{i\in \mathcal{V}} V(\textrm{configuration after transition})\ \hat{a}_i^\dag\hat{a}_i$$
 where $d$ is constant for lattice degree of vertices. The first approximation is using only linear expansion of exponent $e^{-\epsilon\beta V}\approx 1-\epsilon\beta V$, the second that $V$ is smooth - has practically the same values in neighboring vertices. The third approximation is that the dominant eigenvector is nearly constant in neighboring vertices - it becomes more complicated now: requires modification of indexes of annihilation operators ($\hat{a}_i^\dag\hat{a}_j\approx \hat{a}_i^\dag\hat{a}_i$), but it should not essentially change the dominant eigenvector we are interested in. By choosing $V$ we can now get Bose-Hubbard Hamiltonian as required.

To summarize, MERW approach leads to Hamiltonian which is practically equivalent to Bose-Hubbard Hamiltonian if there is no potential/interaction or in continuous limit. In general case they only approximate each other. Bose-Hubbard Hamiltonian seems to be introduced by direct analogy to continuous case, while for MERW it is derived from model which mathematically is nearly Feynman path integrals in imaginary time. Obtained difference suggests to investigate the process of translation continuous models to discrete cases.\\

The next step should be making infinitesimal limit for varying number of particles to get thermodynamical analogue of Quantum Field Theories. There appears technical difficulties for direct approach, like that eigenstates of thermodynamical analogue of momentum operator ($\hbar\nabla$) are exponentially growing or vanishing - are very different from plane waves, which for MERW are unavailable ($\Psi\geq 0$). It suggests using Laplace transform instead of Fourier transform momentum space, but it seems problematic. Future development of such MERW expansions could bring some better intuition about many problematic infinities appearing in QFT. Anyway, from mathematical point of view, the MERW approach is very similar to using quantum mechanics in imaginary time, which is popular while finding the ground state. For example the expansion into Feynman graphs expressed by a sequence of annihilation/creation operators, in imaginary time becomes canonical ensemble of situations described as graph and its parameters - Boltzmann distribution among four-dimensional scenarios what MERW is constructed on.

\section{Conclusions and further perspectives}
There was introduced and discussed basic formalism, properties and intuitive examples of MERW-based thermodynamical modeling. These thermodynamical motion models still require a lot of work to develop it into a mature complementation of standard approaches, but already seems to provide explanation why in some situations standard thermodynamics disagree with experimental and quantum predictions, suggesting the way for correction. Standard way of improving the inconsistencies of stochastic models is by introducing some anomalous diffusion, but usually without proposing an explaining mechanism. Explanation of the presented approach is that the standard models only seem to fulfill thermodynamical principles, while in fact they are often biased: against maximal uncertainty principle, they unknowingly emphasize some possible scenarios without a base for such assumption. Equivalently, instead of using ensembles of static scenarios, for agreement with thermodynamical predictions of quantum mechanics we should consider ensembles of dynamical ones: trajectories, histories.

For MERW Hamiltonian to became the Schr\"{o}dinger operator there was not required fixed noise level, but similarity to quantum formalism for time dependent case suggested to choose $\beta=1/\hbar$. The main source of this fundamental noise seems to be the wave nature of particles, for example as a result of de Broglie's internal clock. This thermodynamical model completely ignores other effects related to the wave nature, which is generally seen as the essence of quantum mechanics, required for interference or orbit quantization. However, surprisingly there already appears the structure of eigenstates of Hamiltonian, but in a different way: as temporary attractors for probability density, which might occur permanent if lower eigenstates are somehow restricted, for example by being occupied by repelling particles. Deep understanding of such thermodynamical analogue of Pauli exclusion principle is one of the most important further line of work. Another essential property ignored by this thermodynamical approach is energy conservation, which while physical deexcitation results in photon production. These lacks could be reduced while considering not nowhere differentable trajectories (diffusive) like presented, but more physical ones - smooth and being perturbations of classical trajectories. Different possible lines of development is for example the infinitesimal limit of Bose-Hubbard-like model to get a better understanding of e.g. problematic infinities of Quantum Field Theories.\\

Richard Feynaman has written that "I think I can safely say that nobody understands quantum mechanics". While recent experiments of Couder (\cite{cou1}) showed that double-slit interference can be observed and understood for classical macroscopic objects having wave-particle duality, the surprising agreement of presented approach with thermodynamical predictions of quantum mechanics seems to complement the picture. There appears new hope to make quantum mechanics not only a tool for calculations, but also a theory with deeply understood foundations. The Schroedinger's cat thought experiment already suggests that quantum mechanics is a theory representing information of subjective observer - while for an external observer the cat may be in quantum superposition, a different observer inside the box may finally get out and show video recording of what has objectively really happened inside the box. The interpretation of de Broglie's and the MERW-based approach suggest that we can see quantum mechanics as not fundamental assumed theory, but emergent effective one - a result of wave nature of particles and thermodynamics representing the most probable behavior for our limited knowledge. While the module of wavefunction describes probability, its argument describes expected relative phase of such particle's internal clock.

While Schr\"{o}dinger equation focuses on wave nature ignoring the corpuscular one, presented approach do exactly opposite - there would be useful to construct some joining them model. For this purpose there is required a model with energy conservation, in which there can naturally appear various number of local particle-like constructs, constantly creating waves of the surrounding field like Couder's walking droplets. Their wave nature would lead to effects typical for quantum mechanics like interference, while MERW-based thermodynamical view suggests additional stochastic shift toward probability densities of quantum eigenstates. Particle's quantum numbers: more or less conserved properties being integer multiplicities, suggests where to search for the corresponding mathematical constructions: topological charges like winding number are also integer multiplicities and are restricted by corresponding conservation laws. Such topological solitons can be created/annihilated in pairs of opposite "charges", there can appear attraction/repulsion for opposite/the same topological charge, they have stored some characteristic minimal rest energy (mass) required to glue such nontrivial boundary conditions, some of them like so called breathers may have the "clock": internal periodic process creating waves around and so on. Skyrme (\cite{sky}) has made popular using topological solitons as effective models of hadrons, but for example Couder's experiments and presented thermodynamical approach suggest to try to use solitons as constructs for all particle. For example there is being developed such electron model (Faber \cite{faber}) in which there naturally appears electromagnetic interaction, but this approach does not include the wave nature. Recently there was introduced model which can be seen as its expansion by a single degree of freedom interpreted as quantum phase, which adds the possibility of internal clock (ellipsoid field \cite{me1}). Its family of topological solitons also grows to became qualitatively similar to known from physics: with three families of solitons resembling correspondingly neutrinos, leptons, mesons, baryons, which can finally combine into nucleus-like structures. Qualitatively it has surprising agreement with expected in physics quantum numbers, interactions, decays and mass hierarchy. To test this looking promising approach, there will be now required a lot of more quantitative considerations.\\

I would like to thank Andrzej Horzela for discussion and help in proofreading of this paper.

\end{document}